\documentclass{aa} 

\usepackage{graphicx}
\usepackage{txfonts}
\usepackage{caption}
\usepackage{url}
\usepackage{natbib}
\usepackage{breqn}
\bibpunct{(}{)}{;}{a}{}{,} 
\usepackage[breaklinks, colorlinks, urlcolor=urlcolor,
linkcolor=linkcolor,citecolor=citecolor,pdfencoding=auto,draft]{hyperref}

\newcommand{\aeff}{a}
\newcommand{\amin}{a_{\mathrm{min}}}
\newcommand{\amax}{a_{\mathrm{max}}}
\newcommand{\ngas}{n_{\mathrm{gas}}}

\newcommand{\nidust}{n_{i,\mathrm{dust}}}
\newcommand{\Tgas}{T_{\mathrm{gas}}}
\newcommand{\Tdust}{T_{\mathrm{dust}}}
\newcommand{\Tidust}{T_{i,\mathrm{dust}}}

\newcommand{\NH}{N_{\mathrm{H}}}

\newcommand{\aalig}{a_{\mathrm{alig}}}
\newcommand{\alarm}{a_{\mathrm{larm}}}
\newcommand{\ud}{\mathrm{d}}
\newcommand{\urad}{u_{\mathrm{rad}}}
\newcommand{\kI}{k_{\mathrm{I}}}
\newcommand{\kQ}{k_{\mathrm{Q}}}
\newcommand{\kV}{k_{\mathrm{V}}}
\newcommand{\Gzero}{G_{\mathrm{0}}}
\newcommand{\Planck}{{\it Planck}}
\renewcommand{\S}{\mathcal{S}}
\newcommand{\PsI}{p}
\newcommand{\StimesPsI}{\S\times\PsI}
\def\POLARIS{\texttt{POLARIS}}
\def\Qgammazeta{Q_{\Gamma}^{\rm ref}}
\def\Qgammaslope{\alpha_Q}
\def\DDSCAT{\texttt{DDSCAT}}
\def\RAMSES{\texttt{RAMSES}}
\def\spinup{\Qgammazeta\,\langle\gamma\rangle\,\langle\cos{\vartheta}\rangle\,U_{\rm rad}/(n_{\rm gas} T_{\rm gas})}
\def\eg{\textit{e.g.}}

\usepackage{ulem}

\begin{document} 

\title{A systematic study of radiative torque grain alignment in the diffuse interstellar medium }
\author{Stefan Reissl\inst{\ref{inst1}}, Vincent Guillet \inst{\ref{inst2},\ref{inst3}}, Robert Brauer\inst{\ref{inst4}}, François Levrier
\inst{\ref{inst5}}, François Boulanger \inst{\ref{inst5}}, \\\and Ralf S. Klessen\inst{\ref{inst1},\ref{inst6}}}
\institute{
\centering \label{inst1} Universität Heidelberg, Zentrum für Astronomie, Institut für Theoretische Astrophysik, Albert-Ueberle-Str. 2, D-69120 Heidelberg, Germany
\and
\centering \label{inst2} Institut d'Astrophysique Spatiale, CNRS, Univ. Paris-Sud, Universit\'{e} Paris-Saclay, B\^{a}t. 121, 91405 Orsay cedex, France
\and
\centering \label{inst3} Laboratoire Univers et Particules de Montpellier, Universit{\'e} de Montpellier, CNRS/IN2P3, CC 72, Place Eug{\`e}ne Bataillon, 34095 Montpellier Cedex 5, France
\and
\centering \label{inst4} CEA Saclay - DRF/IRFU/SAp, Orme des Merisiers, Ba\^{a}t 709, 91191 Gif sur Yvette, France
\and
\centering \label{inst5} Laboratoire de Physique de l'ENS, ENS, Université PSL, CNRS, Sorbonne Université, Université Paris-Diderot, Paris, France
\and
\centering \label{inst6}Universit{\"a}t Heidelberg, Interdisziplin{\"a}res Zentrum f{\"u}r Wissenschaftliches Rechnen, Im Neuenheimer Feld 205, 69120 Heidelberg, Germany
}

\abstract
   {
   The analysis of {\it Planck} data has demonstrated that the grain alignment efficiency is almost constant in the diffuse and translucent ISM.
   }
{
We test if the Radiative Torque (RAT) theory is compatible with these new observational constraints on grain alignment.
}
   {
   We combine a numerical magnetohydrodynamical (MHD) simulation with the state-of-the-art radiative transfer (RT) post-processing code $\POLARIS$ that incorporates a physical dust model and the detailed physics of grain alignment by RATs. A dust model based on two distinct power-law size distributions of spherical graphite grains and oblate silicate grains is designed to reproduce the mean spectral dependence of extinction and polarization observed in the diffuse ISM. From a simulation of interstellar turbulence obtained with the adaptive-mesh-refinement code $\RAMSES$, we extract a data cube with physical conditions representative of the diffuse ISM. We post-process the $\RAMSES$ cube with $\POLARIS$ to compute the grain temperature and alignment efficiency in each cell of the cube. Finally, we simulate synthetic dust emission and polarization observations. 
  }
  { 
  In our simulation the grain alignment efficiency is well correlated with the gas pressure, but not with the radiative torque intensity. Because of the low dust extinction in our simulation, the magnitude of the radiative torque varies little, decreasing only for column densities larger than $10^{22}\,$cm$^{-2}$. Comparing our synthetic maps with those obtained assuming a uniform alignment efficiency, we find no systematic difference and very small random differences. The dependencies of the polarization fraction $p$ with the column density $\NH$ or with the dispersion in polarization angle $\S$ are also similar in both cases. The drop of grain alignment produced by the RAT model in the denser cells of the data cube  does not significantly affect the patterns of the synthetic polarization maps, the polarization signal being dominated by the line-of-sight and beam integration of the geometry of the magnetic field. 
 If a star is artificially inserted at the center of the simulation, the polarization fraction is increased everywhere, with no specific pattern around the star. The angle-dependence of the RAT efficiency is not observed in simulated maps, and only very weakly in the optimal configuration where the magnetic field is artificially set to a uniform configuration in the plane of the sky.
   }
   {
   The RAT alignment theory is found to be compatible with the {\Planck} polarization data for the diffuse and translucent ISM in the sense that both uniform alignment and RAT alignment lead to very similar simulated maps. To further test the predictions of the RAT theory in an environment where an important drop of grain alignment is expected, high-resolution polarization observations of dense regions must be confronted to numerical simulations sampling high column densities ($\NH > 10^{22}\,$cm$^{-3}$) through dense clouds, with enough statistics.
   
  }
 {}
  \titlerunning{A systematic study of radiative torque grain alignment}
  \authorrunning{Reissl et al.}
  \maketitle
%

\section{Introduction}
Polarization of starlight and of dust thermal emission are commonly used as observational tracers of interstellar magnetic field orientation, within the Milky Way as well as in external galaxies~\citep[see e.g.][]{Hiltner1949,Chapman2011,Sadavoy2018,Planck2018XII,Lopez-Rodriguez2019}. This polarization is produced by the dichroism of the solid phase of the interstellar medium (ISM), composed of elongated dust grains that are spinning and precessing around the local magnetic field.

Different mechanisms have been proposed to explain how the spin axis of dust grains can become aligned with interstellar magnetic fields, overcoming the random torques produced by impinging gas particles, which tend to disalign them. Shortly after the discovery of starlight polarization \citep{Hall1949,Hiltner1949}, grain alignment was proposed to result from  magnetic relaxation~\citep[][DG hereafter]{Davis1951}. 

The interstellar magnetic field strength is however too low for the DG mechanism to work in the diffuse ISM.
Furthermore, grain alignment by magnetic relaxation works like a heat engine, which requires a temperature difference between gas and dust.
It must fail in dense cores, where $\Tdust \approx \Tgas$. 
Hence, the DG mechanism cannot account for the observed level of dust polarization on lines of sight (LOS) passing through dense molecular regions. 
\cite{JonesSpitzer1967} 
demonstrated that these limitations of the DG mechanism could be overcome if grains had the superparamagnetic properties that the presence of ferromagnetic inclusions in the grain matrix provides. \cite{Purcell1979} found that the formation of molecular hydrogen on the grain surface might spin-up the grain to suprathermal velocities, allowing for grain alignment even though $\Tdust \approx \Tgas$.
 
The radiative torques (RATs) exerted onto grains by the absorption and scattering of photons can also spin-up and align grains with the magnetic field, provided that grains have a certain asymmetry called helicity \citep{Dolginov1976,Draine_Weingartner1996,DraineWeingartner1997}. Through a number of papers, a model of grain alignment by RATs was constructed \citep{LazarianHoang2007,Lazarian2008,Hoang2008,Hoang2014,HoangLazarian2016,LH18}, opening the path to quantitative comparisons with observations \citep{Bethell2007,Seifried2019}. 
 
A number of studies have looked for the distinctive signatures of the RAT mechanism in polarization observations. The observed variations of the polarization fraction in the optical or in the submillimeter are found to be in qualitative agreement with what is expected from the RAT theory: a strong drop in starless cores \citep{Alves2014,Jones2015}, an increase with the radiation field intensity in dense clouds with embedded YSOs \citep{Whittet2008} or around a star \citep{Andersson2011}, a modulation by the angle between the magnetic field and the direction of anisotropy of the radiation field \citep{Andersson2010,VA15}, or a correlation with the wavelength $\lambda_{\rm max}$ where starlight polarization peaks \citep{AnderssonPotter2007}. For a review of observational constraints favouring grain alignment by RATs, see~\cite{Andersson2015}. On the contrary, studies where the polarization fraction was corrected for the effect of the magnetic field before the analysis do not find any drop in the grain alignment efficiency, whether in the diffuse and translucent ISM \citep{Planck2018XII} or in dense cores \citep{Kandori2018,Kandori2020}. Clearly, more work is needed to solve this discrepancy and reach conclusions that are statistically significant.

The purpose of this paper is 
to confront the predictions of the RAT theory to observations in a quantitative way through synthetic dust polarized emission maps built from a magnetohydrodynamics (MHD) simulation of interstellar turbulence with state-of-the-art grain alignment physics and an accurate treatment of radiative transfer. 
In our new modelling we post-process the MHD simulation of ~\cite{hennebelle_08} used in \cite{Planck2015XX} with the radiative transfer (RT) code \POLARIS\footnote{\tt http://www1.astrophysik.uni-kiel.de/${\sim}$polaris/}~\citep{Reissl2016}, using a physical dust model designed to reproduce the mean extinction and polarization curves observed in the diffuse ISM. The \POLARIS\ tool, which incorporates the detailed physics of the RAT alignment theor, was also applied to predict line emission including the Zeeman effect \citep[][]{Brauer2017A,Brauer2017B,Reissl2018,Pellegrini2019} as well as galactic radio observations \citep{Reissl2019}. Contrary to other dust emission codes, \POLARIS\ is a full Monte-Carlo dust heating and polarization code solving the RT problem in the Stokes vector formalism for dichroic extinction and thermal re-emission by dust, simultaneously. Furthermore, \POLARIS\ keeps track of each of the photon packages in order to simulate the radiation field in complex environments, allowing for the determination of the parameters required by the grain alignment physics. In essence, this paper is a follow-up of \citet{Planck2015XX} in which the modelling was done within the simplifying assumption of uniform grain alignment, and of \cite{Seifried2019} where grain alignment was properly computed with \POLARIS\ but lacked a well-constrained dust model.


In this article, we use the numerical model of the RAT theory outlined by \cite{Hoang2014} to estimate the relative importance of the radiation field properties and of the gas pressure in establishing the level of grain alignment under physical conditions representative of the diffuse and translucent ISM. The alignment of dust grains with the magnetic field by mechanical torques \citep[MAT][]{Lazarian2007C,Das2016,Hoang2018A} is also of great interest for our purpose. However, MAT is not yet a predictive theory like RAT is, and cannot therefore be part of our modeling. Still, we will discuss some   implications of the possible grain alignment by mechanical torques. 

The paper is structured as follows. In Section~\ref{sect:MHDRT}, we introduce the MHD simulation used in this study and the radiative transfer that is applied to it. The modelling of dust  is described in Section~\ref{sect:DustModel} and that of grain alignment in Section~\ref{sect:RATAlignment}.
The output data cubes and maps from the \POLARIS\ modelling are presented in Sects.~\ref{sec:ISRF} and ~\ref{sec:RT_STAR} for two different setups of the radiation field.  Our results are discussed in Section~\ref{sec:discussion} and summarized in Section~\ref{sect:summary}.

\section{MHD simulations and radiative transfer}
\label{sect:MHDRT}
\subsection{The $\RAMSES$ MHD simulation}
\label{sect:MHDSimulation}
\begin{figure*}
\begin{center}
\includegraphics[width=.36\textwidth]{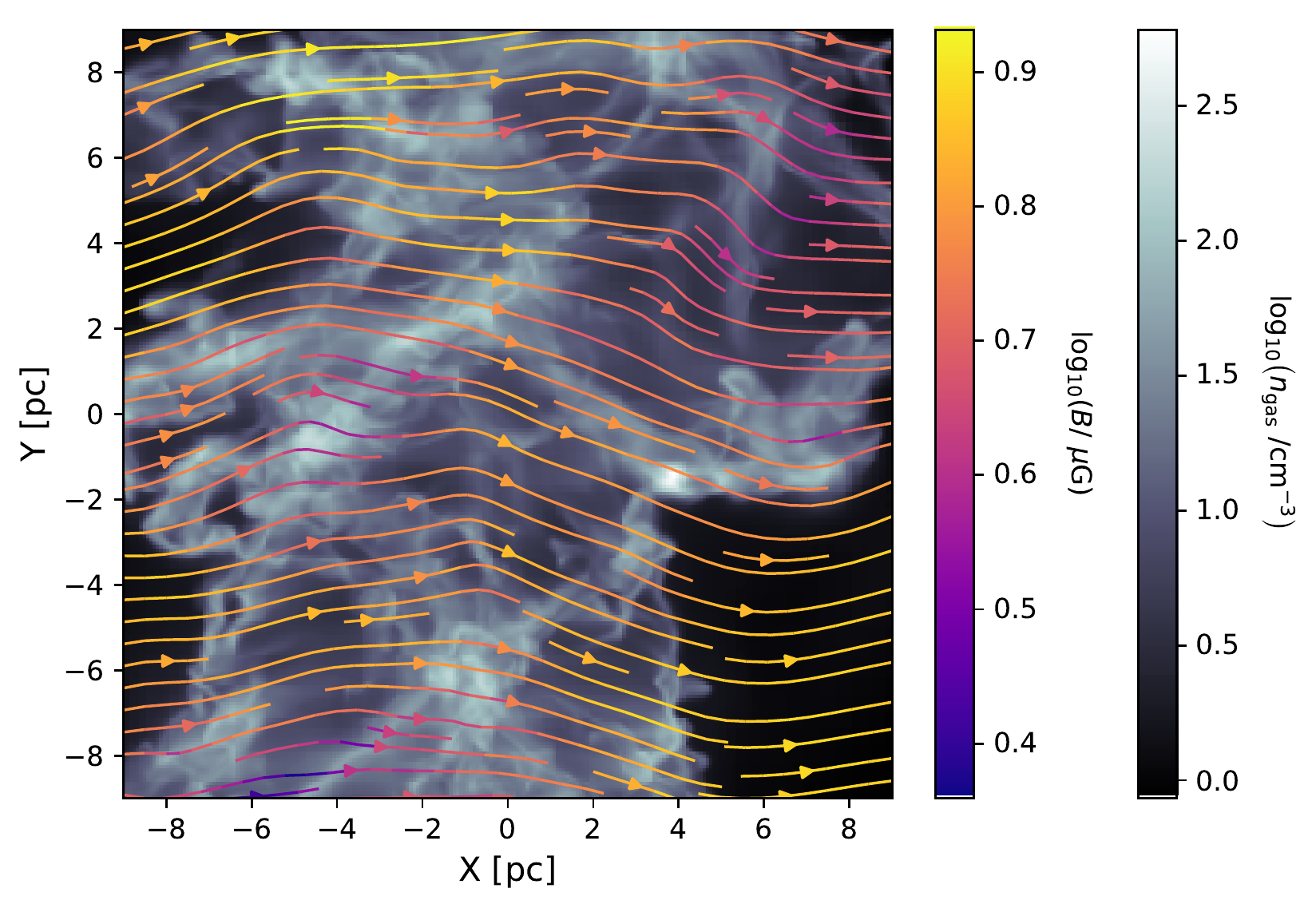}
\includegraphics[width=.31\textwidth]{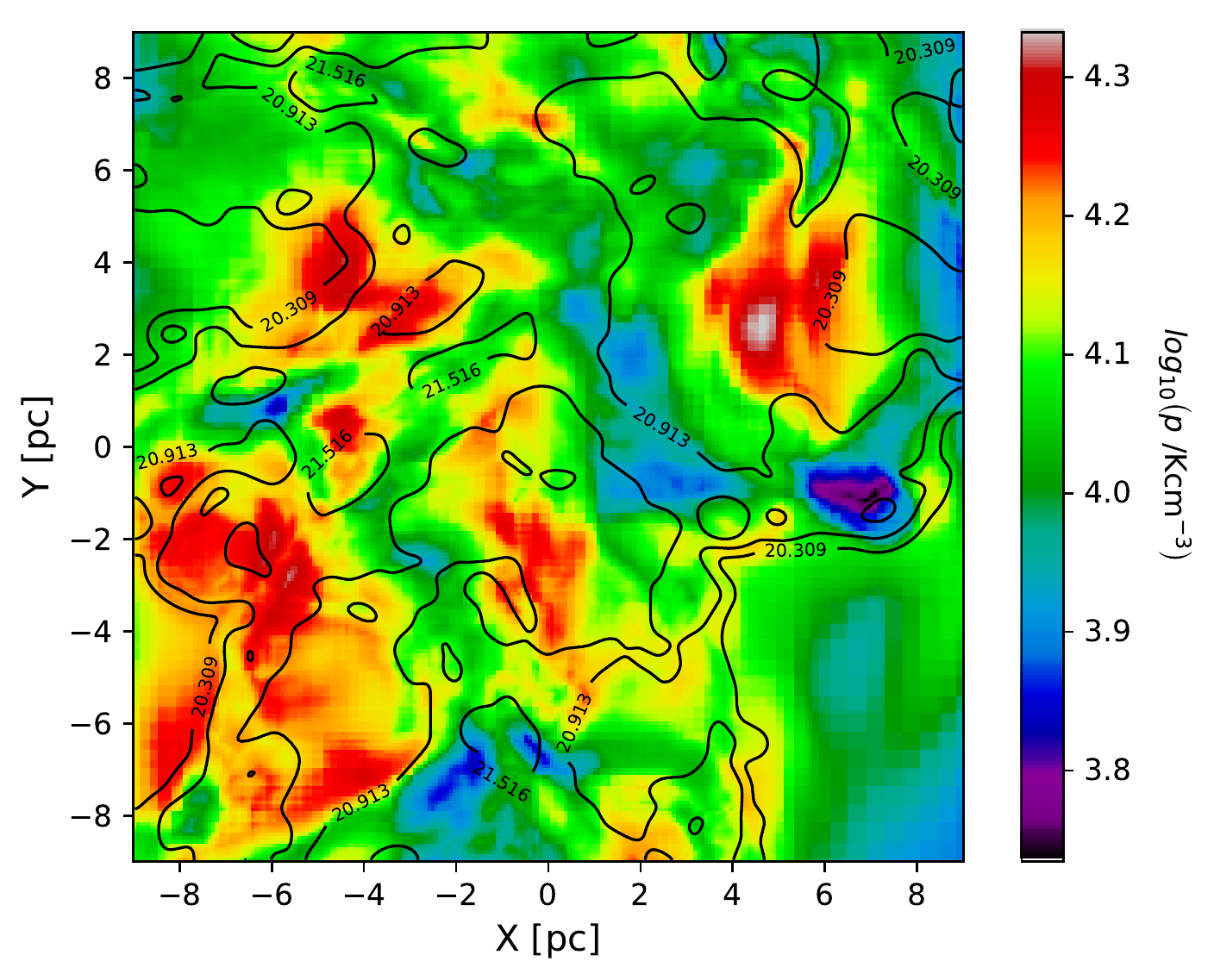}
\includegraphics[width=.31\textwidth]{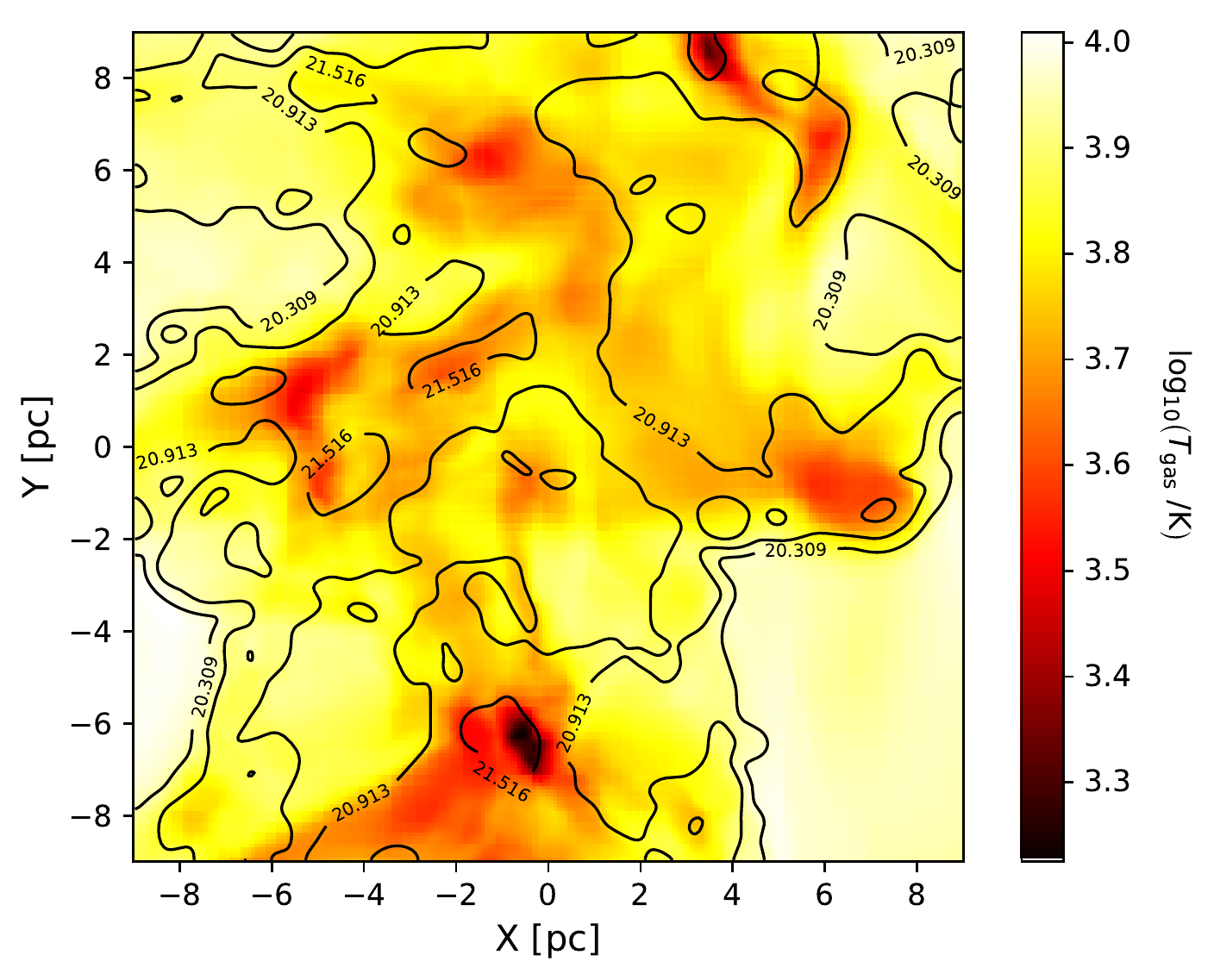}
\end{center}
\caption{$\RAMSES$ MHD parameters averaged cell by cell along the Z axis of the MHD cube: gas density $\ngas$ (left), gas pressure $P_{\mathrm{gas}}$ (center), and gas temperature $\Tgas$ (right). The vector field shows the averaged magnetic $B_{\mathrm{x}}$ and $B_{\mathrm{y}}$ components and the colorbar shows $B=(B_{\mathrm{x}}^2+B_{\mathrm{y}}^2)^{1/2}$. Contour lines indicate the logarithm of column density $\NH$.}
\label{fig:MHD_input}
\end{figure*}

As a model for a volume of neutral ISM material including both diffuse and dense gas on the way to forming molecular clouds, we consider a single snapshot from an MHD simulation computed with the adaptive-mesh-refinement code $\RAMSES$~\citep[][]{Teyssier2002,Fromang2006}. This particular simulation of interstellar MHD turbulence is the same as the one used in \cite{Planck2015XX}, and we refer the reader to that paper and to \cite{hennebelle_08}, where the simulation was originally presented, for more detail. To give its essential characteristics, it follows the formation of structures of cold neutral medium gas (CNM, $\ngas \sim 100\,\mathrm{cm}^{-3}$, $\Tgas \sim 50\,\mathrm{K}$) within head-on colliding flows of warm neutral medium (WNM, $\ngas \sim 1\,\mathrm{cm}^{-3}$, $\Tgas \sim 8000\,\mathrm{K}$). The colliding flow setup provides a convenient way to form such a mixture of diffuse and dense structures reproducing several observational properties of turbulent molecular clouds~\citep[see, e.g.,][]{Hennebelle-Falgarone-2012}, although cloud formation may actually proceed through other mechanisms such as spiral density waves~\citep{Dobbs_2006}. The simulation volume is threaded by a magnetic field that is initially aligned with the direction of the flows. From this simulation, we extract data over a cubic subset $18\ \mathrm{pc}$ along each side, located near the center of the full $50\ \mathrm{pc}$ box. The extracted data comprise total gas density $\ngas$, pressure $P_{\mathrm{gas}}$, and components $B_{\mathrm{x}}$, $B_{\mathrm{y}}$, and $B_{\mathrm{z}}$ of the magnetic field. Unlike in \cite{Planck2015XX}, however, we perform this extraction using the full resolution of the simulation ($0.05\,\mathrm{pc}$ per pixel) instead of the coarser $0.1\,\mathrm{pc}$ per pixel resolution that was used in \cite{Planck2015XX}. The average total gas density in the simulation cube is about $15\ \mathrm{cm}^{-3}$ leading to a total gas mass of $\approx 3400\,\mathrm{M}_{\odot}$, assuming a molecular weight $\mu=1.4$. The components of the magnetic field have a dispersion of $3\,\mu\mathrm{G}$ and a mean value of about $5\,\mu\mathrm{G}$ with a direction that is typically aligned with the flow~\citep[][Figure~15]{Planck2015XX}.

We use this simulation first to allow for a direct comparison of our results with those obtained with the same simulation assuming a uniform alignment of dust grains along the magnetic field lines \citep{Planck2015XX}, and second because it is representative of the diffuse ISM, while still harboring dense cores ($\ngas \sim 10^4$ cm$^{-3}$) where the drop in the grain alignment efficiency may be more pronounced. From this simulation we utilize the gas density $\ngas$, the gas temperature $\Tgas$, and the magnetic field magnitude as well as its direction as input for our subsequent RT post-processing. In Figure \ref{fig:MHD_input} we show the gas density, temperature, pressure as well as the magnetic field direction. The maps show direct, unweighted average quantities over the LOS, i.e., along the  $z$-axis of the simulation cube, for each direction. 

\subsection{Monte-Carlo propagation scheme of $\POLARIS$}
\label{sect:MCPropagation}
The post-processing steps of the MHD data consist of two parts. First, the radiation field is calculated with a Monte-Carlo (MC) approach in order to derive the necessary quantities for dust heating and grain alignment. In a second step we create synthetic dust emission and polarization maps. For all the RT simulations we make use of the RT code \POLARIS\ \citep[][]{Reissl2016}.\\
The local radiation field is determined by the 3D distribution of the dust and of the photon emitting sources. Commonly, the radiation field is quantified by the dimensionless parameter \citep[][]{Habing1968}
\begin{equation}
\Gzero= \frac{1}{5.29\times 10^{-14}\ \mathrm{erg}\ \mathrm{cm^{-3}}}  \int_{6\ \mathrm{eV}}^{13.6\ \mathrm{eV}} u_{E}\,\ud E\, ,
\label{eq:G0}
\end{equation}
where  $u_{E}$ is the spectral energy density of the radiation field within the energy band where photoelectric heating is most relevant.

In this paper we consider two separate setups concerning the sources of radiation. For the setup ISRF we do only use a parametrization of the spectral energy distribution (SED) as presented in \cite{Mathis1983} (see table \ref{tab:Setups}) for the MC sampling of wavelengths.
Note that we keep track of both the wavelength and direction $\hat{k}$ of each photon package per grid cell. For the setup ISRF, photon packages are injected into the MHD simulation from a sphere surrounding the grid with a randomly sampled $\hat{k}$ unit vector. 

Since the considered grain alignment (see Section \ref{sect:RATAlignment}) is sensitive to the radiation field we investigate a second case with an additional source of radiation. For this setup STAR we consider a star (see table \ref{tab:Setups}) at the very center of the grid, in addition to the ISRF radiation, in order to quantify the influence of the radiation field on dust heating and grain alignment. Here, the photon packages start with a random direction  $\hat{k}$ from the very position of the star whereas the wavelengths of the photons are samples from the Planck function. 

We note that the setup STAR is not entirely self-consistent since the star is added in post-processing and does not form in the MHD simulation itself, and so that the magnetic field and the gas do not respond accurately to the stellar feedback. Hence, our model lacks the expected density cavity and the deformation of field lines in the vicinity of the star. Nevertheless, we provide the STAR setup in order to explore the influence of the radiation field and subsequent RAT alignment on ISM polarization patterns in a controlled environment. This way, we can ensure that any deviation compared to the ISRF setup is purely due to radiation since magnetic field and gas properties remain the same.

\begin{table*}[t]
\centering
   \begin{tabular}[]{ |l | c |}
   \hline
      setup & description  \\ 
   \hline  
      ISRF & diffuse and isotropic ISRF with the SED from   \cite{Mathis1983} with $\Gzero=1$\\ 
   \hline
    STAR & ISRF plus one additional star at the very center of the grid with $R_{*}=15\ R_\odot$ and $T_{*}=15000\ \mathrm{K}$\\ 
  \hline
  \end{tabular}
  \caption{Properties of the radiation field setups for the MC dust grain heating and alignment simulations.}
  \label{tab:Setups}
\end{table*}

\begin{table}[b]
\centering
   \begin{tabular}[]{ |l | c |}
   \hline
      alignment & description  \\ 
   \hline  
      FIXED & $\aalig=100\,\mathrm{nm}$\\
   \hline
    RAT & $\aalig$ calculated by RATs (see Section \ref{sect:RATAlignment})\\
  \hline
  \end{tabular}
  \caption{Definition of the considered grain alignment mechanisms.}
  \label{tab:Alignment}
\end{table}

All MC RT simulations are performed with 100 wavelength bins logarithmically distributed  over a spectrum of ${\lambda \in [92\,\mathrm{nm} - 2\,\mathrm{mm}]}$. For the photon package propagation scheme we apply a combination of the continuous absorption technique introduced by \cite{Lucy1999} to keep track of the photons per grid cell and the temperature correction of \cite{BjorkmanWood2001} to ensure the correct spectral shift when a photon package gets absorbed and re-emitted. Assuming thermal equilibrium between the absorbed and emitted energy, the dust temperature per cell can be calculated with \citep[see][for details]{Lucy1999,BjorkmanWood2001,Reissl2016} 
\begin{equation}
\int \overline{C}_{\mathrm{abs},\lambda}\,J_{{\lambda}}\,\ud\lambda = \int \overline{C}_{\mathrm{abs},\lambda}\,B_{\lambda}(\Tdust)\,\ud\lambda\;,
\label{eq:Tdust}
\end{equation}
where $\overline{C}_{\mathrm{abs},\lambda}$ is the size averaged cross section of absorption. Here, we keep track of each of the temperatures corresponding to the distinct grain populations (silicate and graphite) individually as well as an average dust temperature. In detail, we solve Equation~\eqref{eq:Tdust} with  ${\overline{C}_{\mathrm{abs},\lambda}=\overline{C}_{\mathrm{abs},\lambda, \mathrm{silicate}}}$, ${\overline{C}_{\mathrm{abs},\lambda}=\overline{C}_{\mathrm{abs},\lambda, \mathrm{graphite}}}$, and ${\overline{C}_{\mathrm{abs},\lambda}=\overline{C}_{\mathrm{abs},\lambda, \mathrm{silicate}}+\overline{C}_{\mathrm{abs},\lambda, \mathrm{graphite}}}$ separately once the radiation field per cell is known (see also Appendix \ref{app:RTequation}). Consequently, all the dust polarization simulations performed in this work are for an average dust grain and ignore the possible size dependence of the dust temperature. For the MC simulations of the radiation field we consider the dust grains to be spherical since the dust shape and orientation has, for moderate elongations, only a minor influence on the grain absorption and scattering cross-sections \citep[see \eg][Figure~2]{DraineFraisse2009}.

The local spectral energy density, $u_\lambda$, is defined as the sum over the directions of intensity $\vec{J}_{\lambda}$ of all photon packages that crossed a particular grid cell,
\begin{equation}
u_{\lambda} =  \frac{4\pi}{c}\sum \left| \vec{J}_{\lambda} \right|\;,
\end{equation}
and the spectra anisotropy factor of the radiation field, $\gamma_\lambda$, can be defined as the vector sum of all radiation normalized by the total energy density
\begin{equation}
\gamma_{\lambda} =  \frac{\left|\sum  \vec{J}_{\lambda} \right|}{\sum \left|  \vec{J}_{\lambda} \right|}\, . 
\label{eq:gamma_lambda}
\end{equation}
Consequently, $\gamma_{\lambda}=1$ stands for an anisotropic radiation per wavelength i.e., a plane wave and $\gamma_{\lambda}=0$ for a totally diffuse i.e. fully isotropic radiation field.

The total energy density per cell, $\urad$, is then defined by
\begin{equation}
\urad=\int u_{\lambda}\,\ud\lambda\,,
\end{equation}
where we integrate over the full wavelength range, from which we derive the normalized quantity $U_{\rm rad} = u_{\rm rad}/u_{\rm ISRF}$, where $u_{\rm ISRF}=8.64\times 10^{-13}\ \mathrm{erg\ cm}^{-3}$ is the total energy of the ISRF per unit volume in our solar neighborhood as introduced by \cite{Mezger1982}.
For later analysis and discussion, we define for each position in the MHD cube the average anisotropy factor of the radiation field \citep[e.g.][]{Bethell2007,Tazaki2017} to be
 \begin{equation}
 \left\langle \gamma \right\rangle  =  \frac{1}{\urad} 
  \int{ \gamma_\lambda  \, u_{\lambda} \,\ud\lambda }
 \end{equation}
and the average $\cos{\vartheta}$ as
 \begin{equation}
 \left\langle \cos{\vartheta} \right\rangle  = \frac{1}{\urad}  \int{ \cos{\vartheta_\lambda}  \,u_{\lambda}  \,\ud\lambda}\,, 
 \label{eq:AvgCos}
 \end{equation}
where ${ \vartheta_\lambda = \angle \left(\vec{k}_\lambda,\vec{B} \right) }$ is the angle between the direction $\vec{k}_\lambda$ of the radiation field per wavelength bin and the direction $\vec{B}$ of the local magnetic field lines. 

The quantification of polarized radiation can be done in the Stokes vector formalism with $S=(I,Q,U,V)^T$ where $I$ is the total intensity, $Q$ and $U$ are the components of the linear polarization, and $V$ is the circularly polarized part. For the subsequent ray-tracing we make use of the full set of RT equations in the Stokes vector formalism in order to carry the full information of dust emission and extinction including polarization through the grid. We use a Runge-Kutta solver to project the rays onto a detector that stores each of the Stokes components as well as optical depth and column density. For the intensity $I$ we handle the contribution of silicate and graphite grains separately. Finally, the fraction of linear polarization is defined to be 
\begin{equation}
\PsI=\frac{\sqrt{Q^2+U^2}}{I}\, .
\end{equation}
The orientation angle of the polarization vectors can be derived by
\begin{equation}
\psi = \frac{1}{2}\mathrm{atan2} \left(U,Q\right)
\label{eq:PolAng}
\end{equation}
in the IAU convention for angles, as in \cite{PlanckXIX2015}.

We use the polarization angle dispersion function $\S$ introduced by 
\cite{Serkowski1958} and \cite{Hildebrand2009}. This function is a
measure of the local dispersion of magnetic field orientations within an annulus $\vec{\delta}$ around each position $\vec{r}$. The dispersion function reads
\begin{equation}
\S(\vec{r},\delta) = \sqrt{ \frac{1}{N} \sum_{i=1}^N 
\left[\psi(\vec{r}) - \psi(\vec{r}+\vec{\delta}_{\mathrm{i}})\right]^2}
\label{eq:PolCorAngle}
\end{equation}
 where $N$ is the number of pixels within the annulus. Following \cite{Planck2015XX}, we place the MHD cube at a distance of $D=100\ \mathrm{pc}$, use ${\delta ={\rm FWHM}/2}$, here with a FWHM of $5'$, and a pixellisation of 3 pixels per beam.

\subsection{Radiative transfer post-processing of the $\RAMSES$ simulation}
\label{sect:RTPostProcessing}
Since the MC method is based on stochastic sampling, derived quantities are inherently prone to noise \citep{Hunt1995}, a qualitative analysis of the noise is provided in Appendix \ref{app:MCNoise}. Hence, we perform the MC simulation with $5\times 10^8$ photon packages per wavelength for the ISRF setup. For the STAR simulation we apply $5\times 10^8$ photon packages per wavelength for the ISRF and $2\times 10^7$ photon packages for the star in the very center of the MHD cube constituting a balance between noise reduction and run-time. 

The quantity $\Gzero$ seems to be particularly sensitive to the number of photons. For low photon numbers $\Gzero$ stays far below unity, whereas $G_0=1$ is expected at the borders of the MHD grid considering the \cite{Mathis1983} ISRF. These photons are emitted towards the computational domain from a sphere of radius twice the sidelength of the MHD cube. Only this combination of amount of photons and sphere radius guarantees a $\Gzero \approx 1$ and an anisotropy factor $\gamma_\lambda \approx 0$ on average over all photons entering the simulation domain. Photons permanently scatter or become absorbed and are subsequently re-emitted in the $\POLARIS$ RT simulations. Photons newly injected into the grid may be deflected out of the grid already after a few such events and cannot carry their energy deeper inside. Consequently, the average energy density $\urad$ is about $2-5\ \%$ lower than the expected $u_{\rm ISRF}$ towards the center of the $\RAMSES\ $ MHD domain. As we will discuss below, such a loss of energy will only result in a modification of the polarization fractions by a fraction of a per cent.





\section{The dust model}
\label{sect:DustModel}

\subsection{Grain properties}\label{sec:grainprop}

Dust models compatible with observational constraints in the diffuse ISM in extinction, emission and polarization  require distinct dust populations \citep{DraineFraisse2009,Siebenmorgen2014,Guillet2018}: one population of very small grains to reproduce the UV bump and the mid-IR emission bands, one population of non-spherical silicate grains to account for the observed polarization in the optical, in the mid-IR silicate bands and in the FIR and sub-millimeter, and one population of carbonaceous grains (graphite or amorphous carbon, spherical or not) to complete the fit.

To confront the predictions of the RAT theory \citep{LazarianHoang2007,Hoang2014} to the statistics of dust polarized emission at $353\ \mathrm{GHz}$ obtained by the \Planck\ collaboration \citep{Planck2015XX,Planck2018XII}, we use a simplified dust model adapted to the \POLARIS\  code.
It is composed of two distinct size distributions of graphite ($\rho_{\rm G} = 2.24$ g.cm$^{-3}$) and silicate ($\rho_{\rm S} = 3.0$ g.cm$^{-3}$) grains. Graphite grains are assumed to be spherical, while silicate grains are spheroidal with an oblate shape. We note $a_\parallel$ (resp. $a_\perp$) the size of the oblate silicate grain along (resp. perpendicular to) its symmetry axis, and $s = a_\parallel/a_\perp = 0.5$ its aspect ratio. The sphere of equal volume has a radius $\aeff = a_\parallel^{1/3}\,a_\perp^{2/3}$.


Each size distribution follows a power-law of index $q$, with cut offs at $\amin$ and $\amax$, and mass per H $m$ [g/H]: 
\begin{equation}
\frac{\text{d}n\left(\aeff\right)}{\text{d}a} = \frac{3m\left(q+4\right)}{4\pi\rho\left(\amax^{q+4}-\amin^{q+4}\right)} \,\aeff^q, 
\end{equation}
where $\aeff$ is the radius of the grain for spherical grains, and the radius of the sphere of equal volume for spheroidal grains. 


The absorption, scattering and polarization coefficients of spheroidal grains are calculated with the \DDSCAT\ 7.3 code \cite[][]{DraineFlatau2013}. \DDSCAT\ provides the differential cross sections for extinction, absorption, and circular polarization required for an all-encompassing radiative transfer (RT) scheme, but has numerical limitations for large dust grains and small wavelengths. For this reason, we do not calculate those cross-sections for $\lambda < 0.25\,\mu$m, a domain of the UV that is not of interest for our study.


The absorption and scattering  coefficients 
for spherical dust grains are calculated on the fly at all wavelengths by \POLARIS\  itself with Mie theory, based on the refractive indices of the silicate and graphite grains \citep[][]{Weingartner2001}. 

Following the RAT theory as outlined in \cite{LazarianHoang2007} and \cite{Hoang2014}, we will assume that grains larger than a certain threshold-size, $\aalig$, are aligned along magnetic field lines, while other grains are not aligned, \textit{i.e.} they do not present any preferred orientation. The value of $\aalig$, which depends on the local physical conditions, will be determined using the RAT theory implemented in the \POLARIS\ code. However, to be able to compare our results with those obtained ignoring variations in the grain alignment efficiency \citep{Planck2015XX}, we also define a FIXED alignment setup 
in which $\aalig=100\ \mathrm{nm}$ for silicate grains throughout the cube, independently of the local physical conditions (see table \ref{tab:Alignment}). We outline the physics of grain alignment later in Section \ref{sect:RATAlignment} in more detail.

\subsection{Radiative transfer with spherical grains}

A few arguments favour using the optical properties of spherical grains, and not that of spheroidal grains, to compute the radiation transfer and dust temperature with \POLARIS\ (see Section \ref{sect:RTPostProcessing} and \ref{sect:RATAlignment} for details). First, the dust shape and orientation has, for moderate elongations, only a minor influence on the grain absorption and scattering cross-sections \citep[see \eg][Figure~2]{DraineFraisse2009}. Second, only silicate grains are spheroidal in our model, contributing to $\sim50\%$ of the dust extinction in the optical \citep{Weingartner2001}. Third, we are not able to compute correctly the radiation transfer by oblate grains in the far-UV ($\lambda<0.25\,\mu\mathrm{m}$) because the dust cross-sections for oblate grains could not be calculated for large grains at these wavelengths (see Section \ref{sec:grainprop}), and would impose an extrapolation of those cross-sections down to the geometrical limit. Fourth, in the RAT theory the properties of the radiation field in each cell must be known to determine the grain alignment efficiency. As a consequence, the \POLARIS\ code must first make an assumption on the grain alignment - random or perfect alignment for example - when computing the radiation transfer and dust temperature. Therefore, for the radiation transfer calculations leading to the determination of the characteristic of the radiation field and dust temperature in each cell, and for this only, we will replace oblate silicate grains by spherical grains of the same equivalent size.



\subsection{Fitting extinction and polarization curves in the optical}\label{sec:fitdustmodel}

\begin{figure}
\includegraphics[width=0.49\textwidth]{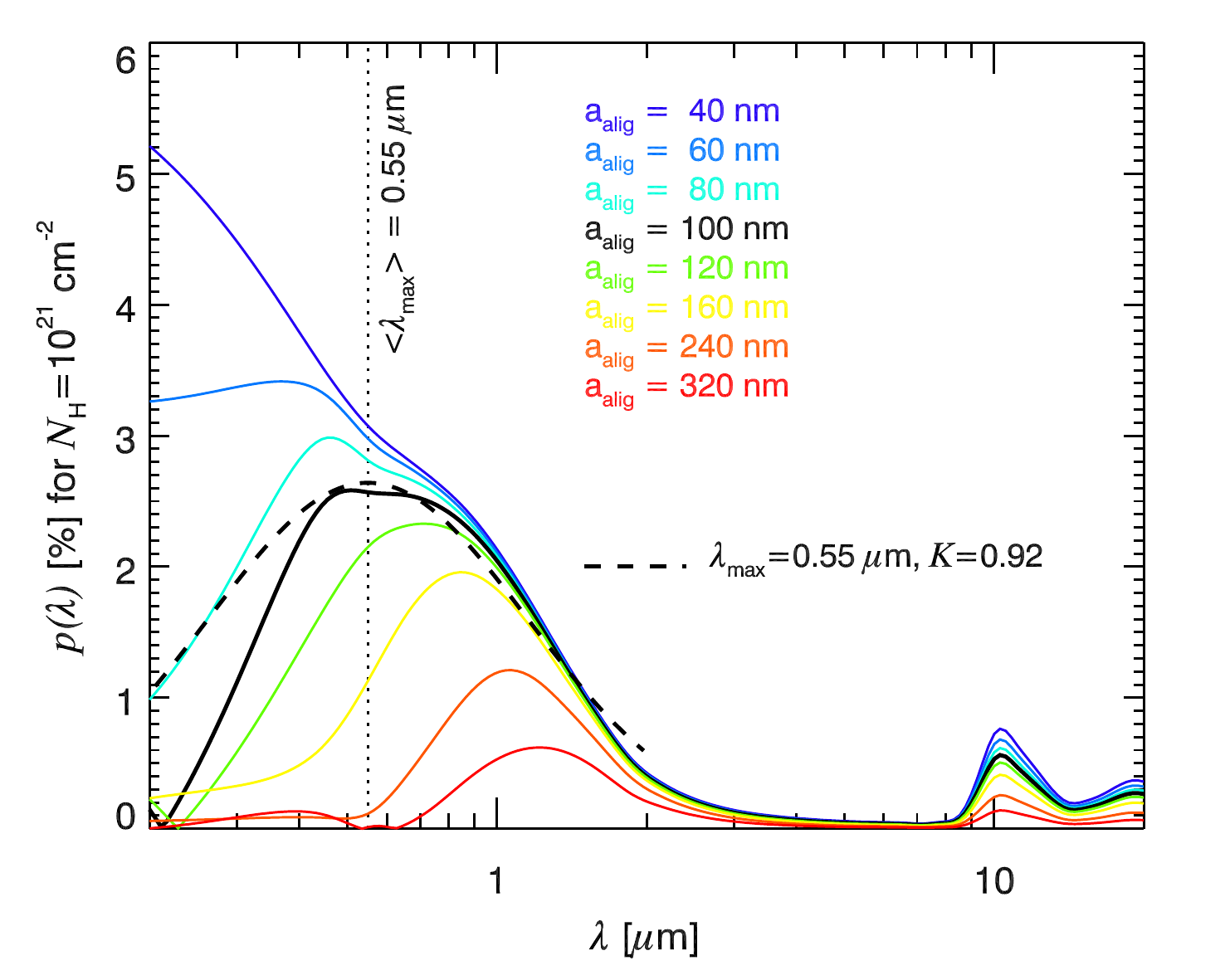}
\caption{Starlight polarization percentage for a column density $N_{\rm H} = 10^{21}\,$cm$^{-2}$, as a function of the wavelength, for increasing values of the alignment parameter $\aalig$. Silicate grains are assumed to be aligned, while graphite grains are not. The dashed curve is the mean polarization curve observed in the diffuse ISM ($A_V\sim 1$), which constrains both the upper limit $a_{\rm max}^{\rm S}$ of the silicate size distribution and the minimal size of aligned grains $\aalig$. Using our model, it is best fitted in the optical and NIR for $\aalig = 100\,\mathrm{nm}$ (see Section \ref{sec:fitdustmodel} for a discussion of the poor fit in the UV). }
\label{fig:p_lambda}
\end{figure}

Our dust model must reproduce the mean extinction and polarization curves observed in the diffuse interstellar medium for moderate extinction ($A_V \sim 1$). Polarization curves in extinction are usually modelled by the Serkowski law \citep{Serkowski1975,Whittet1992},
\begin{equation}
p(\lambda) = p_{\rm max}\,\exp\left\{ -K[\ln({\lambda/\lambda_{\rm max}})]^2 \right\}\;,
\end{equation}
where $\lambda_{\rm max}$ is the wavelength at which $p(\lambda)$ peaks, and $K$ controls the width at half maximum of the curve. The mean values observed in the diffuse ISM are  $\lambda_{\rm max}=0.55\,\mu$m, and $K=0.92$ \citep{Whittet1992}. The maximal value of $p_{\rm max}/E(B-V)$ was long considered to be of 9\% \citep{Serkowski1975}. It has been recently reevaluated to at least 13\% \citep{Planck2018XII,Panopoulou2019}, corresponding to a polarization fraction $p_V/\tau_V \simeq 4.5\%$, with $\tau_V=1.086\,A_V$.

For simplicity, lacking more constraints, we fix the power-law index of the silicate size distribution to $q_{\rm S} = -3.5$ \citep{Mathis1977}. The value of $\lambda_{\rm max}$ severely constrains the minimal size of aligned grains $\aalig$, while the value of $K$ provides a looser constraint on the upper cut-off $a_{\rm max}^{\rm S}$ of the silicate size distribution, which is here the only aligned population. We adapt $a_{\rm max}^{\rm S}$ and $\aalig$ to reproduce the overall shape of Serkowski's curve. Figure~\ref{fig:p_lambda} shows that a reasonable fit is obtained for $a_{\rm max}^{\rm S}=400\,$nm and $\aalig = 100\,$nm. A better correspondence could not be obtained in the UV ($\lambda < 400\,$nm) using oblate grains if we imposed a peak close to $550\,$nm. This weakness is however of little importance for our investigation, first because we do not study this part of the polarization spectrum, and second because the weak UV polarization in the mean polarization curve is known to be entirely produced by large ($a \ge 0.1\,\mu$m) aligned grains \citep{KimMartin1995}, whose abundance is constrained by the optical and NIR part of the spectrum. Therefore, we do not expect any change in our conclusions with a better fit of the UV polarization spectrum using more complex size distributions as per \cite{DraineFraisse2009}, or a power-law size distribution with prolate grains replacing oblate grains as per \cite{Guillet2018}.

\begin{figure}
\centering
\begin{minipage}[c]{1.0\linewidth}
   \begin{flushleft}
      \includegraphics[width=1.0\textwidth]{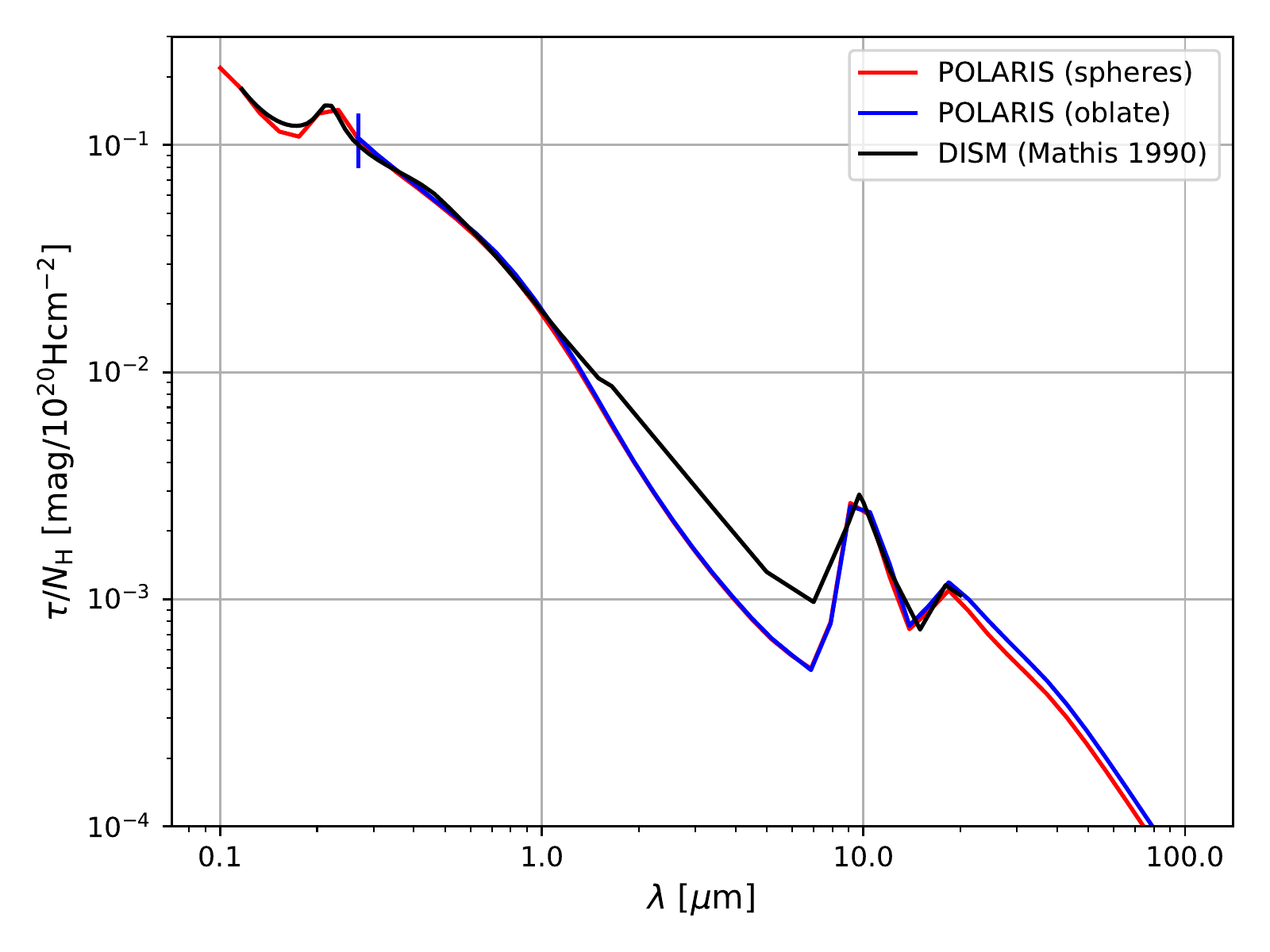}
 \end{flushleft}
\end{minipage}
 \caption{Extinction curve from the UV to the mid-IR of the applied \POLARIS\ dust models of spherical (red line) and non-spherical (blue line) grains in comparison with the measurements (black line) of \cite{Mathis1990}.}
\label{fig:DiffuseDust}
\end{figure}

The remaining parameters can be constrained with the mean extinction curve observed in the diffuse ISM. We use the \cite{Mathis1990} extinction curve per hydrogen, between 0.1 and 1 $\mu$m with $a_{\rm amin}^{\rm S}= 8\,$nm, $a_{\rm amin}^{\rm S}= 400\,$nm, $a_{\rm amin}^{\rm G} = 10\,$nm, $a_{\rm amax}^{\rm G} = 170\,$nm, $q_{\rm G}=-3.9$, $m_{\rm S} = 0.0034$ and $m_{\rm G}=0.0021$. This makes a total dust mass to gas mass ratio ${m_{\mathrm{dust}}/m_{\mathrm{gas}} = 0.55\ \%}$. Figure~\ref{fig:DiffuseDust} shows a comparison of the resulting output of the \POLARIS\ code with the \cite{Mathis1990} mean extinction curve per H. The fit is correct in the UV and optical, but not in the NIR, as expected in a silicate-graphite model with power-law size distributions. 
Replacing spheres by oblate grains only marginally affects the resulting extinction curve (Figure~\ref{fig:DiffuseDust}). In Section \ref{sec:sizedist}, we further discuss the potential impact on our results of the extinction curve in the NIR used for this particular MHD simulation. 

\begin{figure}
\includegraphics[width=0.49\textwidth]{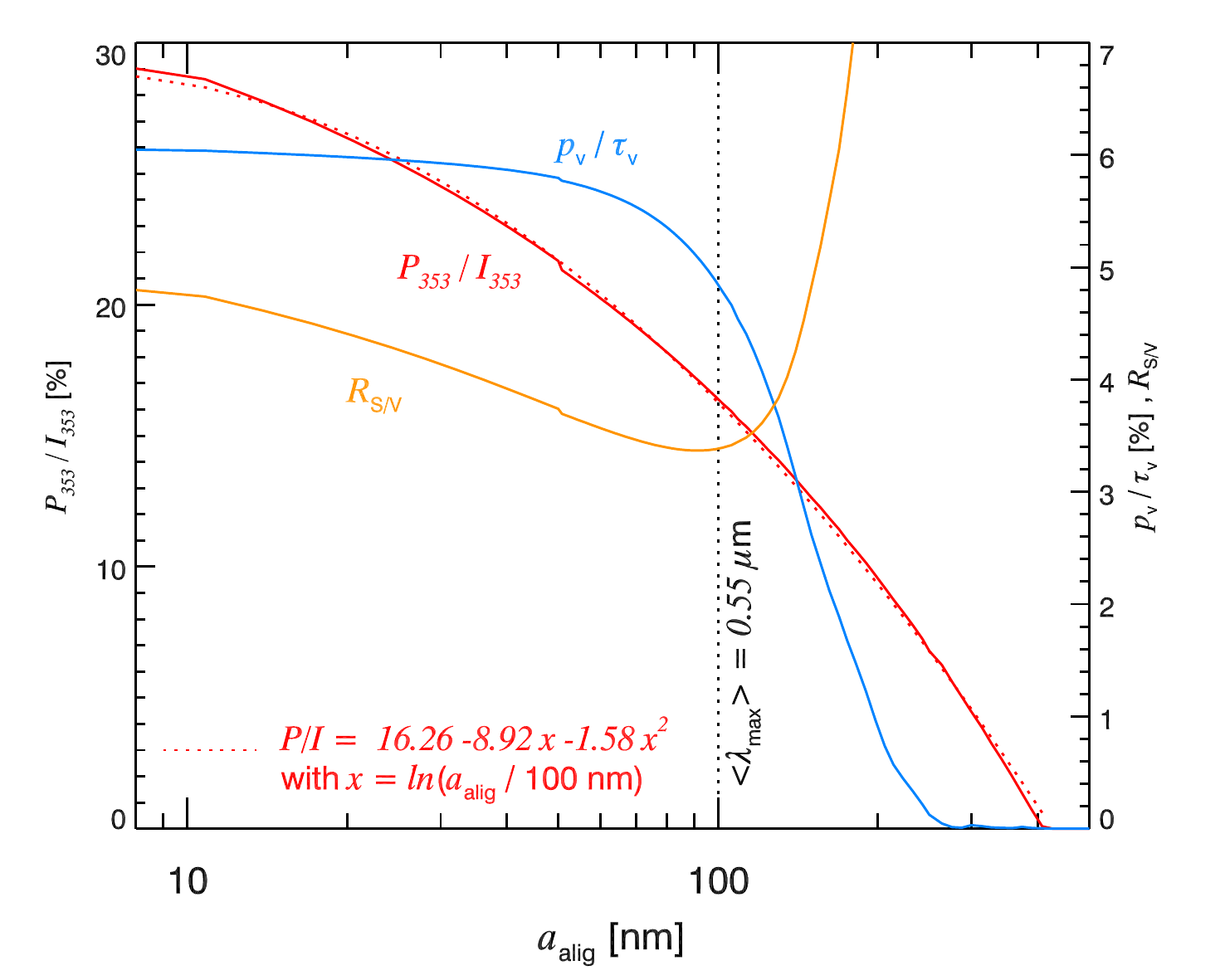}
\caption{(Left axis) Polarization fraction at $353\ \mathrm{GHz}$, $P/I$, as a function of the alignment parameter $\aalig$. (Right axis) Same for the polarization fraction in the V band, $p_V/\tau_V$, and the polarization ratio $R_{\rm S/V} = P/I/(p_V/\tau_V)$. The vertical dotted line indicates the value of $\aalig$ needed to reproduce the mean value of the Serkowski's parameter $\lambda_{\rm max}$ of $0.55\,\mu$m observed in diffuse and translucent LOS. An empirical fit to the dependence of $P/I$ on $a_{\rm alig}$ is provided for convenience.}
\label{fig:pst_PsI_lambda}
\end{figure}
Figure~\ref{fig:pst_PsI_lambda} presents the resulting dependence of the polarization fraction in the optical (V band) and at $353\ \mathrm{GHz}$ as a function of our alignment parameter, $\aalig$.
Figure~\ref{fig:p_lambda} demonstrated that the mean value value of $\lambda_{\rm max}$ observed in the diffuse ISM is obtained with our model for $\aalig \simeq 100$\,nm. According to Figure~\ref{fig:p_lambda}, the observed range of variation of $\lambda_{\rm max}$ through the ISM ($0.4 \ge \lambda_{\rm max} \ge 0.8\, \mu m$, \citealt{Whittet1992,Vosh2016}) translates into a range of values for $\aalig$, between 75 and 150 nm. Between these two values $\lambda_{\rm max}$, we expect of drop of the polarization fraction by a factor of 2 in the optical, and only by a factor 1.4 at $353\ \mathrm{GHz}$. This figure makes it clear that the dependence of the polarization fraction on the grain alignment efficiency differs in emission at $353\ \mathrm{GHz}$ and in extinction in the optical. A drop in grain alignment would therefore be easier to observe in the optical than at $353\ \mathrm{GHz}$, because of a steeper dependence on $\aalig$.   
For the mean $\lambda_{\rm max} = 0.55\,\mu$m, our model predicts $p_ V/\tau_V = 4.8\%$, thereby reproducing the highest polarization fraction $p/E(B-V) > 13\%$ observed in the optical \citep{Planck2018XII,Panopoulou2019}. The situation is different in emission where, with $P/I=16.3\%$, our model is 20\% below the highest polarization fraction at $353\,\mathrm{GHz}$ \citep[$p_{\rm max} = 20-22\%$,][]{PlanckXIX2015,Planck2018XII}. Correspondingly, the value of the polarization ratio $R_{S/V} = P/I/(p_V/\tau_V)\simeq 3.4$, is weaker than  the value $R_{S/V} \simeq 4.2$ observed in the diffuse and translucent ISM \citep{Planck2018XII}, but as expected within the range of the values obtained with compact astrosilicates \citep[see][for a discussion of dust optical properties adapted to \Planck\ observations]{Guillet2018}. 
These small discrepancies will have no consequence on our conclusions as we will not attempt to the reproduce the absolute value of the polarization fraction at $353\ \mathrm{GHz}$, but its relative variations with the environment, and especially with the column density.

\section{Insights on the radiative spin-up model}
\label{sect:RATAlignment}
Simulating dust polarization by means of extinction and emission not only requires non-spherical dust grains but also a detailed knowledge of the grain alignment efficiency with the magnetic field orientation. Here, we focus explicitly on the RAT alignment physics as it is outlined in \cite{Hoang2014}.

\subsection{The fiducial radiative torque (RAT) physics}\label{sec:RATphysics}
Abandoning the idea of perfectly aligned dust grains requires to model the physics of non-spherical spinning dust grains having their minor principle axis precessing around the magnetic field direction. In the RATs framework, a non-spherical irregular dust grain of equivalent radius $a$ 
can gain angular momentum through the torques $\Gamma_{\mathrm{rad}}$ exerted by an anisotropic radiation field  \citep{Hoang2014}:
\begin{equation}
\Gamma_{\mathrm{rad}} = \,\pi\aeff^2\int \left(\frac{\lambda}{2\pi}\right)\,  \gamma_\lambda\, \cos(\vartheta_\lambda) \,Q_\Gamma(\aeff,\lambda)\, u_\lambda\,\ud\lambda\, ,
\label{eq:JRAT}
\end{equation}
where $\gamma_\lambda$ is the spectra anisotropy factor (Equation~(\ref{eq:gamma_lambda})), and $Q_\Gamma$ is the RAT efficiency \citep{Draine_Weingartner1996,Hoang2014}:
\begin{equation}
 Q_{\mathrm{\Gamma}} = \begin{cases} \Qgammazeta   &\mbox{if } \lambda \leq 1.8\, \aeff \\ 
		    \Qgammazeta \times\left(\frac{\lambda}{1.8\, \aeff} \right)^{\Qgammaslope}  & \mbox{otherwise} \end{cases}\, .
 \label{eq:QGamma}
\end{equation}
$\Qgammazeta$ and $\Qgammaslope$ are parameters that depend on the grain shape and grain material. We note that the exact values of the parameters $\Qgammazeta$ and $\Qgammaslope$ are not well constrained. Numerical calculations show that $\Qgammazeta$ can present a range of values comprised between 0.01 and 0.4, and that $\Qgammaslope$ is between $-2.6$ and $-4$  \citep{Hoang2014,Herranen2019}. We take an average of $\Qgammaslope = -3$ as a reference value, and the exact value of $\Qgammazeta$ is determined for our dust model in Section \ref{sect:Calibrating}.




Spinning grain tends to be disaligned by the random momentum transferred in collisions with gas particle \citep{Davis1951}, as well as by the emission of IR photons \citep{DraineLazarian1998}. The gas drag on dust grains acts on a characteristic timescale of 
\begin{equation}
\tau_{\mathrm{gas}} = \frac{3}{4\sqrt{\pi}}\frac{I_{\mathrm{||}}}{\mu m_{\mathrm{H}}\,\ngas\,v_{\mathrm{th}} \,\aeff^4 \,\Gamma_{\mathrm{||}}} 
= \frac{2\sqrt{\pi}}{5} \frac{\rho s^{-2/3}}{\mu m_{\mathrm{H}} \,\Gamma_{\mathrm{||}}} \frac{\aeff}{\ngas\,v_{\mathrm{th}}}\;,
\label{eq:taugas}
\end{equation}
where $v_{\mathrm{th}}=(2k_{\mathrm{B}}\Tgas/(\mu m_{\mathrm{H}}))^{1/2}$ is the thermal velocity of the gas particles (of mean mass $\mu m_{\mathrm{H}}$), $\Gamma_{\mathrm{||}}\approx 1.1$ is a geometrical factor for oblate grains of aspect ratio $s=0.5$, and ${ I_{\mathrm{||}}=\frac{8\pi}{15}\rho_{\rm S} a_\parallel a_\perp^4 = \frac{8\pi}{15}\rho_{\rm S} s^{-2/3}\,\aeff^5 }$ is the moment of inertia of the oblate grain with respect to the minor axis. The drag timescale $\tau_{\rm FIR}$ by IR photon emission can be accounted for by a single parameter $\mathrm{FIR} \equiv \tau_{\rm gas}/\tau_{\rm FIR}$ \citep[][]{DraineLazarian1998,LH18}:
\begin{equation}
\mathrm{FIR} = 0.4\left(\frac{0.1\,\mu{\rm m}}{\aeff}\right)\left(\frac{30\,{\rm cm}^{-3}}{\ngas}\right)\sqrt{\frac{100\,{\rm K}}{\Tgas}}\left(\frac{\urad}{u_{\rm ISRF}}\right)^{2/3}\;,
\label{eq:FIR}
\end{equation}
Combining gas drag and FIR photon emission, this leads to a total drag timescale for the dust grain of 
\begin{equation}
\tau_{\mathrm{drag}}=\frac{\tau_{\mathrm{gas}}}{1+\mathrm{FIR}}\,.
\end{equation}
The alignment of dust grains with their minor axis parallel to the magnetic field direction is closely connected to overcoming the randomization of the rotation axis by gas bombardment and emission of IR photons. In the absence of any aligning torques, dust grain rotation is at thermal equilibrium with the gas, leading to a grain angular momentum of 
\begin{equation}
J_{\mathrm{th}}=\sqrt{k_{\mathrm{B}}\Tgas I_{\mathrm{||}}} \propto a^{2.5}\sqrt{\Tgas}\,.
\label{eq:Jth}
\end{equation}
We note that the magnitude $J_{\mathrm{th}}$ becomes constant as the dust grains thermalize with the gas and the orientation remains randomized over time. 

In order to ensure the alignment of dust with the magnetic field direction, the spin-up process by RATs needs to dominate over gas collision and IR photon emission and bring grains to suprathermal rotation \citep{Hoang2014}. Following \citet{Hoang2008}, we will assume that dust grains are aligned in a stable configuration for 
\begin{equation}
\frac{J_{\mathrm{rad}}}{J_{\mathrm{th}}} = \frac{\tau_{\mathrm{drag}}\,\Gamma_{\mathrm{rad}}}{J_{\mathrm{th}}} \ge 3\,.
\label{eq:JradoverJth}
\end{equation}
This condition defines the minimal grain size $\aalig$ for dust grains to be aligned.
If we use the approximate expression $Q_\Gamma \propto a^{-2.7}$ \citep{Hoang2014}, and momentarily restrict our study to the case where the disaligning effect of IR photon emission can be neglected with respect to collisions with gas particles ($\mathrm{FIR} \ll 1$), this minimal grain size follows the scaling :
\begin{equation}
\aalig^{\mathrm{FIR}\ll 1} 
\propto \left(\frac{\Qgammazeta\,\langle\gamma\rangle\,\langle\cos\vartheta\rangle\,U_{\rm rad}\,}{\ngas\,\Tgas}\right)^{-1/3.2} \,.
\label{eq:aalig_powerlaw}
\end{equation}
This expression shows that the grain alignment radius is a slowly varying function of the ratio between the gas pressure and an effective intensity $\Qgammazeta\,\langle\gamma\rangle\,\langle\cos\vartheta\rangle\,U_{\rm rad}$ of the anisotropic component of the radiation field.

The final condition for grains to be aligned with the magnetic field direction requires a stable Larmor precession around the magnetic field. This condition can be estimated by comparing the Larmor precession timescale $\tau_{\rm larm}$ \citep{LazarianHoang2007}
\begin{equation}
\tau_{\rm larm} \propto \frac{a^2\,s^2\,\rho\,\Tdust}{\chi\,B}\;,
\end{equation}
accounting for the interplay of field strength and the paramagnetic properties of the grain, with the gas drag timescale $\tau_{\rm gas}$ (Equation~(\ref{eq:taugas})). If the grain can complete its precession before any gas-grain interaction significantly affects its angular momentum, it can be considered to be aligned with the magnetic field direction. Consequently, for $\tau_{\rm larm}  < \tau_{\rm gas}$ a grain is considered to be aligned with the magnetic field direction, which defines the maximal grain radius $a_{\mathrm{larm}}$ where grains ceased to be aligned along the magnetic field lines \citep{LazarianHoang2007}
\begin{equation}
\alarm=2.71\times 10^5 \frac{s^2\,\left( \chi/10^{-4}\right)\left(B/5\,\mu\mathrm{G}\right)} 
{\left(n_{\mathrm{gas}}/30\,\mathrm{cm}^{-3}\right)\left(T_{\mathrm{dust}}/15\, \mathrm{K}\right)  \left(\sqrt{T_{\mathrm{gas}}/100\, \mathrm{K}}\right)}\, \mathrm{cm}\, .
\label{eq:TimeLarmor}
\end{equation}
Here, $\chi$ is the paramagnetic susceptibility of the grain material. Graphite grains have a magnetic susceptibility of about $\chi=9.6\times 10^{-10}$ \citep[][]{Weingartner2006} whereas for ordinary paramagnetic silicate we have $\chi=4.2\times 10^{-4}$ \citep[][]{Hunt1995,Hoang2014A}. In essence, graphite can barely perform a stable Larmor precession for the range of parameters of the $\RAMSES$ simulation (see Equation \ref{eq:TimeLarmor}) due to this difference of about six orders of magnitude in.

Laboratory experiments also suggest that most of the iron is bound within the silicate component of the ISM dust \citep[see e.g.][]{Davoisne2006,Demyk2017}, providing super-paramagnetic properties to the silicate populations for which a much better alignment is predicted \citep{JonesSpitzer1967}, if not perfect \citep{Lazarian2008,HoangLazarian2016}. In our model we will therefore assume that only silicate grains are aligned with the magnetic field direction, a choice that is consistent with observations of dust polarization \citep[][]{Mathis1986,Costantini2005,DraineFraisse2009,Vaillancourt2012}.

The nutation of the grain during its precession will tend to reduce polarization. This can be quantified by the Rayleigh reduction factor ${R=\left\langle Q_{\mathrm{J}}Q_{\mathrm{X}} \right\rangle}$ \citep[][see also Appendix \ref{app:RTequation}]{Greenberg1968,Roberge1999}.  $Q_{\mathrm{J}}$ characterizes the degree of alignment of the angular momentum $J$ with the magnetic field direction, whereas $Q_{\mathrm{X}}$ describes the internal degree of alignment between the minor principle axis $a_{||}$ of the dust grain and $J$. The average is then done over the time on the distribution function of the angle between the spin axis and the minor axis (for $Q_{\mathrm{x}}$) and between the spin axis and the magnetic field (for $Q_{\mathrm{J}}$). Radiative torques can align grains with the magnetic field $\vec{B}$ in two distinct attractor points. One is characterized by $J \gg J_{\mathrm{th}}$ (highJ hereafter) and the other one where $J$ is of the same order as $ J_{\mathrm{th}}$ (lowJ), respectively. While highJ corresponds to a perfect alignment, meaning $Q_{\mathrm{J}}\approx 1$ and $Q_{\mathrm{X}}\approx 1$, the lowJ attractor point is less well constrained. For paramagnetic materials such as pure silicate without iron inclusions, the fraction of highJ to lowJ alignment, together with the values for $Q_{\mathrm{J}}$ and $Q_{\mathrm{X}}$ in the lowJ case, are not well determined by the RATs theory \citep{Hoang2014}.
As discussed in \cite{HoangLazarian2016}, a significant fraction of dust grains in the lowJ attractor would prevent the model from reproducing the highest polarization fractions observed by {\it Planck} in the diffuse ISM \citep[$p_{\mathrm{max}}\simeq 20\%$,][]{PlanckXIX2015}. 
Alternatively, the polarization fraction could also be increased by introducing larger grains \citep[][]{Bethell2007} because this would increase the mass fraction of aligned grains. However, the presence of a significant fraction of large grains would prevent the dust model introduced in Section \ref{sect:DustModel} from reproducing the mean Serkowski's law and extinction curve of the diffuse ISM.

Thus, we make the assumption that silicate grains have ferromagnetic inclusions. Consequently, silicate grains align only at the highJ attractor point, and the Rayleigh reduction factor for RAT alignment is: 
\begin{equation}
R(\aeff) = \begin{cases} 1 & \mbox{if } \aalig < \aeff < \alarm   \\0 & \mbox{otherwise}   \end{cases}\, .
\label{eq:RayleighRAT}
\end{equation}
Assuming that silicates settle only at highJ would also prevent the so called wrong alignment, that is the alignment of the minor principal grain axis with the magnetic field direction \citep[see][]{LazarianHoang2007}. Thus, we do not model or discuss the implications of a possible wrong alignment of dust grains. 
Nevertheless, we note that $\POLARIS$ is in principle able to calculate the internal alignment efficiency \citep[][]{Reissl2016} at lowJ. Furthermore, $\alarm$ is only of minor relevance for silicate grains in our dust model since $\amax \ll \alarm$ even for ordinary paramagnetic grains let alone superparamagnetic ones (see also appendix \ref{app:Alignment}).

Altogether, the exact parametrization of $R(\aeff)$ (i.e. with or without internal alignment) is of minor relevance for the  dust polarization calculations presented in this article since more sophisticated assumptions would scale down the overall degree of linear polarization, without affecting the polarization patterns at the smaller scales \citep[see e.g.][]{Brauer2016}.



\subsection{Calibrating the RAT efficiency $Q_\Gamma$ on observational data}
\label{sect:Calibrating}
Our Equation~\eqref{eq:QGamma} involves the physical parameter $\Qgammazeta$ that controls the efficiency of radiative torques. The higher $\Qgammazeta$, the better grains are aligned. The value of $\Qgammazeta$ must be determined using numerical tools like \DDSCAT\ by calculating  the radiative torques efficiency for a particular grain shape and material that constitutes the aligned dust population, here oblate silicate grains of axis ratio $s=1/2$. \cite{Herranen2019} did the most recent and extensive study of the dependence of $Q_\Gamma$ on the ratio $\lambda/a$, for various grain shapes and materials. Although $Q_{\Gamma}$ is not strictly constant at low $\lambda/a$ according to these calculations, a constant value for $\Qgammazeta$ between 0.05 and 0.4 appears to be a reasonable model owing to the scatter in the calculations presented for different shapes  \citep[][Figure~20]{Herranen2019}.

This theoretical value for $\Qgammazeta$ can be compared to the value that is needed to obtain an alignment parameter $\aalig$ of $100\,$nm, the value necessary for our model to reproduce the mean Serkowski's curve observed in the diffuse and translucent ISM (see Section \ref{sect:DustModel}). Using our \RAMSES\ simulation with \POLARIS, we find $\Qgammazeta=0.14$, a value that we will use from now on.


\subsection{Phase diagram for the grain alignment efficiency}\label{sec:phasediagram}

It has been long established that the suprathermal rotation can allow for grain alignment \citep{Purcell1975,Purcell1979}. In the RAT modeling of \cite{Hoang2014}, grains are assumed to be aligned if the local physical conditions make them rotate three times faster than in thermal equilibrium with the gas. Given any efficient spin-up process \citep[][]{Lazarian2015}, this necessary prerequisite allows for dust grains to align with the magnetic field direction because of paramagnetic effects acting on a microscopic level \citep[see e.g.][for details]{Barnett1915,Davis1951,JonesSpitzer1967,Purcell1979}. 

Figure~\ref{fig:aalig_modelHL14} presents a synthetic view on the dependence of $\aalig$ on the local physical conditions in the \cite{Hoang2014} RAT theory, in the form of a phase-diagram for the diffuse ISM. 
The $y$ axis is the spin-up parameter $\spinup$ (see Eq.~\eqref{eq:aalig_powerlaw}). The $x$ axis is the FIR ratio (Eq.~\eqref{eq:FIR}) calculated for the reference value $\aeff=100\,$nm. 
This phase diagram allows to estimate the grain alignment radius predicted by RATs for any physical conditions, as long as the wavelength-dependence of the radiation field can be reasonably described by the ISRF with a scaling factor $U_{\rm rad}$. At low FIR ratio, $\tau_{\mathrm{drag}}\simeq \tau_{\mathrm{gas}}$, and the alignment radius becomes independent of the FIR ratio. 
At high FIR ratio, $\tau_{\mathrm{drag}} \simeq \tau_{\mathrm{FIR}}\propto a^2\,U_{\rm rad}^{-2/3}$, and the alignment efficiency $\aalig$ becomes independent of the gas density.
Arrows indicate how $\aalig$ varies when the corresponding parameter increases by a factor 10. Because of the exponent $-1/3.2$ in Equation~\eqref{eq:aalig_powerlaw}, variations by orders of magnitude of any of these parameters are needed to significantly affect the value of $\aalig$. 




\section{Grain alignment in the translucent and diffuse ISM}\label{sec:ISRF}

In this section, we present the results of our calculations with \POLARIS. Our MHD cube is representative of the diffuse and translucent ISM. We start by presenting the statistics of the radiation field in the MHD cube, which controls the radiative torque efficiency. Then, we look for the physical variables that, under these conditions, control the variations of the grain alignment efficiency under the RAT theory.
Finally, we compare the dust polarization maps when grains are aligned by radiative torques, and when the grain alignment is uniform, to test if the alignment model leaves some imprint in the polarization maps calculated with this MHD simulation.
\begin{figure*}
\begin{center}
\includegraphics[width=.45\textwidth]{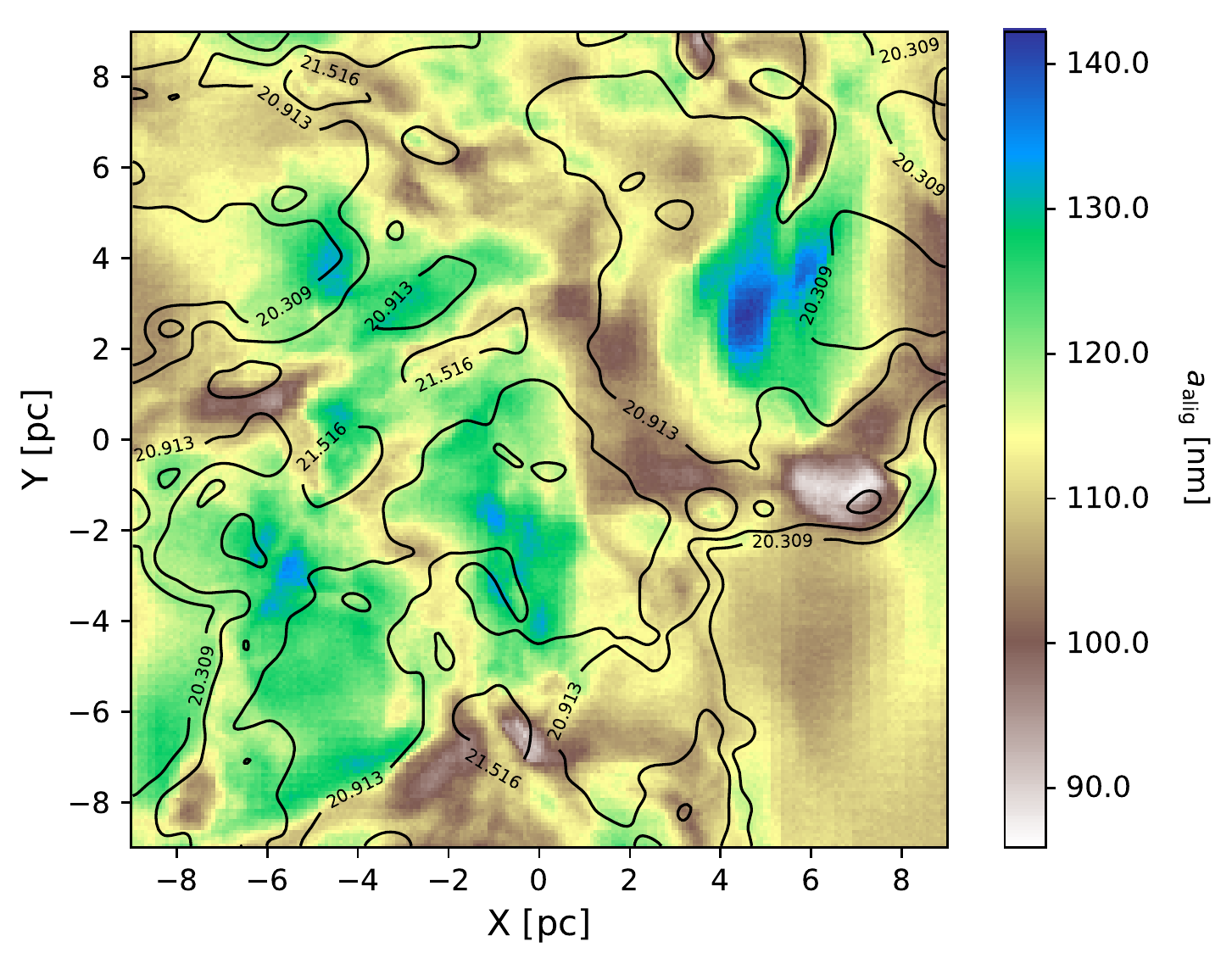}
\includegraphics[width=.45\textwidth]{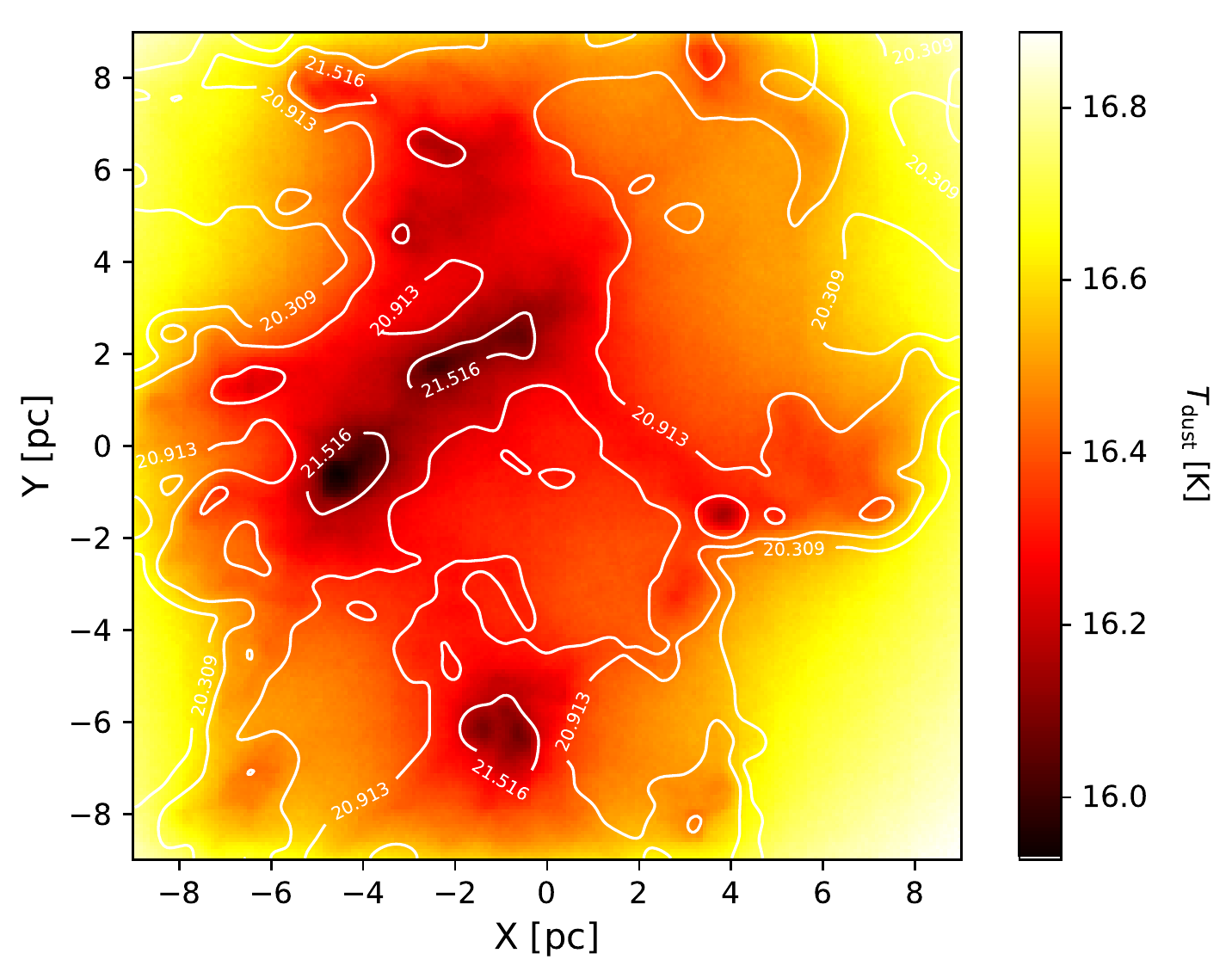}
\includegraphics[width=.45\textwidth]{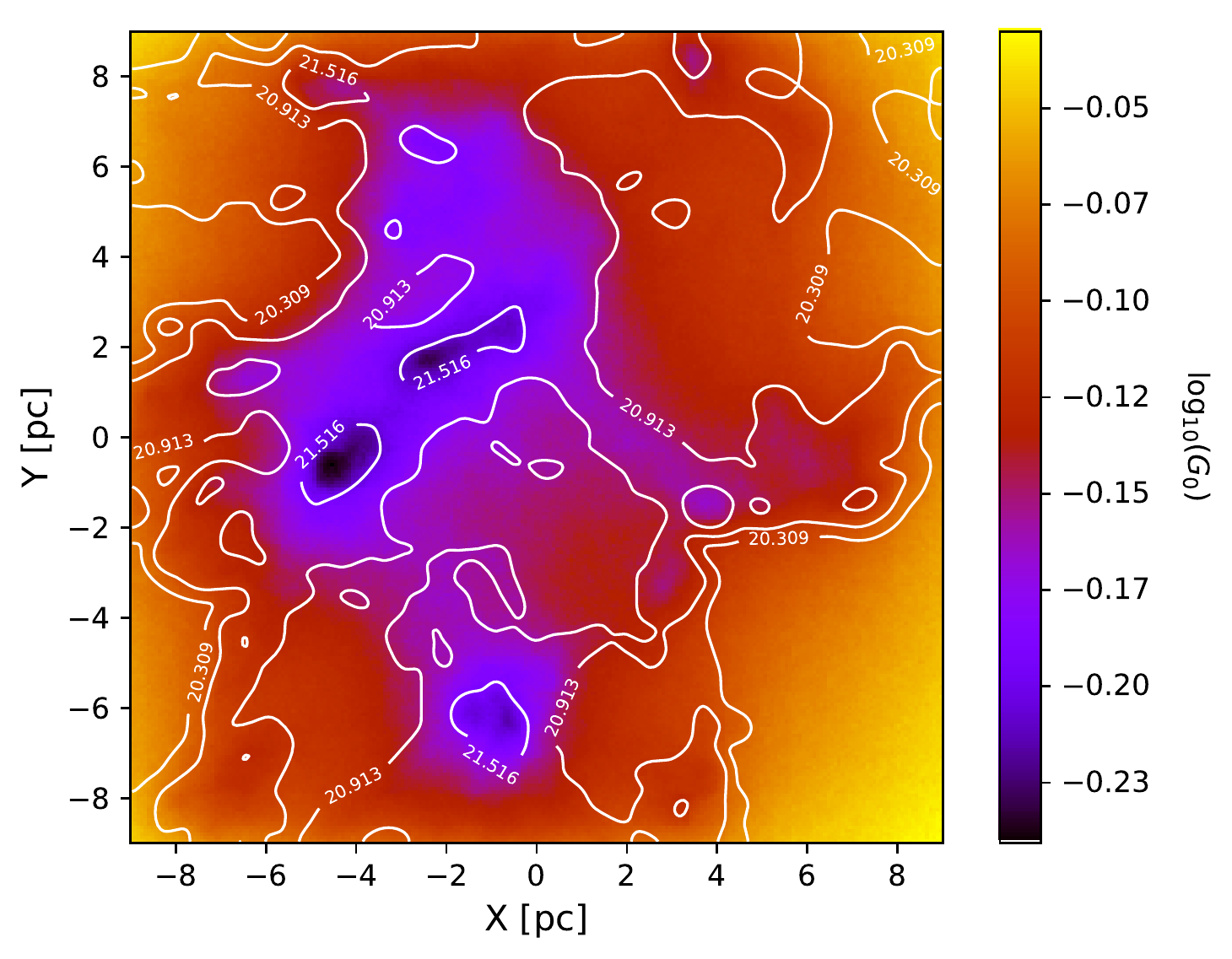}
\includegraphics[width=.45\textwidth]{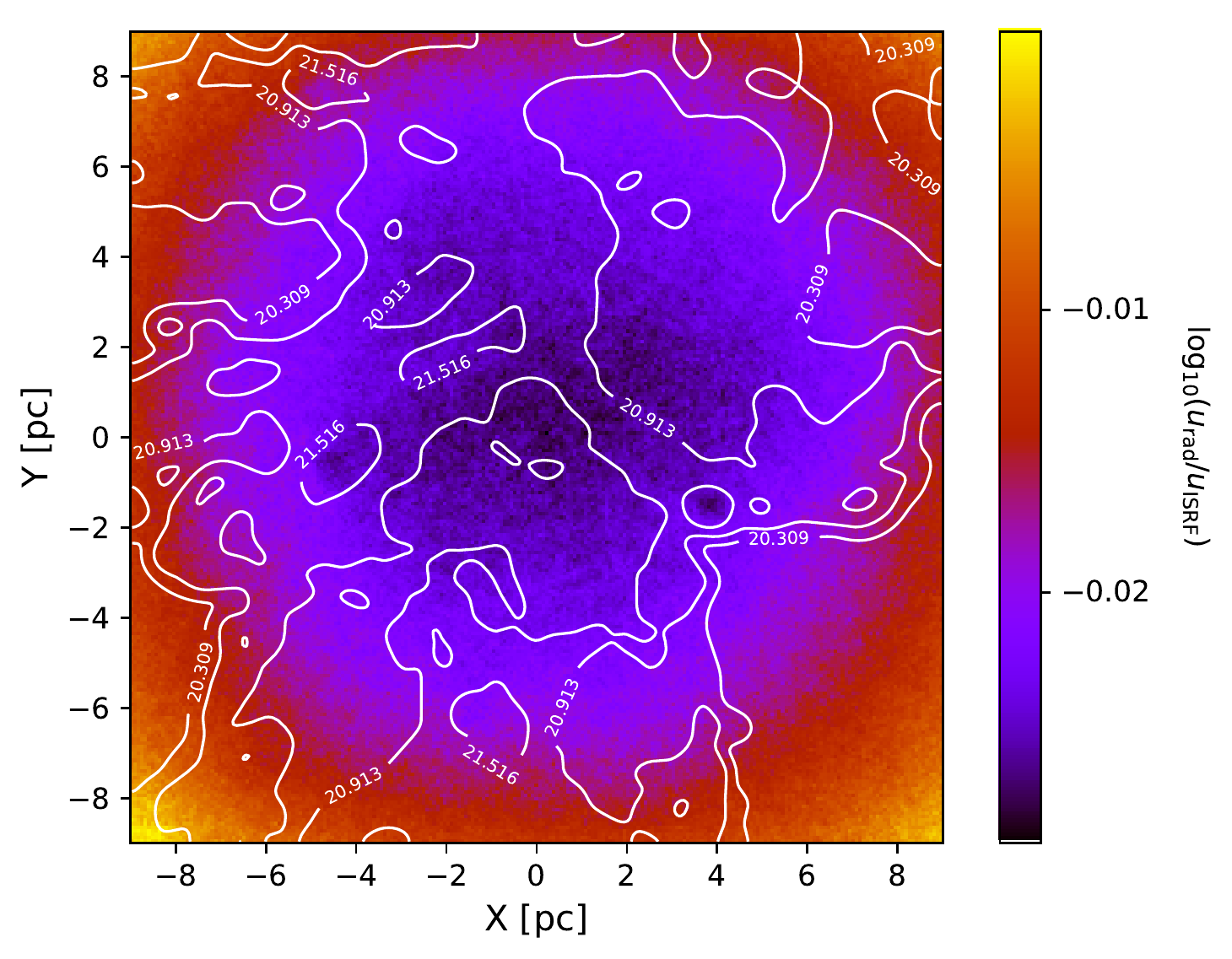}
\includegraphics[width=.45\textwidth]{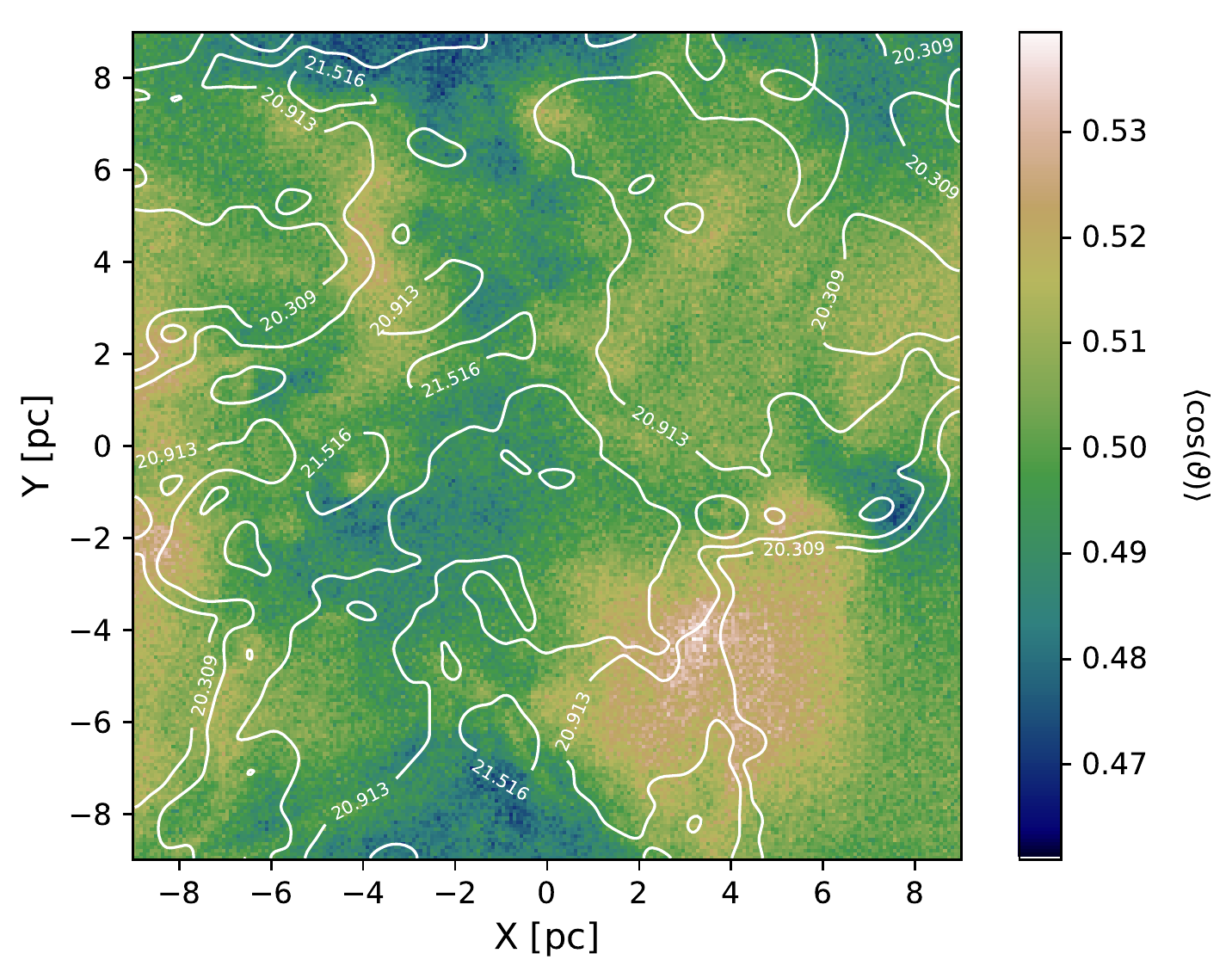}
\includegraphics[width=.45\textwidth]{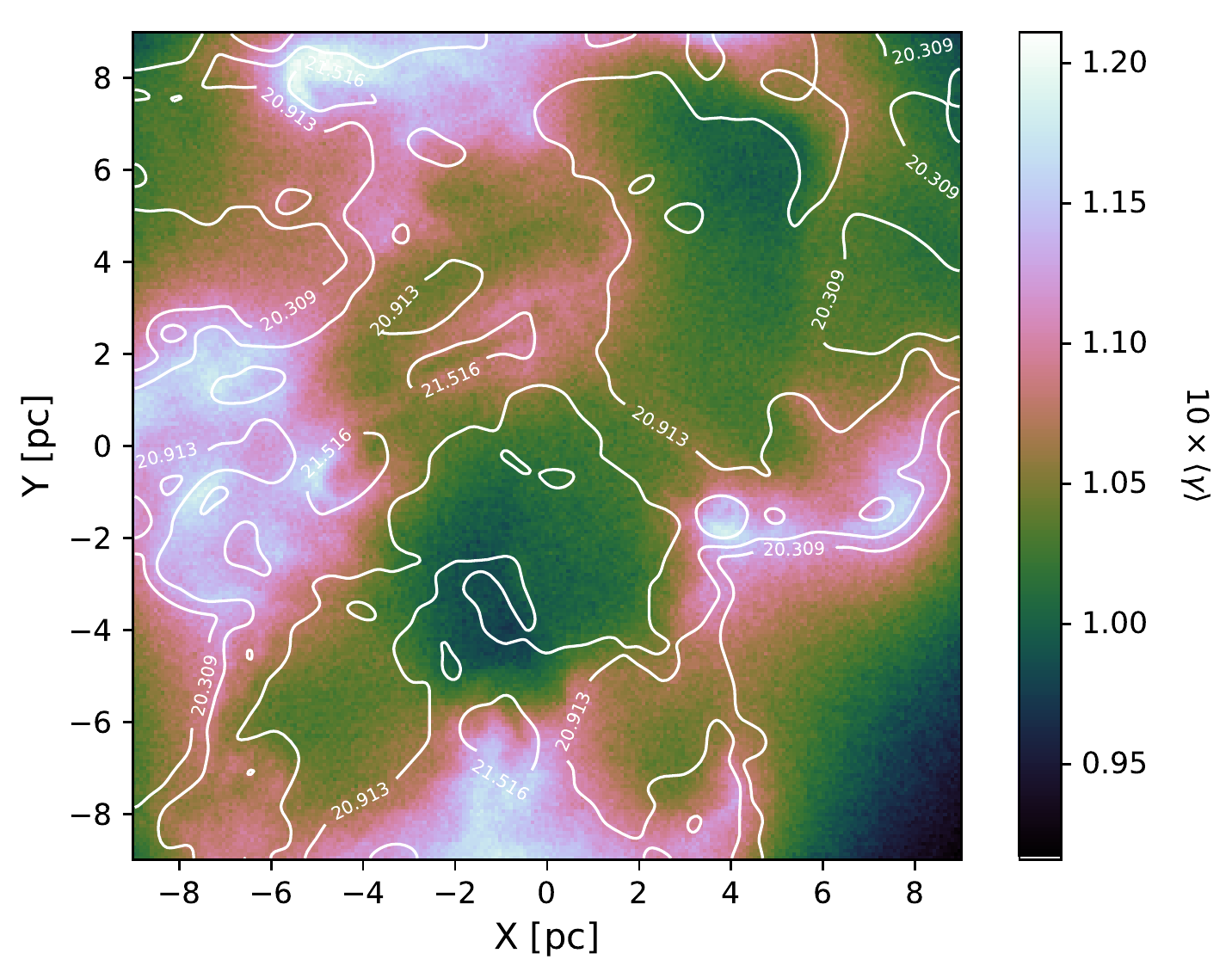}
\end{center}
\caption{Quantities averaged along the LOS derived by \POLARIS\ MC simulations for the ISRF-RAT setup. Here, direct averages are done, without any weighting by another parameter. The individual panels show the alignment radius $\aalig$ (top left), average dust temperature $\Tdust$  (top right), $\Gzero$ (middle left), the radiation field $U_{\rm rad}$ (middle right), the average angle $\left\langle\cos{\vartheta}\right\rangle$ (bottom left), and the anisotropy factor $\left\langle \gamma \right\rangle$ (bottom right), respectively.}
\label{fig:POLARIS_ISRF_RAT}
\end{figure*}

\subsection{Characteristics of the radiation field}
\label{sec:RT_ISRF}

Figure~\ref{fig:POLARIS_ISRF_RAT} presents the set of derived MC quantities for the case ISRF-RAT. All maps show the average of grid cells along the $z$ axis of the MHD cube i.e. along the LOS (histograms over the complete 3D domain are provided in Appendix \ref{app:3DDistribution}). For clarity regarding the  parameters characteristic for the radiation field, the maps are not weighted by any quantity e.g. by density. Otherwise characteristic features of the model ISRF would get modulated by the weighting, making it harder to discuss the different quantities on an individual basis. The map of the alignment radius $\aalig$ has a range of ${80\,\mathrm{nm} - 145\,\mathrm{nm}}$. The map of $\aalig$ clearly correlates with the pressure map presented in Figure \ref{fig:MHD_input}, not with the density map. 

The dust temperature is rather uniform in the entire simulation, between 16\,K and 17\,K. 
Here, we show the combined temperatures averaged over the materials of silicate and graphite (individual temperatures are provided in Appendix \ref{app:3DDistribution}). Even with such a small temperature variation, a correlation with the column density stands out. 
Dust grains in the densest regions are colder because of the shielding by the surrounding dust. 
As expected, the small variations of $\Gzero$ coincide with the column density structure, as
photons become more likely absorbed in dense regions. 
In contrast to $\Gzero$, the total energy density $\urad$ is  integrated over the entire spectrum and should be almost constant independently of the density, because the total energy within the system remains conserved while being shifted towards longer wavelength by dust emission. Still, we see that $\urad$ is slightly (2\% at most) smaller than $u_{\mathrm{ISRF}}$. This is a small artefact of the MC method associated with the loss of photons (see Section \ref{sect:MCPropagation}).

The average angle $\left\langle \cos{\vartheta} \right\rangle $ between the radiation field and the magnetic direction draws the same  picture of a totally diffuse radiation field. The values of $\left\langle \cos{\vartheta} \right\rangle $ in Figure~\ref{fig:POLARIS_ISRF_RAT} cluster around a value of $0.5$. We acknowledge that the quantity  $\left\langle \cos{\vartheta} \right\rangle $ does not strictly correspond to a particular angle $\vartheta$ but represents an average over an ensemble of angles weighted by the cosine function and the radiation field (see Equation \ref{eq:AvgCos}). However, we note that a $\cos{\vartheta}=0.5$  would correspond to an angle of $\vartheta = 60^\circ$. This is exactly the value one obtains when averaging over a large ensemble of pairs of randomly orientated vectors. Hence, a value of $0.5$ is consistent with a mostly isotropic radiation field (see also Appendix \ref{app:3DDistribution}).

Finally, the anisotropy factor $\left\langle \gamma \right\rangle $ has a trend with higher values in denser regions and amounts to an average value of $0.11$ comparable with the value of $0.1$ usually given in the literature \citep[see e.g.][]{LazarianHoang2007,Hoang2014} for the ISM. We run simulations with no dust at all
i.e. a ratio of ${m_{\mathrm{dust}}/m_{\mathrm{gas}} = 0\ \%}$ in the $\RAMSES$ cube. These test simulations show that $\left\langle \gamma \right\rangle > 0$ (see Appendix \ref{app:3DDistribution}). Even with more photons and for different radii of the source sphere, the anisotropy factor cannot be pushed below $\left\langle \gamma \right\rangle < 0.045$. We speculate that this may be a numerical limitation of the applied MC techniques.

\subsection{What drives the variations of the grain alignment parameter $\aalig$ ?}\label{sec:whatdrivesRAT}

\begin{figure}
\includegraphics[width=.49\textwidth]{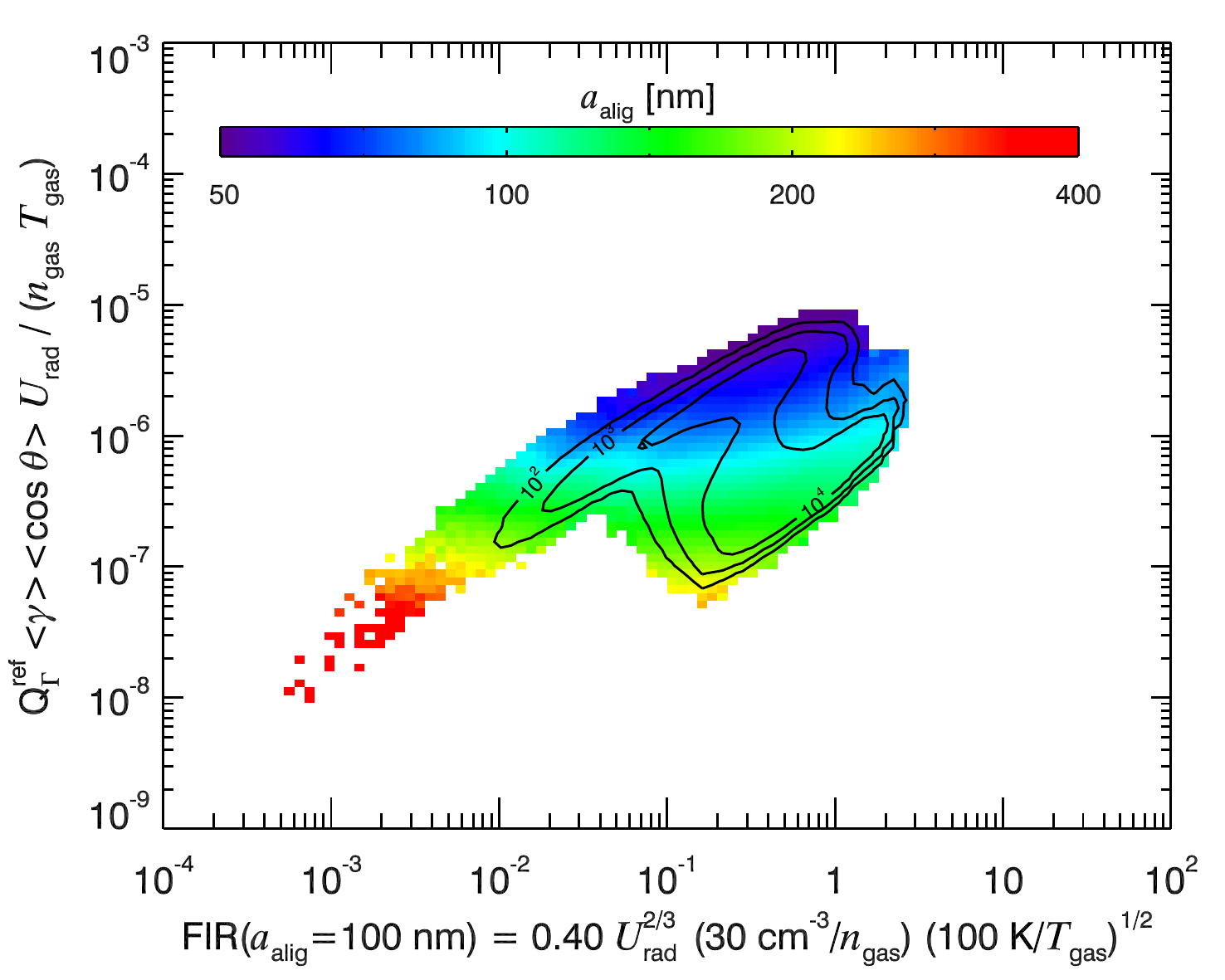}
\caption{Alignment parameter $a_{\rm alig}$ calculated by POLARIS for the ISRF-RAT simulation, in the phase diagram of Figure~\ref{fig:aalig_modelHL14}. Contour lines indicate the density of points, delimitating two valleys of points corresponding to the cold (CNM, upper branch) and warm (WNM, lower branch) phases in the simulation.}
\label{fig:aalig_POLARIS_ISRF}
\end{figure}

\begin{figure}
\includegraphics[width=.49\textwidth]{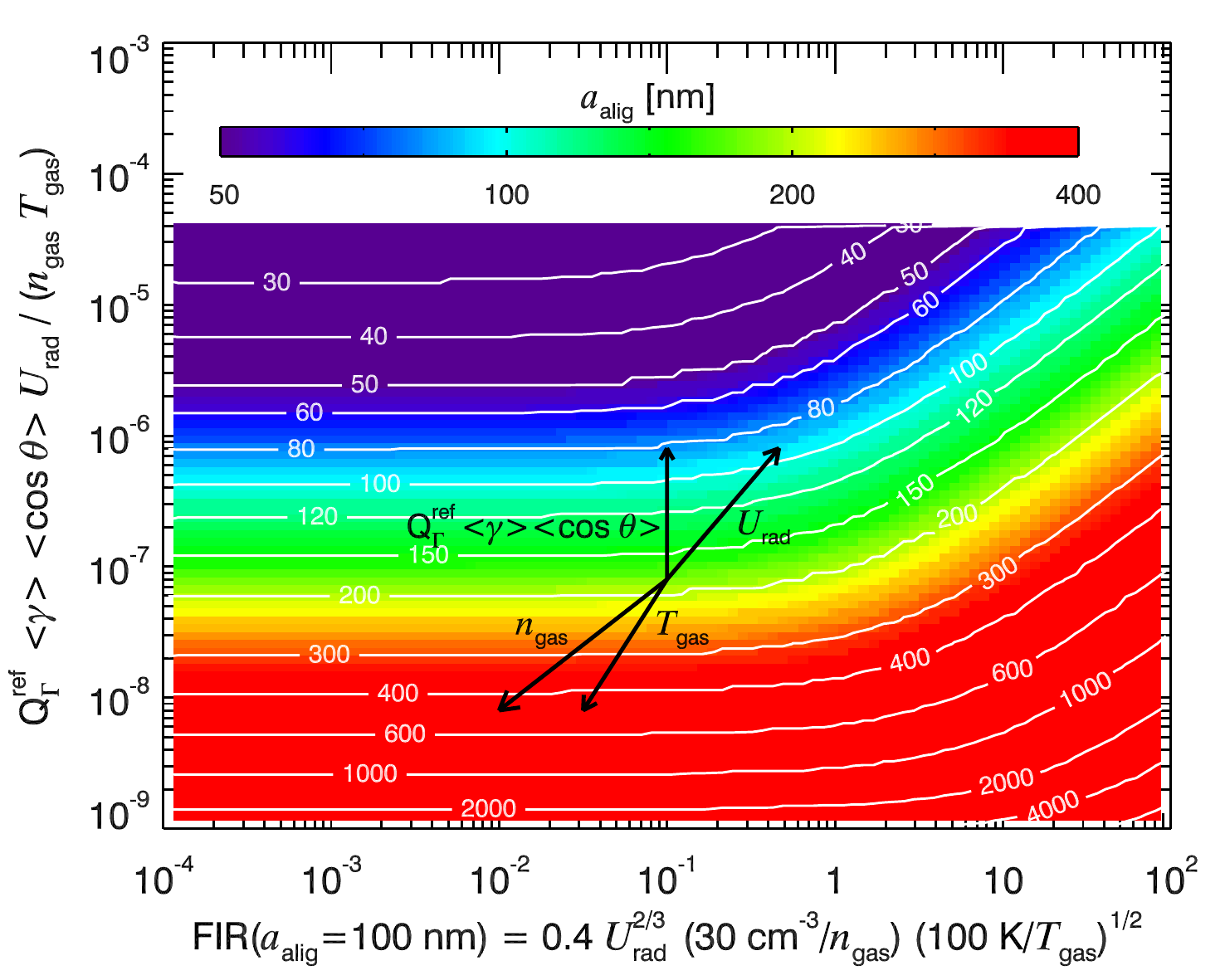}
\caption{Alignment parameter $a_{\rm alig}$ in $\mathrm{nm}$, as a function of the spin-up parameter $\spinup$ (see Eq.~\eqref{eq:aalig_powerlaw}) and of the FIR ratio at $a=100\,\mathrm{nm}$ (Eq.~\eqref{eq:FIR}), following \cite{Hoang2014}. Black arrows indicate the displacement in that frame when the corresponding physical quantity increases by a factor 10.}
\label{fig:aalig_modelHL14}
\end{figure}

Figure~\ref{fig:aalig_POLARIS_ISRF} presents how the alignment parameter calculated by \POLARIS\ for our \RAMSES\ simulation depends on the local physical conditions, using the same phase diagram as in Figure~\ref{fig:aalig_modelHL14}. The density, temperature, and radiation field characterizing this simulation only occupies a small surface in our phase diagram. The density of points in this phase diagram allows to separate the WNM phase (high temperature, low density) from the CNM phase (high density, low temperature) where grains are not well aligned in a small fraction of cells (red points). 
Comparing Figure~\ref{fig:aalig_modelHL14} and Figure~\ref{fig:aalig_POLARIS_ISRF}, we see that our simple analytic derivation of $\aalig$ (see Section~\ref{sect:RATAlignment}) reproduces quite well the numerical results of \POLARIS.

\begin{figure}
\includegraphics[width=0.49\textwidth]{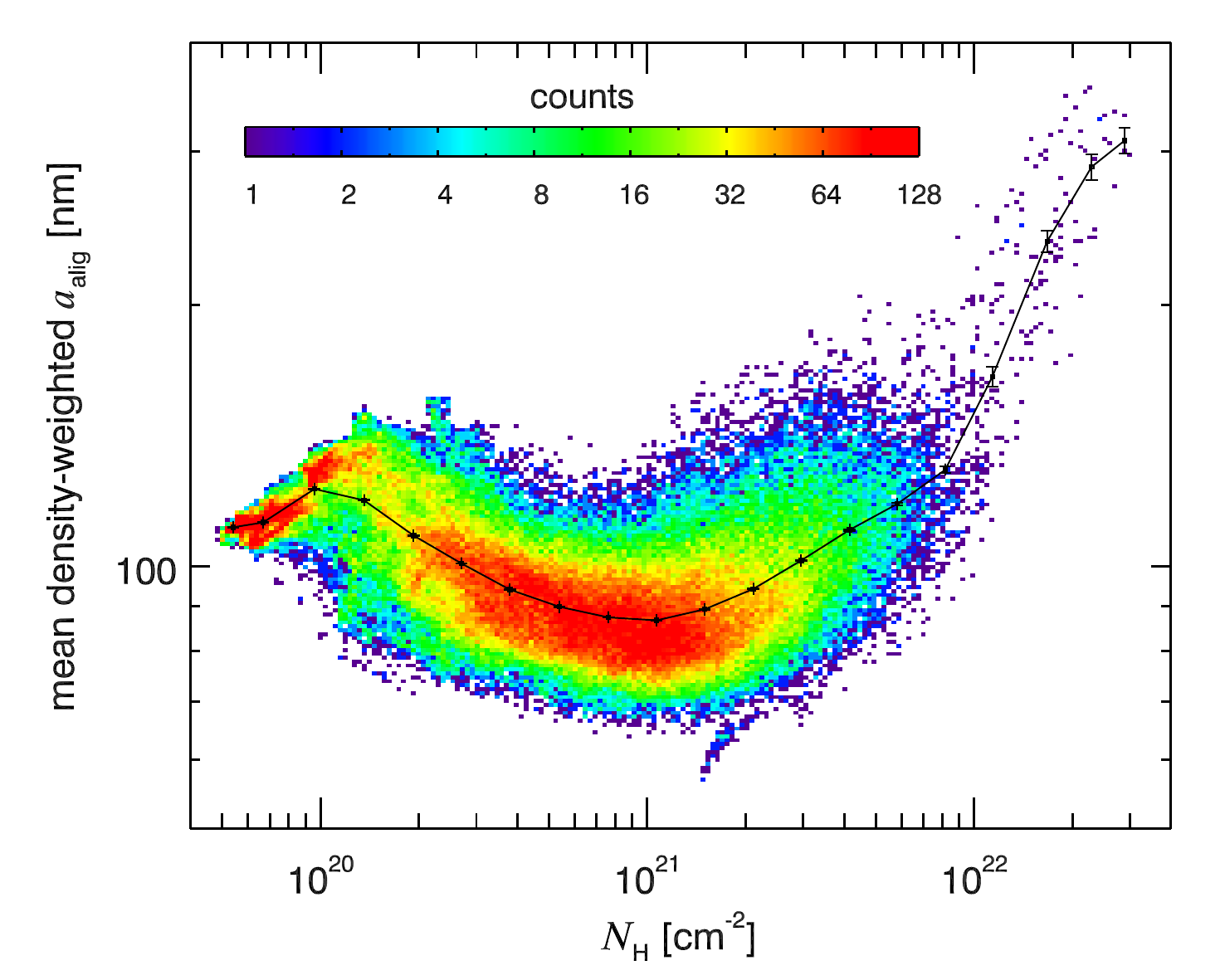}
\caption{Mean, density-weighted, $a_{\rm alig}$ parameter for our ISRF case, as a function of the column density for our simulated cube, combining viewing angles along $x$, $y$, and $z$.}
\label{fig:aalig_NH}
\end{figure}

Figure~\ref{fig:aalig_NH} shows the dependence of the mean, density-weighted, $\aalig$ parameter, as a function of the column density, for any LOS along the three axes of the cube. The value of $\langle\aalig\rangle$ is rather uniform on a large range of column densities, from $4\times 10^{20}$ to $2\times 10^{21}$ cm$^{-2}$, but increases at the lowest and highest column densities. A trend of similar shape is reported in \cite{Seifried2019} for the dependency of the alignment radius $\aalig$ on gas density $\ngas$. However, their MHD data set has about a one order of magnitude lower gas densities and temperate and a $\Gzero>0$ leading to values of $\aalig$ up a factor of 6.5 smaller than ours.

\begin{figure}
\includegraphics[width=0.49\textwidth]{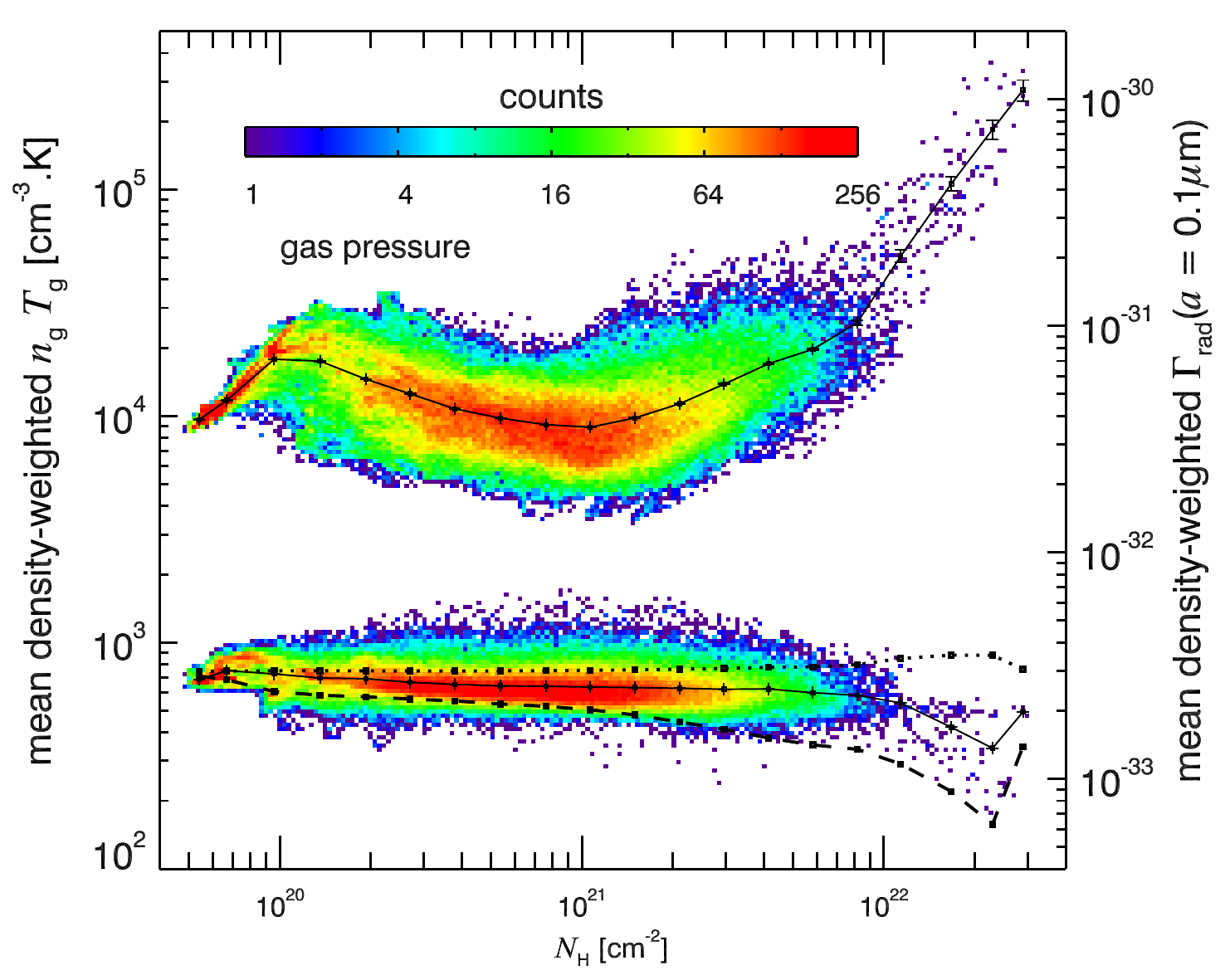}
\caption{
Mean, density-weighted, gas pressure $\ngas\,\Tgas$ (left axis) and radiative torque $\Gamma_{\mathrm{rad}}$ (right axis), as a function of the column density for our simulated cube, combining viewing angles along $x$, $y$, and $z$. To avoid a biased comparison, both axes share the same amplitude in log. This figure is to be compared with Figure~\ref{fig:aalig_NH}. A comparison with our simple model using $\urad$ (dotted) and $G_0$ (dashed) for the calculation of $\Gamma_{\mathrm{rad}}$ is overplotted. 
}
\label{fig:nT_RATs_NH}
\end{figure}

To understand what drives grain alignment, we plot on Figure~\ref{fig:nT_RATs_NH} how the mean, density-weighted, gas pressure $n_{\rm gas}\,T_{\rm gas}$ (responsible for grain disalignment) and radiative torque $\Gamma_{\rm rad}$ (responsible for grain alignment)  calculated for a grain size $a=0.1\,\mu$m 
depend on the column density.

The comparison of Figure~\ref{fig:nT_RATs_NH} with Figure~\ref{fig:aalig_NH} makes it clear that, unlike what is usually assumed, it is the variations in the gas pressure that drive the variations of grain alignment, and not the variations of the radiative torques through dust extinction. The latter is almost constant, slightly decreasing with $N_{\rm H}$. The decrease of the radiative torques intensity cannot therefore be invoked to explain the decrease of the alignment efficiency within the range of column densities present in our simulation.

\begin{table}
    \centering
    \begin{tabular}{c|c}
         $G_0$&  0.21 \\
         $1\,/\,\left(n_{\rm gas}\,T_{\rm gas}\right)$& 0.91 \\
         $G_0 \,/\,\left(n_{\rm gas}\,T_{\rm gas}\right)$ & 0.86 \\
$\left\langle\gamma\right\rangle\, \left\langle\cos{\vartheta}\right\rangle\, G_0 \,/\,\left(n_{\rm gas}\,T_{\rm gas}\right)$ & 0.90 \\
    \end{tabular}
    \caption{Pearson coefficients for the correlation of $\log(\aalig)$ with the $\log$ of different physical quantities. In the diffuse and translucent ISM, $\aalig$ is primarily driven by the gas pressure, not by the characteristics of the radiation field (direction, anistropy factor, or intensity). An increasing intensity of the radiation field even tends to disalign grains by increasing FIR photon emission.}
    \label{tab:pearson}
\end{table}

Table.~\ref{tab:pearson} quantifies this interpretation by presenting the value of the Pearson correlation coefficient between the alignment parameter $\aalig$ and different physical quantities characterizing the local ISM in our simulation, such as the density, temperature, gas pressure, radiation field intensity. A positive (resp. negative) correlation coefficient means that an increase of the quantity tends to increase (resp. decrease) $\aalig$, and therefore to disalign (resp. align) grains. 
The correlation between the radiation field intensity as measured by $G_0$ and the grain alignment parameter $\aalig$ is weak but, surprisingly, positive. This results from two competing effects of the radiation field on the RATs efficiency. These are the spin-up effect of radiative torques, expressed by Equation~\ref{eq:JradoverJth}, and the disaligning effect of FIR emission, described by Equation~\ref{eq:FIR}. In the WNM phase of the diffuse ISM, where the gas temperature is high and dust extinction remains weak everywhere, it is the latter effect that dominates over the former. This implies that in the WNM, an increase in the radiation field intensity makes the grain alignment efficiency decrease, not increase, due to the damping of grain rotation by the emission of IR photons.

 Grain alignment in the diffuse ISM is therefore primarily driven by gas pressure, and therefore by disalignment, while the alignment capacity of RATs is almost constant. 
The anisotropy of the radiation field $\gamma$, or the cosine of the angle $\vartheta$ between the radiation field anisotropy and the magnetic field
only act as secondary factors which are not able to produce any significant patterns in the correlation of $\aalig$ with $\NH$.

\subsection{Statistical analysis of dust polarization maps}

\begin{figure}
\includegraphics[width=0.49\textwidth]{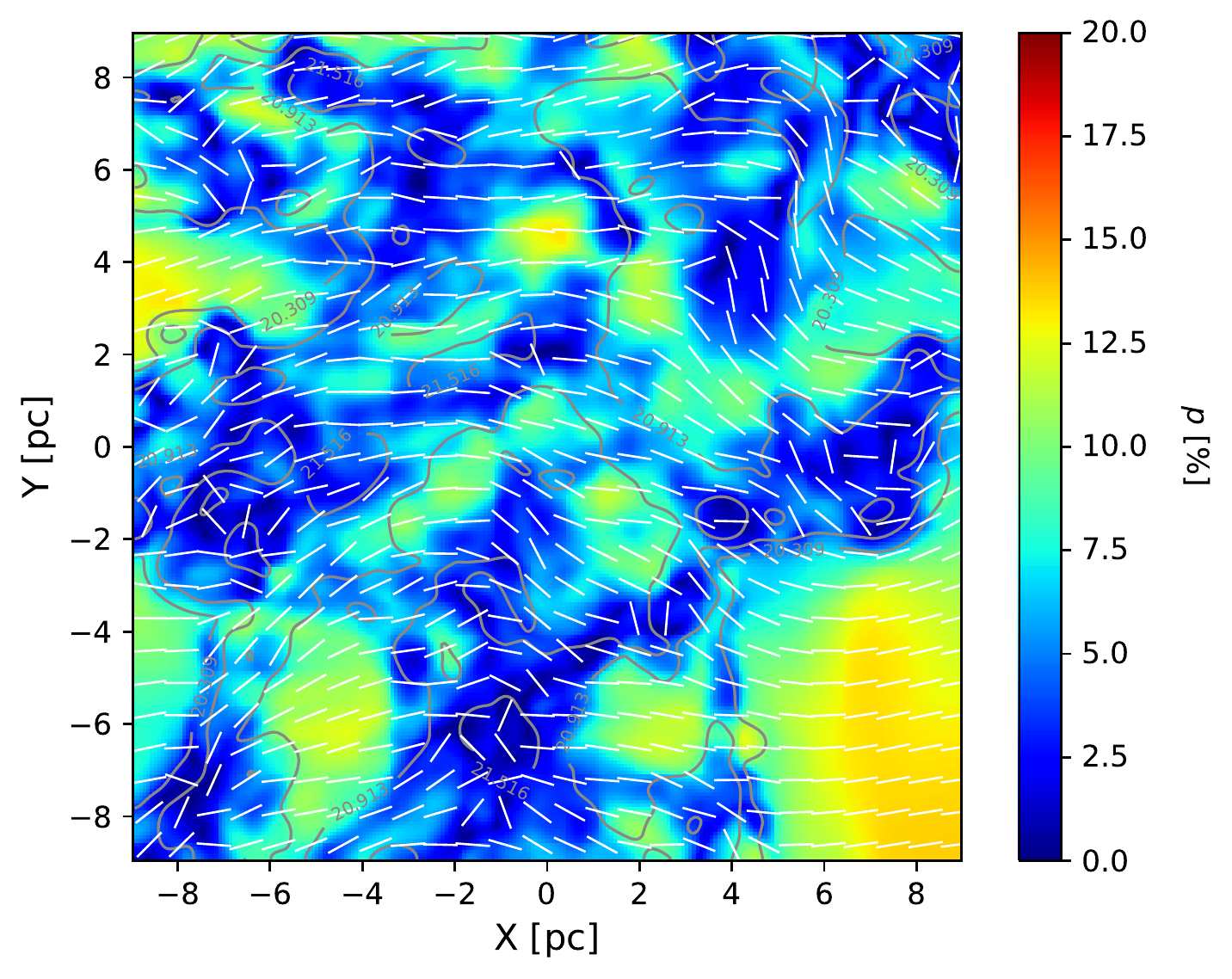}
\includegraphics[width=0.49\textwidth]{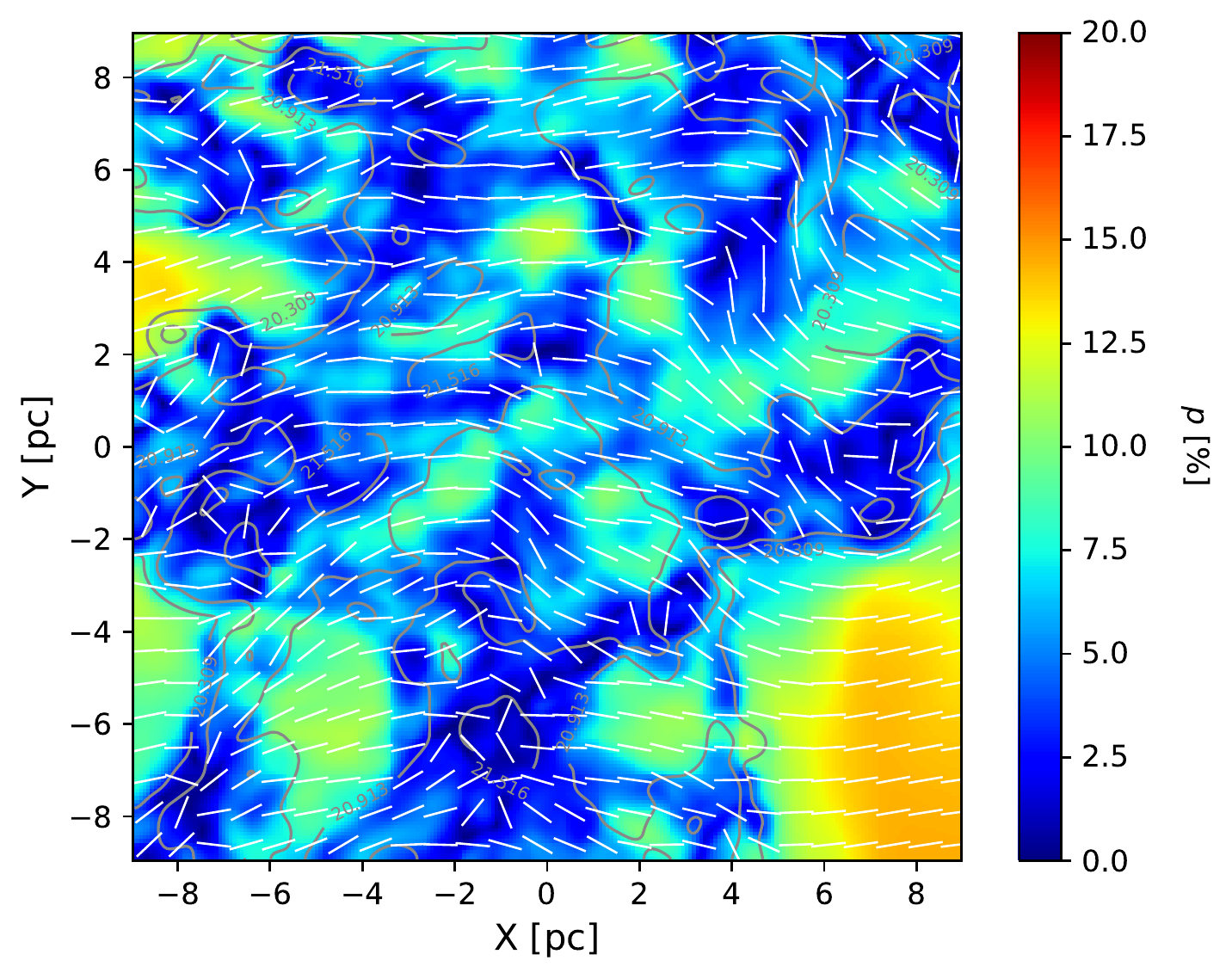}
\caption{Simulated maps of the polarization fraction at $353\ \mathrm{GHz}$ for the RAT (top) and FIXED (bottom) alignment cases. The contour lines show the column density. The white segments give the orientation of the magnetic field derived from the polarization angle.}
\label{fig:PsI_map_ISRF}
\end{figure}

In Figure~\ref{fig:PsI_map_ISRF} we show the resulting polarization maps for the ISRF setup with RAT alignment, and for the FIXED alignment setup. The general polarization pattern resembles the maps presented in \cite{Planck2015XX}, with peak values about $10\ \%$ lower. 
The cases RAT and FIXED are almost identical, with some minor amplification in the overall magnitude of polarization fraction $p$ in the latter case. The characteristic hallmarks of RAT alignment (angular dependency of $p$ with the radiation and magnetic field direction as well as the increase of $p$ with a higher radiation) seem not to cause any signature in the polarization signal shown in Figure~\ref{fig:PsI_map_ISRF}. 
A comparison of the polarization vectors (rotated by 90$^\circ$) with the averaged magnetic field orientation presented in Figure \ref{fig:MHD_input} shows that they do not perfectly match over the entire map. This demonstrates that dust polarization patterns cannot be simply interpreted as a projection of the magnetic field direction onto a plane. Hence, quantitative interpretation requires modeling by means of RT simulations including proper dust alignment physics. 

\begin{figure}
\includegraphics[width=0.49\textwidth]{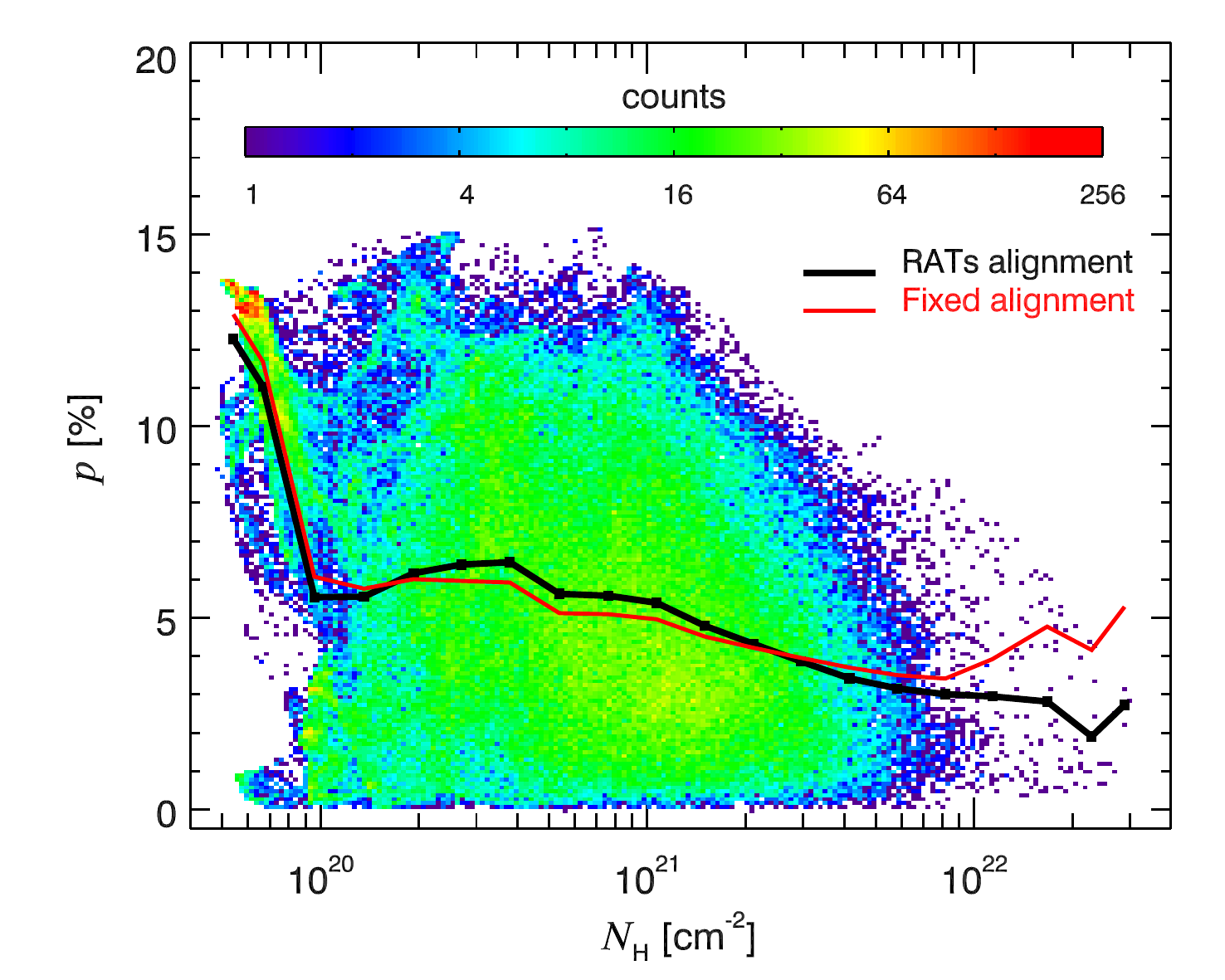}
\caption{
Polarization fraction $p$ at $353\ \mathrm{GHz}$at 5 arcmin of resolution, as a function of the column density for the ISRF-RAT simulation. The mean trend is overplotted for RATs alignment case (black) and for the FIXED alignment case (red).
}
\label{fig:PsI_NH}
\end{figure}

Figure~\ref{fig:PsI_NH} presents how the polarization fraction varies with the column density in our simulation, for all LOS along the three axes of the cube. The mean trend is compared for the RAT and FIXED alignment cases. The RAT case starts to depart from the FIXED case for $\NH > 2\,10^{21}$\,cm$^{-2}$ (or $A_V=1$) predicting systematically lower polarization fractions. As discussed in Section~\ref{sec:whatdrivesRAT}, this is not due to dust extinction, but to the higher pressure encountered in denser environments. This departure is however quite small in the range of column densities covered with a sufficient statistics by our simulation. 

\begin{figure}
\includegraphics[width=0.49\textwidth]{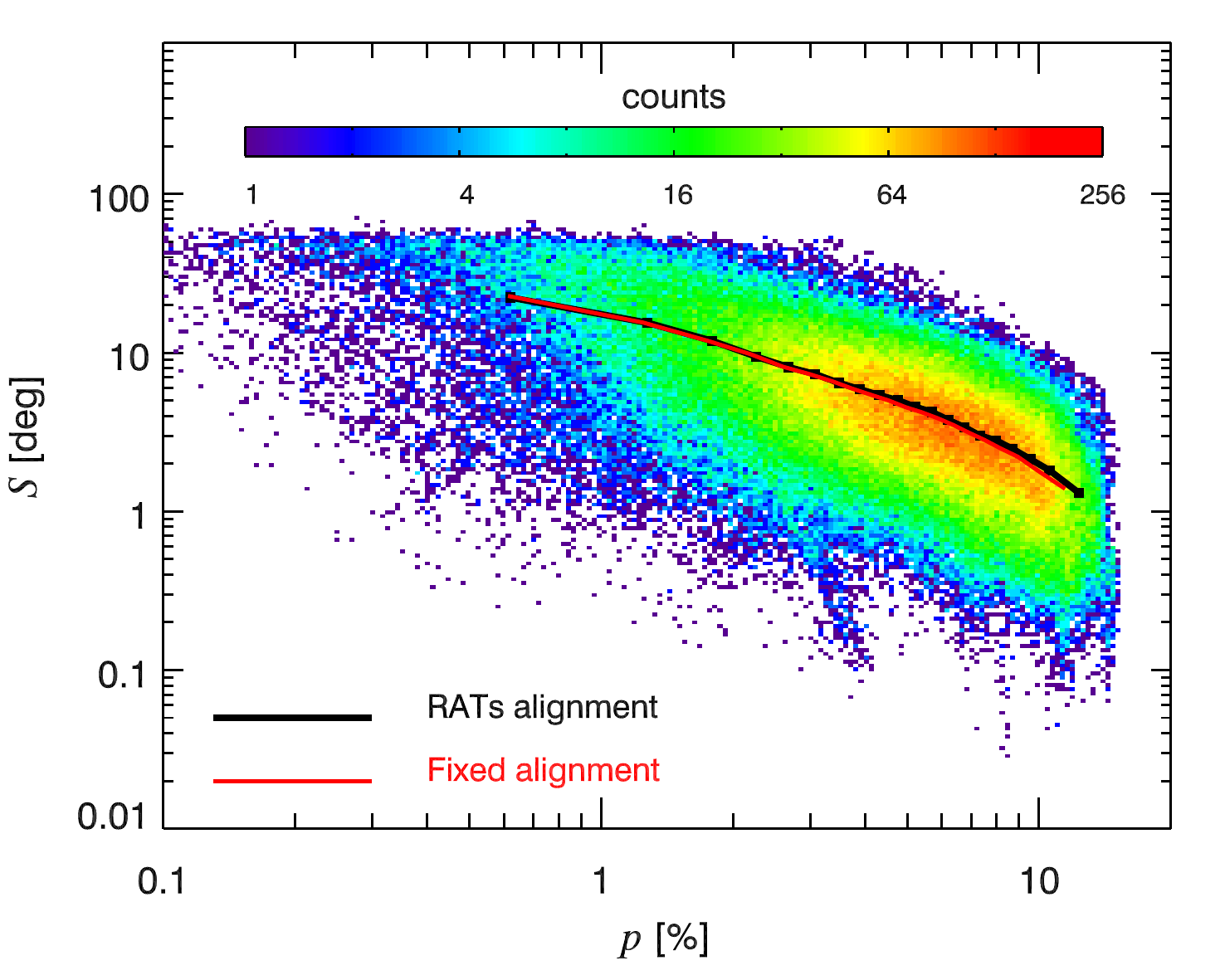}
\includegraphics[width=0.49\textwidth]{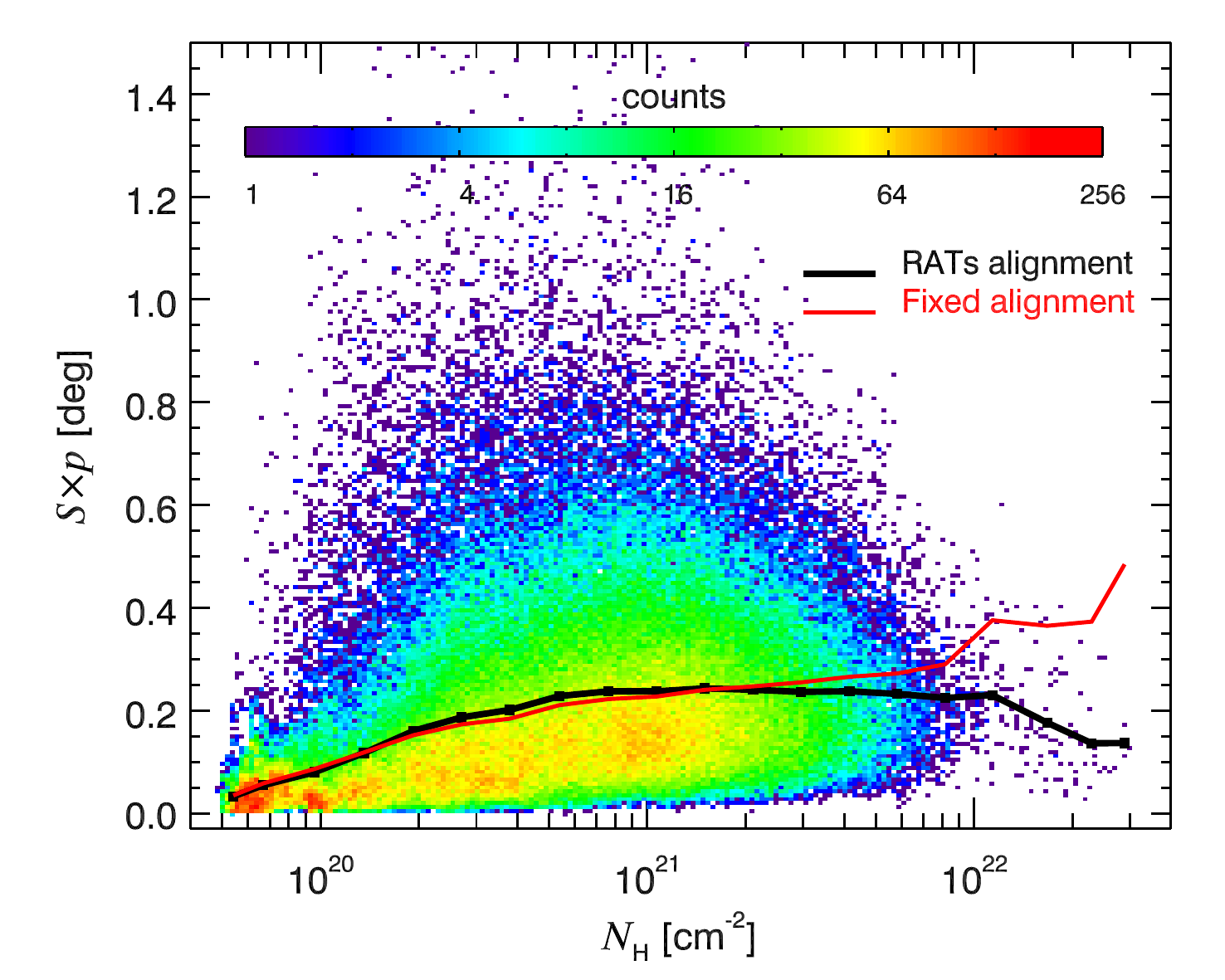}
\caption{Top: Dispersion of polarization angles $\S$ as a function of the polarization fraction $p$, all taken at $353\ \mathrm{GHz}$, combining viewing angles along $x$, $y$, and $z$. 
Bottom panel: Same for the product $\S\times p$ considered as a tracer of grain alignment efficiency. Mean trends for RAT alignment (black) and FIXED alignment (red) are overplotted.
}
\label{fig:Sp_NH}
\end{figure}

Figure~\ref{fig:Sp_NH} allows to extend our analysis by studying the product $\StimesPsI$ which was proposed by \cite{Planck2018XII} as a tracer of the grain alignment efficiency. The top panel shows that $\S$ and $\PsI$ are anti-correlated, whether we align grains uniformly or following the RATs model. The bottom panel, which presents the variations of the $\StimesPsI$ product with the column density, confirms that the grain alignment predicted by RATs decreases with $\NH$ in our simulation from $\NH=2\,10^{21}$\,cm$^{-2}$. The value of the mean trend of $\StimesPsI$ is however harder to interpret. As discussed in \cite{Planck2015XX}, this particular \RAMSES\ simulation does not reproduce perfectly the observed inverse correlation $\S\propto 1/\PsI$. As a consequence, and unlike in \Planck\ data \citep[see][]{Planck2018XII}, we do not observe a constant $\StimesPsI$ with $\NH$. 

\section{Looking for signatures of RATs}
In this section, we modify the physical conditions in the cube so as to favour the observation of characteristic signatures of RATs, such as its dependence on the radiation field intensity and its angle-dependence \citep{LazarianHoang2007}.

\label{sec:RT_STAR}
\subsection{Results with a star at the center of the MHD simulation}
\begin{figure*}
\begin{center}
\includegraphics[width=.45\textwidth]{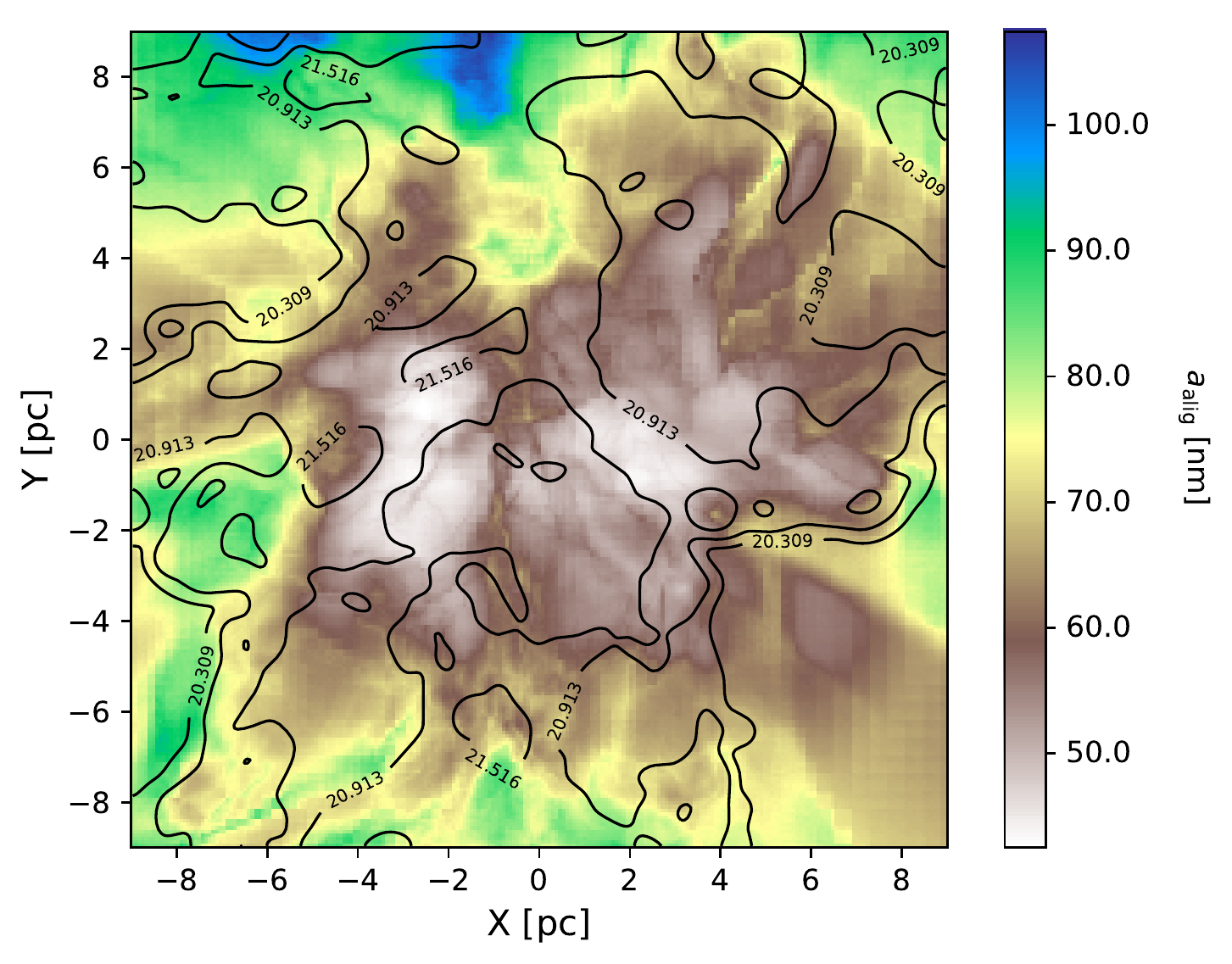}
\includegraphics[width=.45\textwidth]{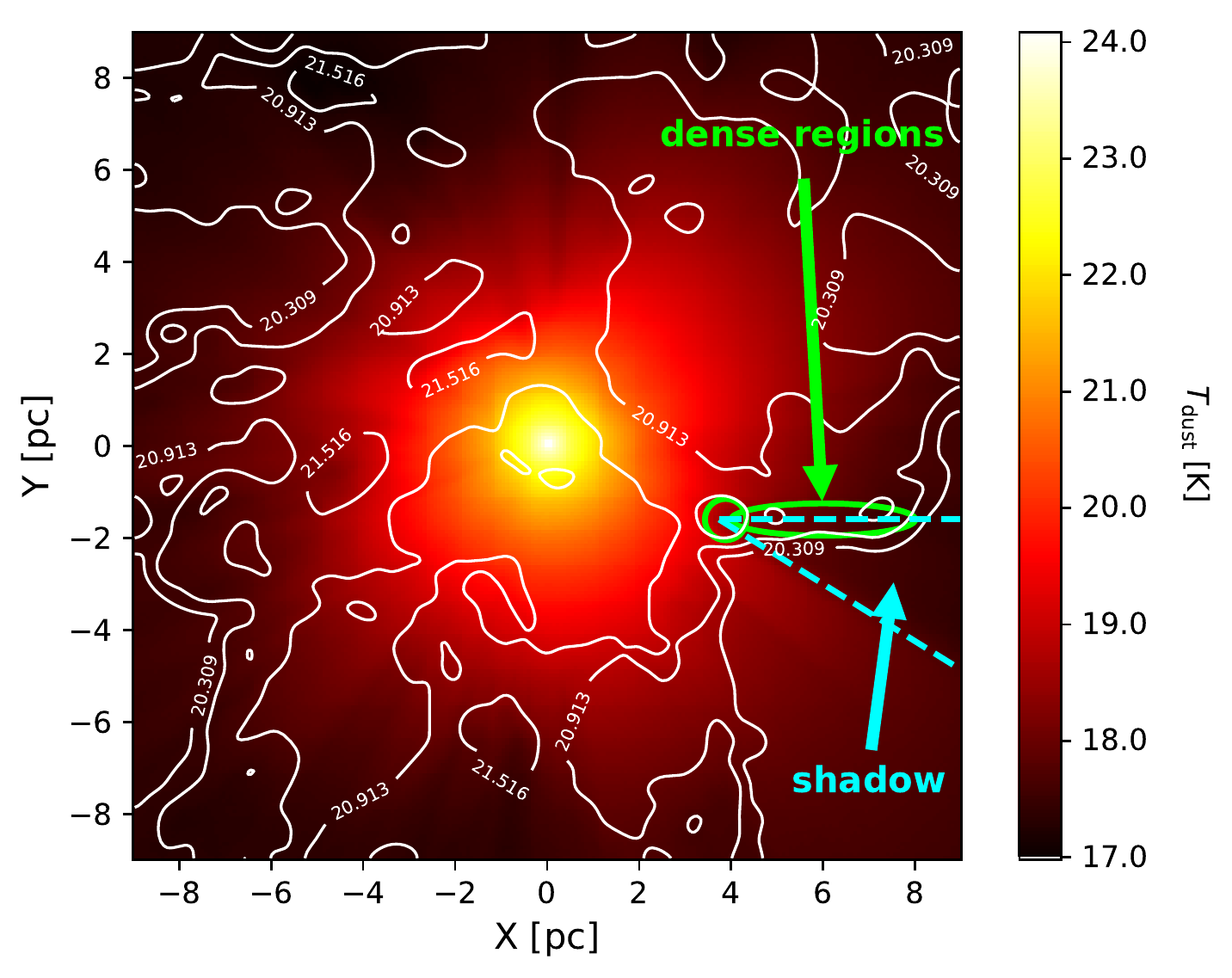}
\includegraphics[width=.45\textwidth]{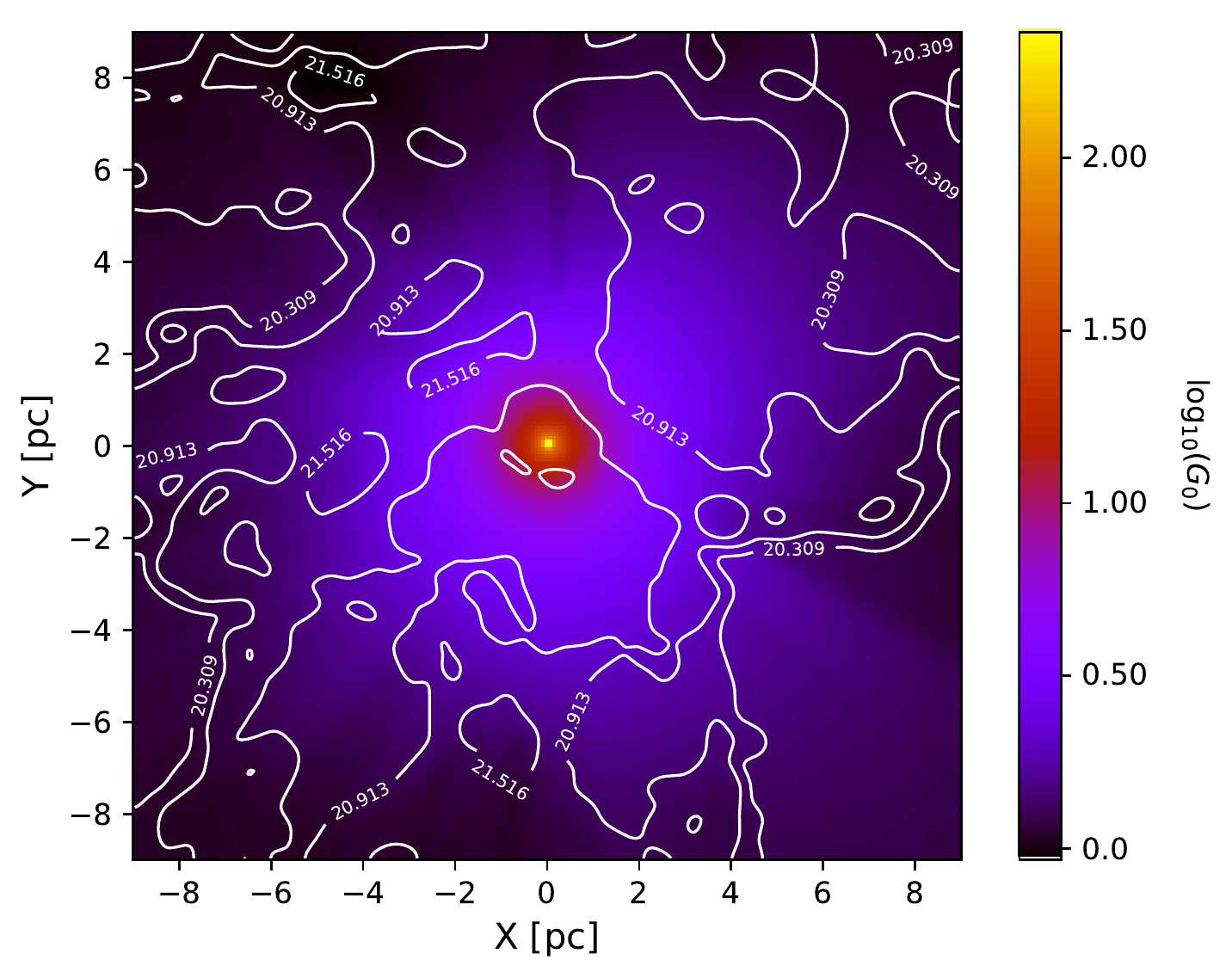}
\includegraphics[width=.45\textwidth]{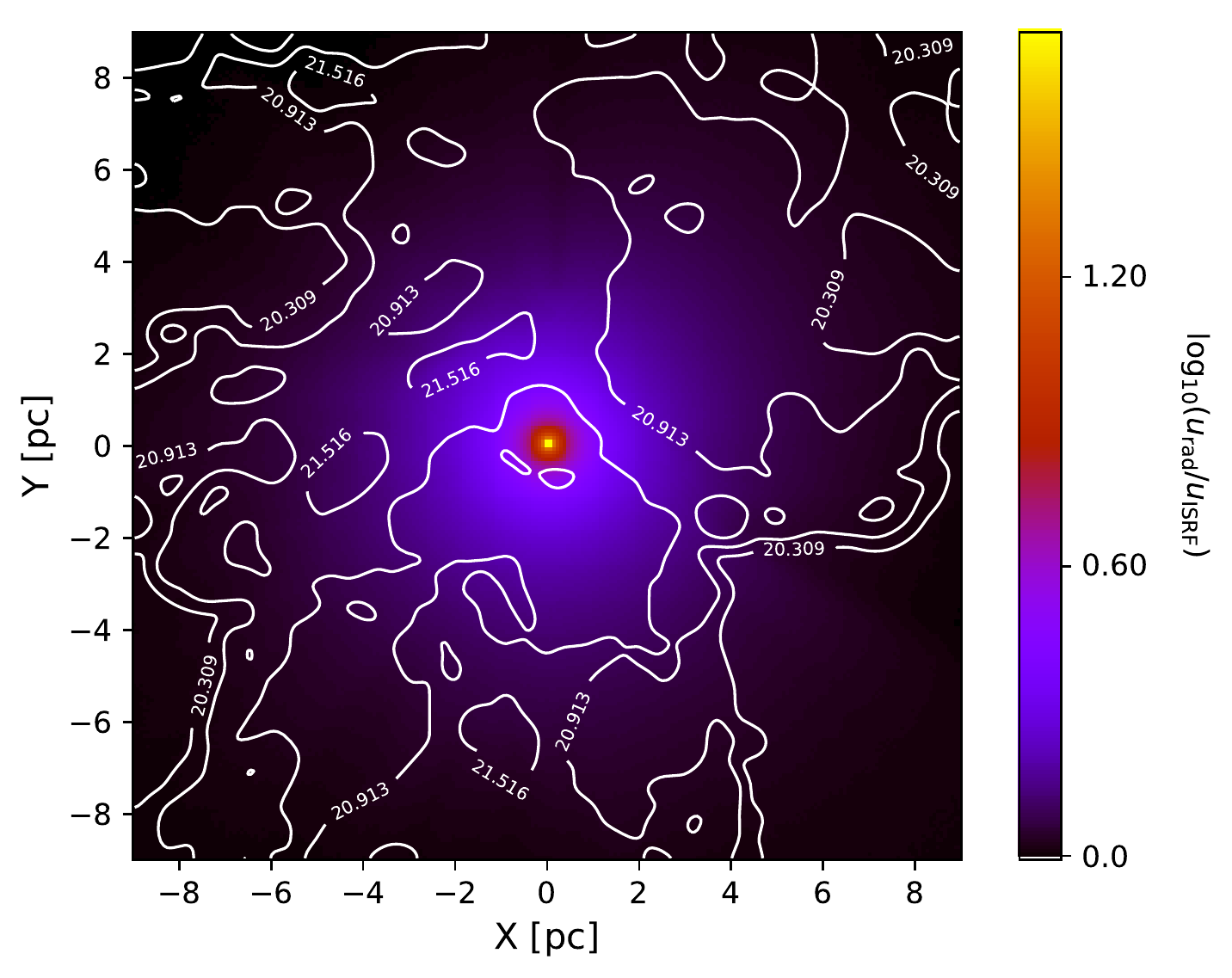}
\includegraphics[width=.45\textwidth]{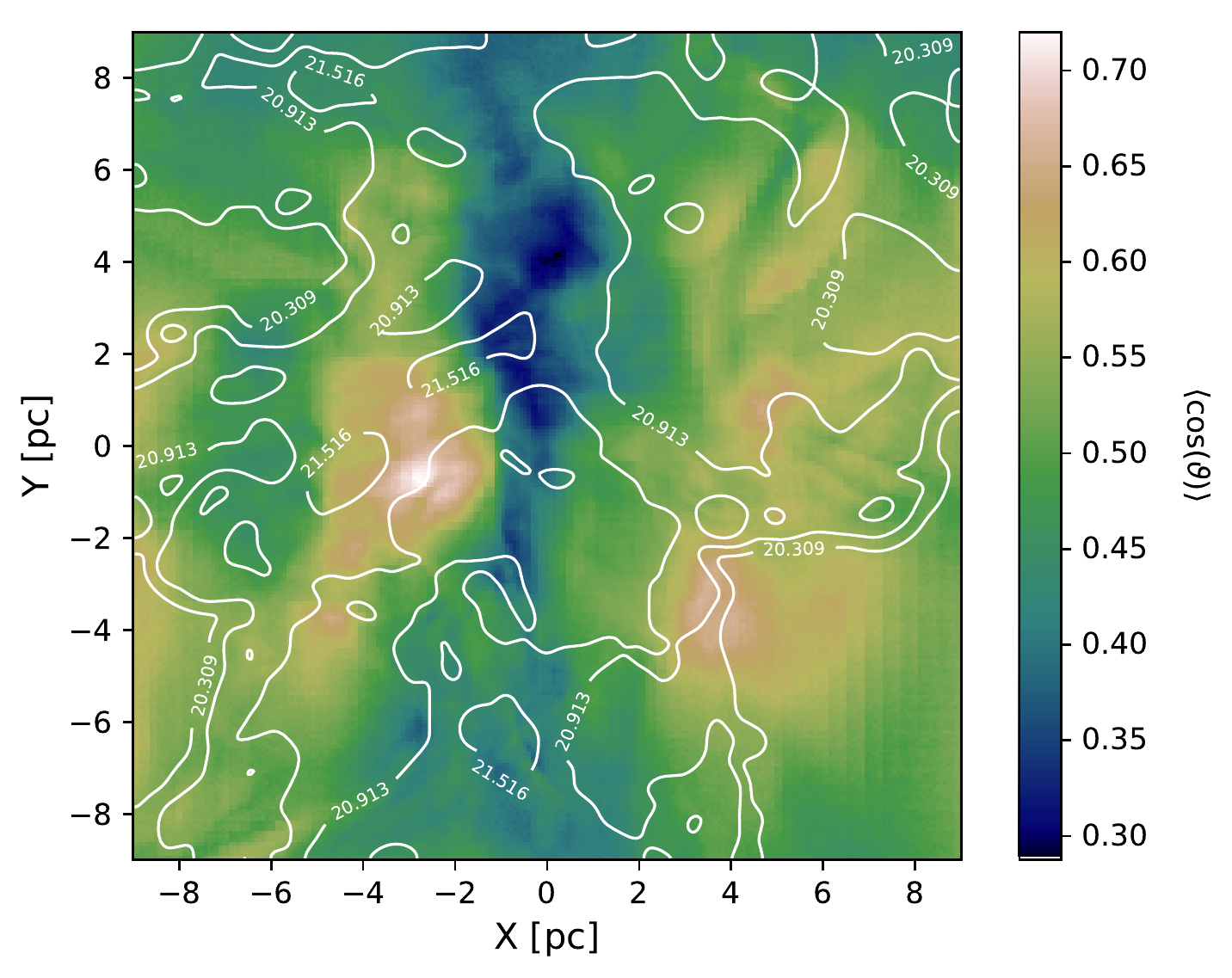}
\includegraphics[width=.45\textwidth]{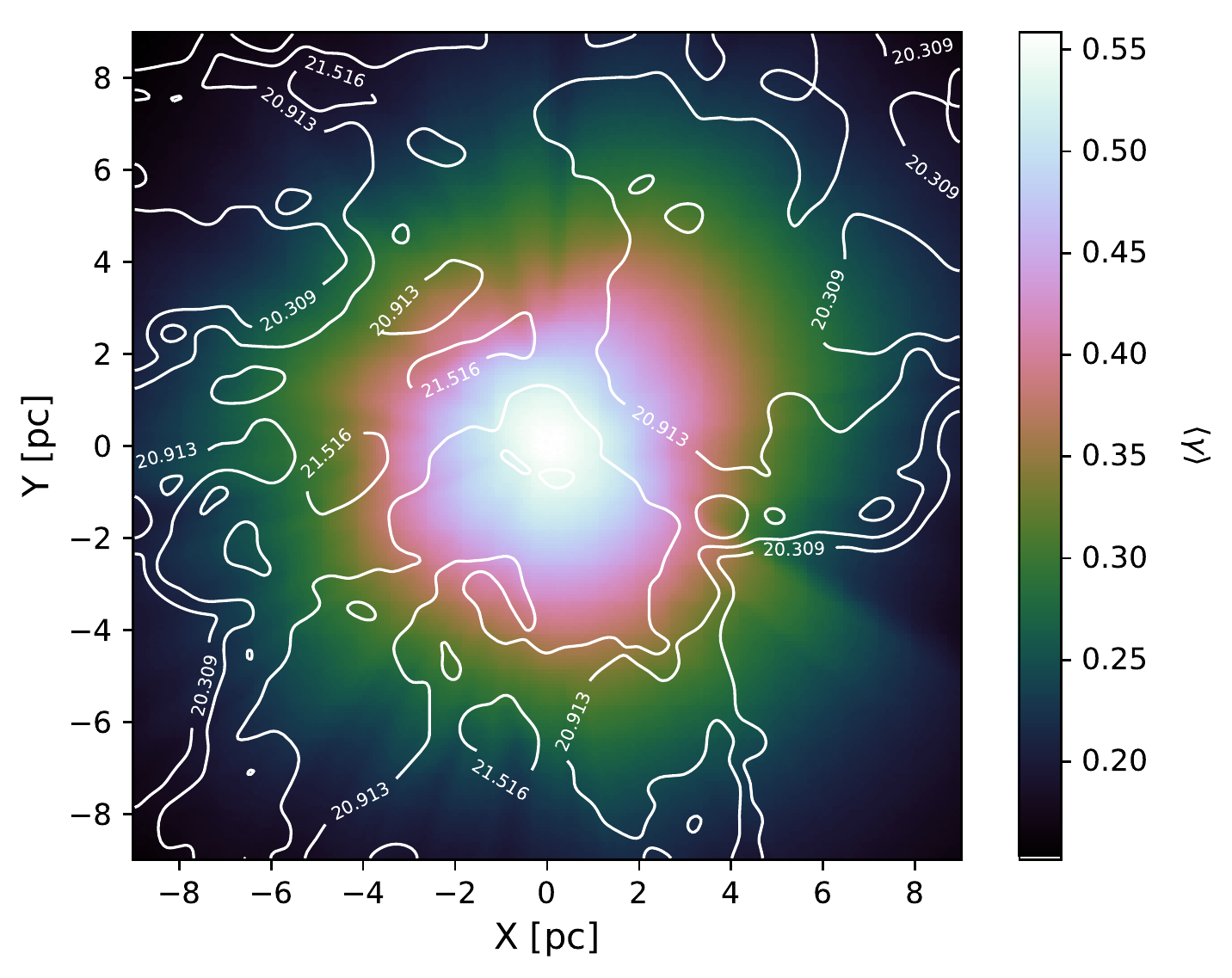}
\end{center}
\caption{The same as Figure \ref{fig:POLARIS_ISRF_RAT} for the  STAR-RAT setup.}
\label{fig:POLARIS_STAR_RAT}
\end{figure*}

In Figure~\ref{fig:POLARIS_STAR_RAT} we show the output of our MC simulation for the STAR setup where a star is introduced at the center of the cube without changing the MHD simulation (see Section \ref{sect:RTPostProcessing}). For the STAR setup the radiation field is clearly dominated by the central star, both in magnitude and direction. Consequently, the RAT alignment is most efficient in the center of the MHD cube with a minimum of the alignment parameter $\aalig$ down to $45\ \mathrm{nm}$, a maximum of about $250\ \mathrm{nm}$ and an average of $55\ \mathrm{nm}$. 

Here, the averaged map of $\aalig$ barely shows any resemblance to the  gas distribution. The only exception is at $X=4\ \mathrm{pc}$ and $Y=-2\ \mathrm{pc}$ where the clump with the highest density within the $\RAMSES\ $ simulation is situated. However, this effect is a result of the radiation from the star being shielded by the clump. Here, we note a lane of minimal grain alignment size starting at this clump going radially outwards. Such a shadowing effect is due to extinction of radiation in the densest regions of the cube (compare with Figure~\ref{fig:MHD_input}).

This shadowing is even more obvious for the average dust temperature $\Tdust$ map of Figure~\ref{fig:POLARIS_STAR_RAT}. Here, we highlight the densest regions and the resulting shadow by lines and arrows. As for the alignment efficiency we report a decreased dust temperature in regions that are shielded from radiation. Several similar features can be observed e.g.  directly above the star. This shadowing effect can also be seen in the  maps of $\Gzero$, $\urad$, and $\left\langle\gamma \right\rangle$, respectively. In detail, the anisotropy factor $\left\langle\gamma \right\rangle$ reaches values up to $0.56$ meaning that the radiation field has a stronger unidirectional component compared to the ISRF setup where we have $\left\langle\gamma \right\rangle \approx 0.1$ in the center of the map. 
The same is true of the quantity $\left\langle \cos(\vartheta) \right\rangle$: on average the alignment angles cluster around $\vartheta \approx 60^\circ$ but the STAR setup has much smaller values of $\left\langle \cos(\vartheta) \right\rangle$ along the Y-axis through the center where the radiation is perpendicular to  the direction of the large scale magnetic field. Hence, radiation and magnetic field direction are not randomly oriented with respect to each other in that region, with an anisotropic radiation field that is much stronger at the center. This configuration of the STAR setup represents a significantly different set of parameters regarding the  radiation field compared with the ISRF setup. 


\begin{figure}
\centering
\begin{minipage}[c]{1.0\linewidth}
      \includegraphics[width=1.0\textwidth]{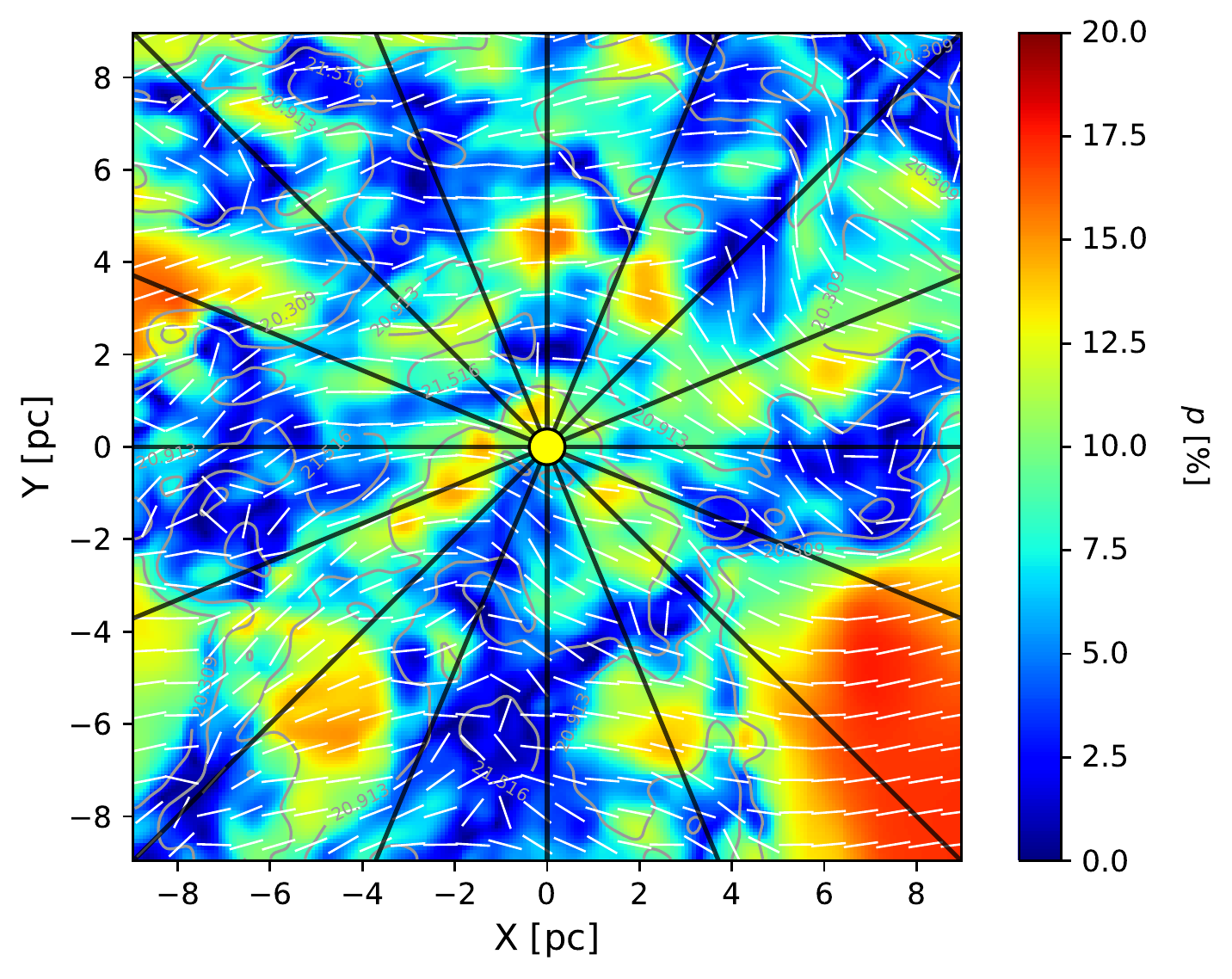}
\end{minipage}
\caption{Map of the polarization fraction in the STAR-RAT case, to be compared with the ISRF setup (Figure~\ref{fig:PsI_map_ISRF}, top panel). Black solid lines are the projected direction of the radiation field $\vec{k}$ while the central yellow dot indicates the position of the star and the contour lines represent the column density $\NH$.}
\label{fig:PsI_map_STAR}
\end{figure}

In Figure~\ref{fig:PsI_map_STAR} we show the resulting polarization maps for the STAR setup with RAT alignment. The map shows the idealized direction of the radiation  field drawn on it for later analysis (see Section \ref{sect:RAT_test}). 
Regarding the polarization pattern, Fig.~\ref{fig:PsI_map_STAR} does not significantly differ from Fig.~\ref{fig:PsI_map_ISRF}, or from the one presented in \cite{Planck2015XX}.
We compared polarization angles pixel by pixel between all combinations of ISRF and STAR setups with RAT or FIXED alignment (see Tables \ref{tab:Setups} and \ref{tab:Alignment}). Despite a significant change in the radiation field and subsequent RAT alignment between all these setups, the resulting polarization angles only differ by about $2^\circ$ on average. There is also no variation in $p$ that can be attributed to the shadowing effect observed in Figure \ref{fig:POLARIS_STAR_RAT}.
For the radiation field coming from the STAR setup the magnitude of the polarization $p$ increases only by about $3\ \%$. However, this increase is a general trend throughout the $p$ map and not only limited to the center region where the star is situated. We analyse and discuss this phenomenon in the following sections in further detail.

\begin{figure}
\includegraphics[width=.49\textwidth]{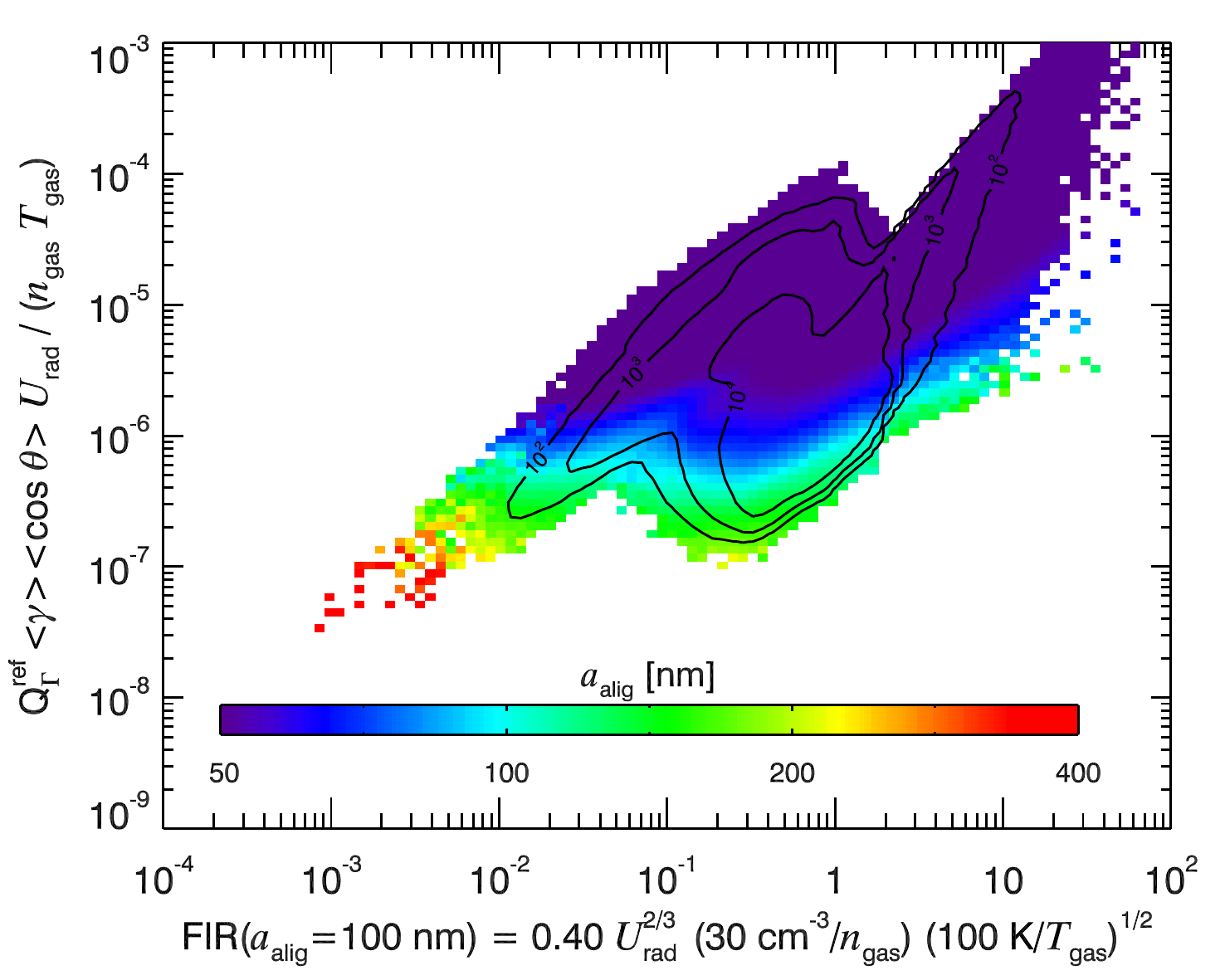}
\caption{Same as Figure~\ref{fig:aalig_POLARIS_ISRF} for the STAR-RAT case.}
\label{fig:aalig_POLARIS_STAR}
\end{figure}

Figure~\ref{fig:aalig_POLARIS_STAR}, similarly to Figure~\ref{fig:aalig_POLARIS_ISRF}, illustrates how the alignment parameter $\aalig$ varies in the phase diagram of Figure~\ref{fig:aalig_modelHL14}. With a star illuminating the cube, the whole physical quantities are driven toward the top right corner of the phase diagram. As a consequence, the alignment efficiency is globally increased everywhere in the cube, increasing the mean value of the polarization fraction on any LOS (Figure~\ref{fig:PsI_map_STAR}) without modifying the patterns observed for the ISRF case (Figure~\ref{fig:PsI_map_ISRF}).

\begin{figure}
\centering
\includegraphics[width=.49\textwidth]{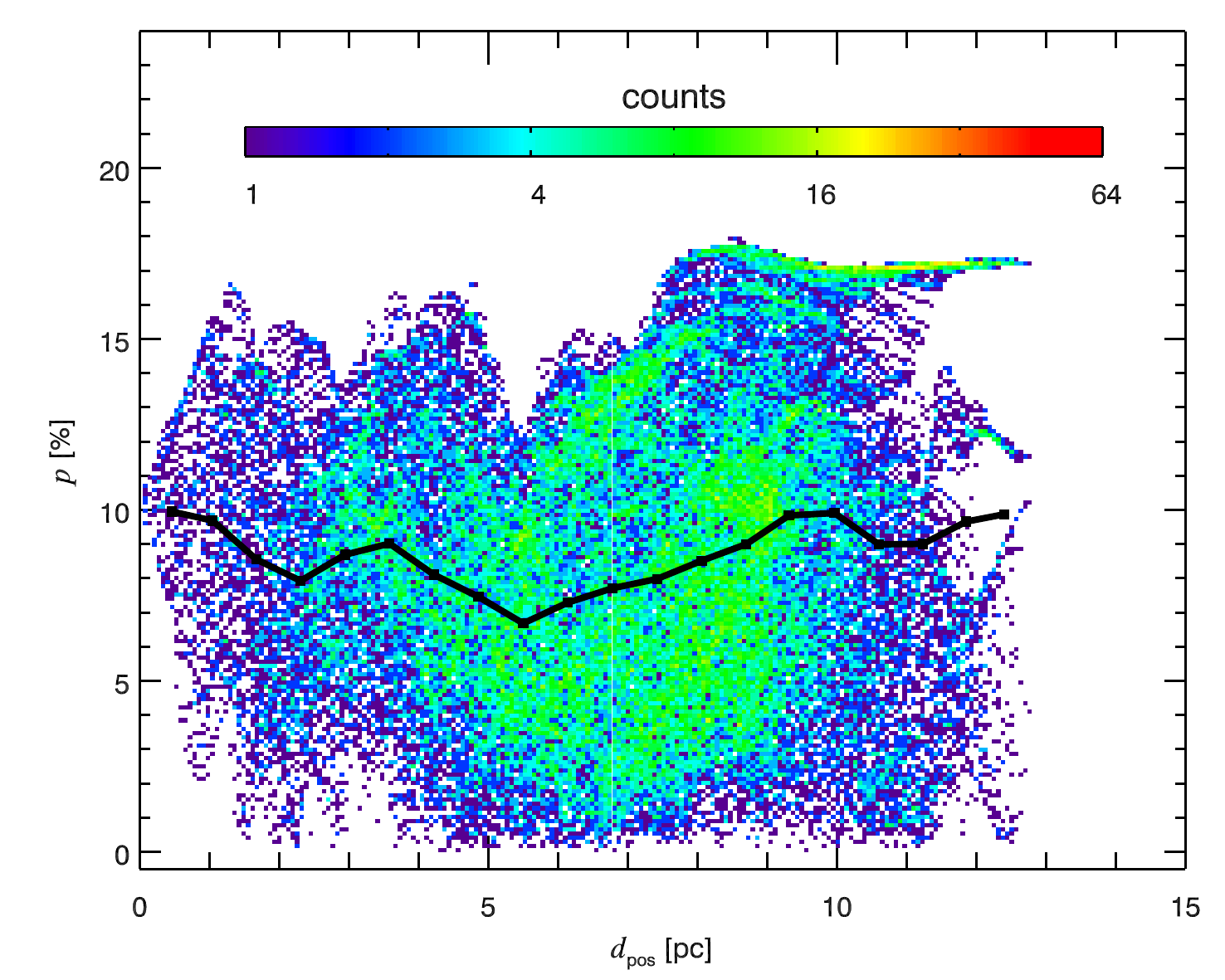}
\includegraphics[width=.49\textwidth]{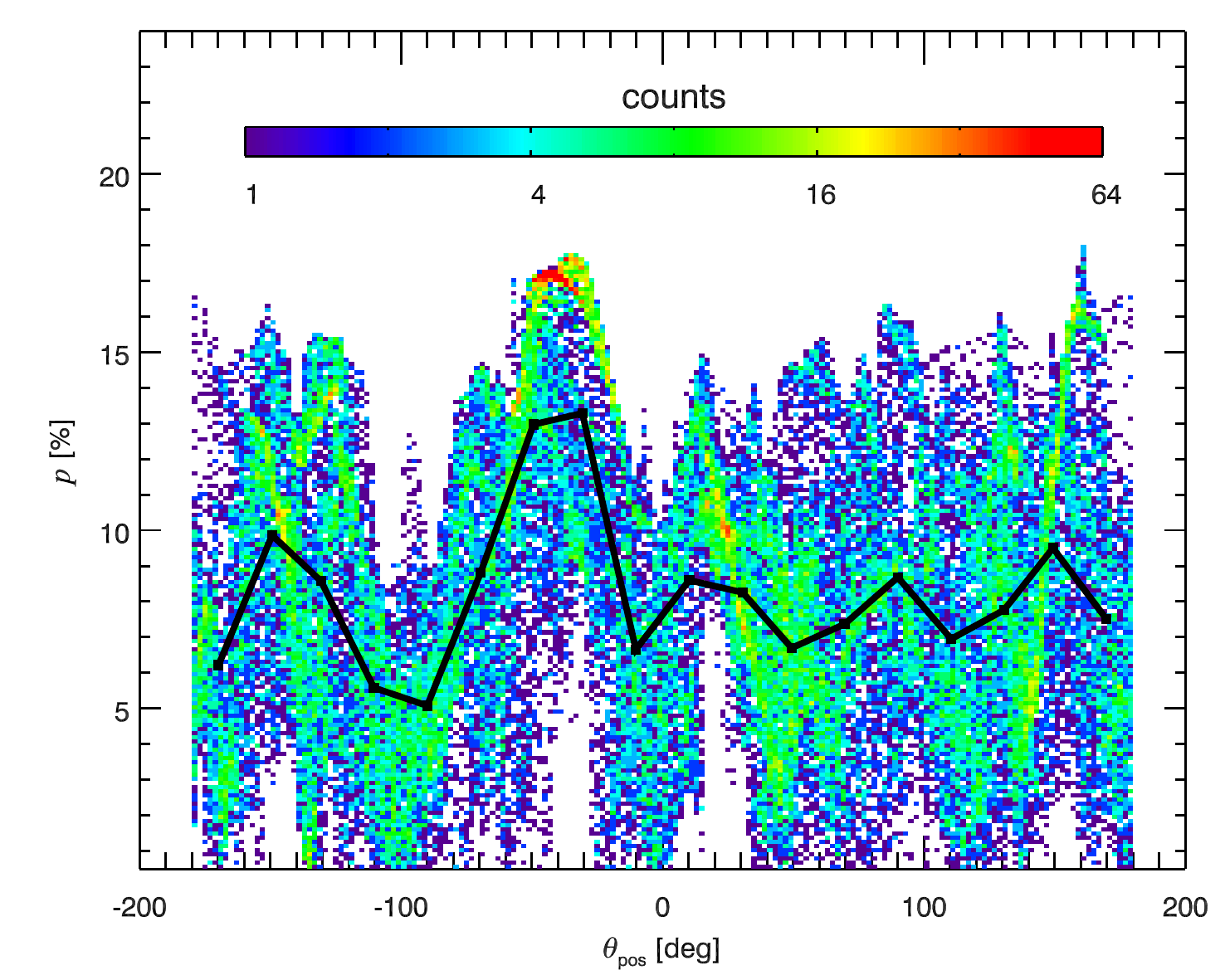}
\caption{Polarization fraction as a function of the distance to the star projected on the plane of the sky (top) and of the angle between the projected magnetic field and starlight direction (bottom), for the STAR-RAT setup. 
}
\label{fig:PsI_STAR_VG}
\end{figure}
 

\begin{figure}
\includegraphics[width=0.49\textwidth]{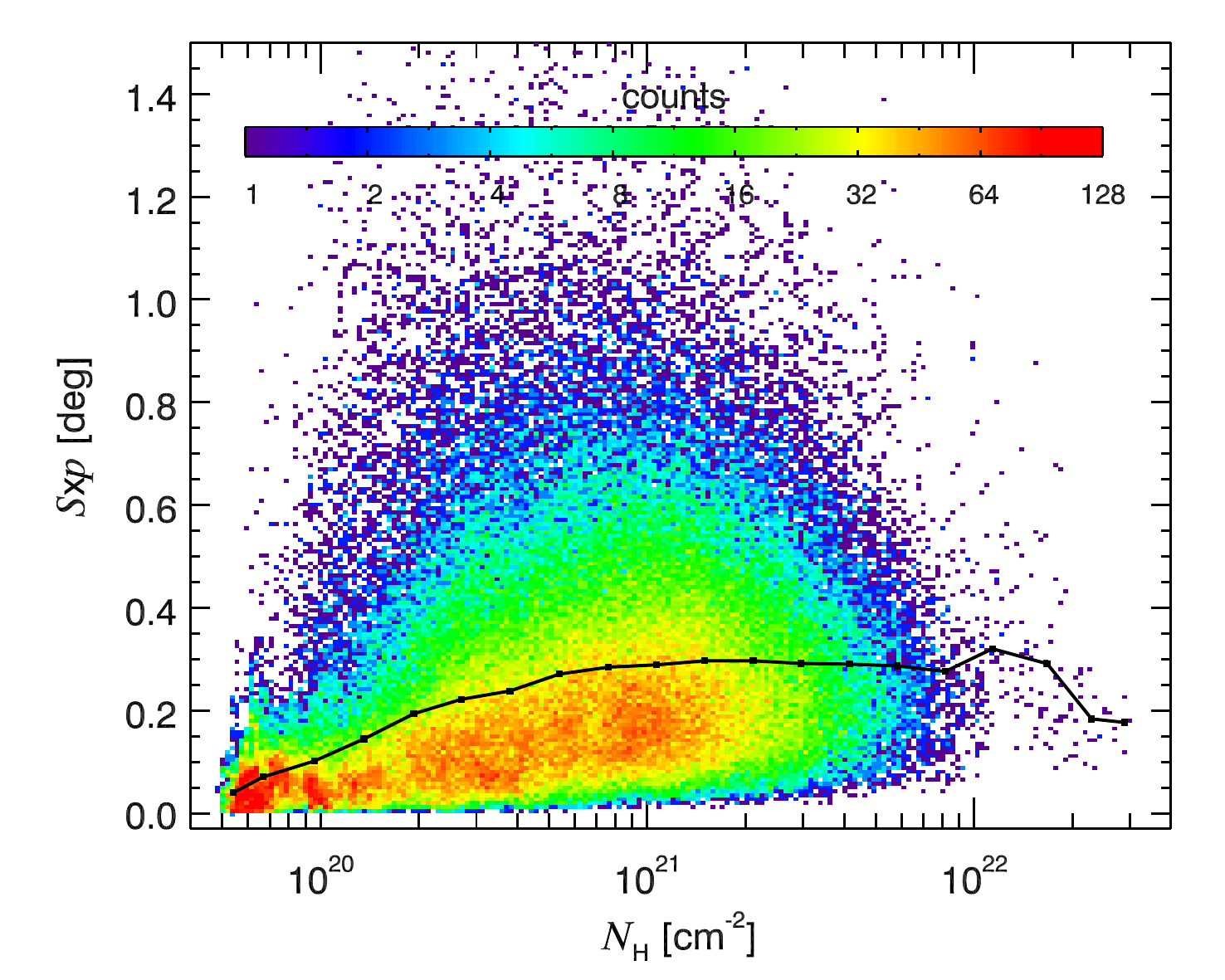}
\caption{$S\times p$ for the STAR case, considered as a tracer of grain alignment efficiency, as a function of the column density, with its mean trend overplotted.
}
\label{fig:Sp_NH_STAR}
\end{figure}

Despite the presence of a strong radiation field emitted by the star at the center of the cube, Fig.~\ref{fig:PsI_STAR_VG} show that we do not observe any systematic relation expected from the RATs theory, namely, a decrease 
of the polarization fraction with the distance to the star, or a sinusoidal dependence of the polarization fraction on the 2D-angle $\theta_{\mathrm{pos}}= \angle(\vec{k},\vec{B})$ 
between the projected directions of the magnetic field $\vec{B}$ (estimated from the rotated polarization vectors) and the assumed radiation field $\vec{k}$.
This is explained by two main factors. First, the physical quantities that characterize RATs alignment, namely the intensity, the direction and the anisotropy of the radiation field, do not vary at small scales by a factor that is strong enough to dominate over the other factors affecting the polarization fraction : the structure of the magnetic field on the line sight and within the beam, and the grain alignment randomization by gas collisions. Second, when the alignment is very efficient (as is the case when grains are irradiated by a star: $\aalig \sim 10$\, nm, see

Figure~\ref{fig:aalig_POLARIS_STAR}), strong variations in $\aalig$ do not produce a corresponding strong variation in the intrinsic polarization fraction of dust polarized emission because the dependence of $\PsI$ on $\aalig$ is not steep when $\aalig$ is small (see Figure~\ref{fig:pst_PsI_lambda}). As a consequence, $\PsI$ at $353\ \mathrm{GHz}$ does not trace the alignment efficiency very well even though the alignment is very efficient (low value of $\aalig$). 

The polarization fraction does not reflect only the variations in the alignment efficiency, but also the structure of the magnetic field. Studying the statistics of $\StimesPsI$ instead of $\PsI$, we can get rid of the influence of the magnetic field structure \citep{Planck2018XII}. However, as presented in Figure~\ref{fig:PsI_STAR_VG}, the mean dependency of $\StimesPsI$ with the distance and 2D-angle $\theta_{\mathrm{pos}}$ does not show any of the expected systematic trends either. 

We conclude that, under normal circumstances, the angle-dependence or distance-dependence of dust polarization with respect to a star is not present in simulated observations. However, this only holds true for the diffuse ISM case presented here whereas models of molecular clouds \citep[][]{Bethell2007,Hoang2014,Reissl2016} and circumstellar disks \citep[][]{Tazaki2017} do indeed reproduce the telltale signs for the presence of ongoing RAT alignment.

\subsection{Optimal configuration for detecting the angle dependence of RATs}
\label{sect:OptimalConfiguration}

\begin{figure}
\centering
\includegraphics[width=.49\textwidth]{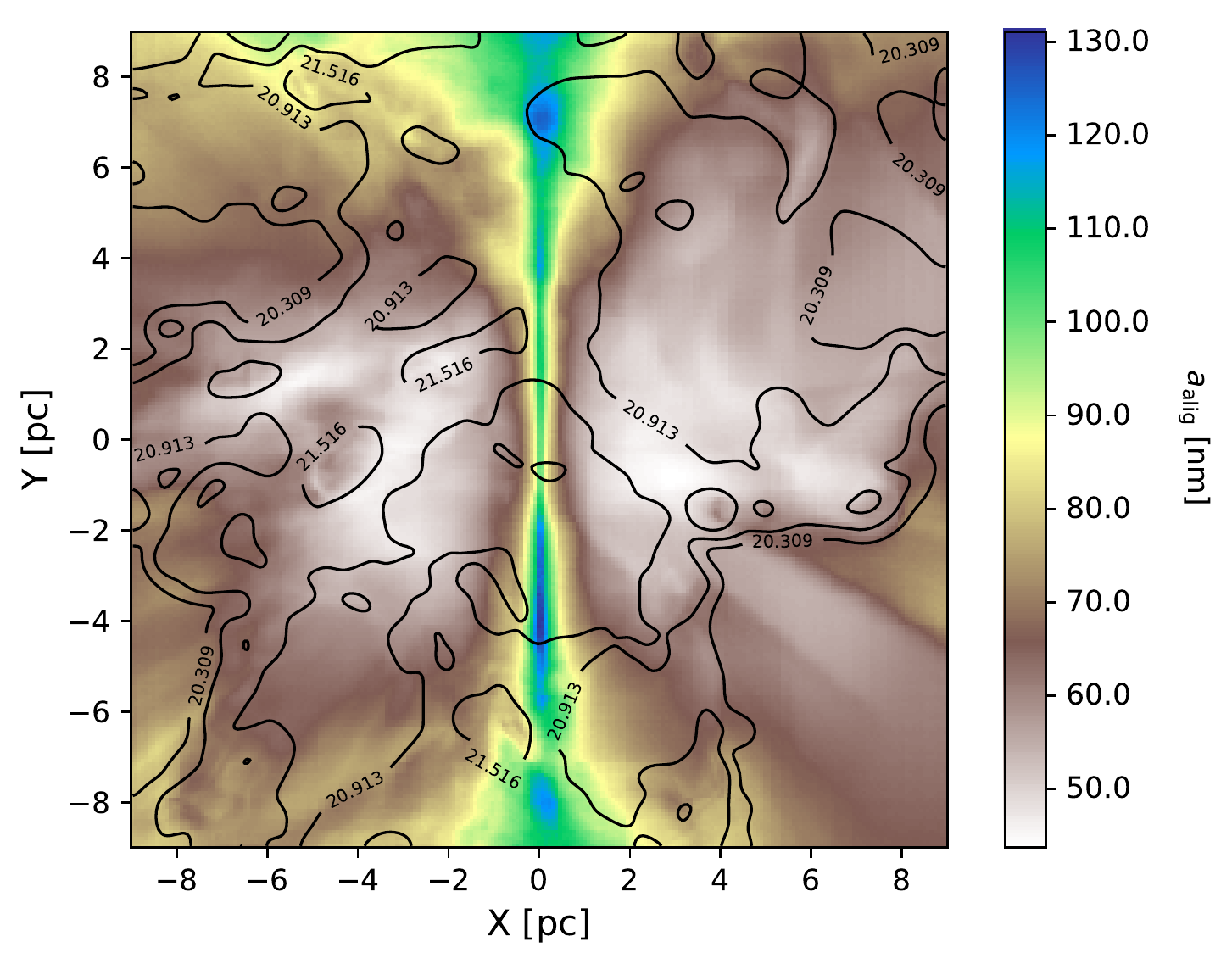}
\includegraphics[width=.48\textwidth]{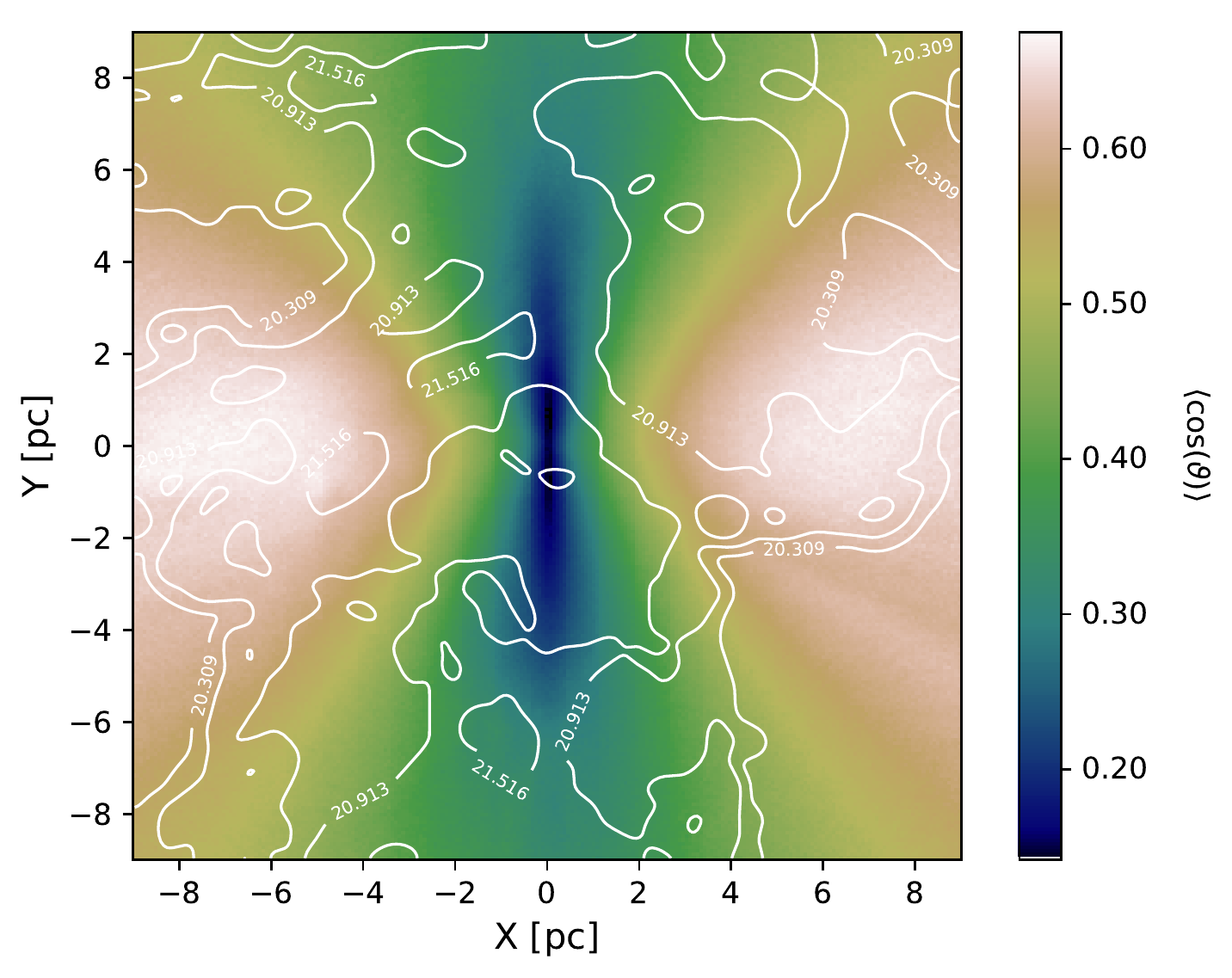}
\includegraphics[width=.49\textwidth]{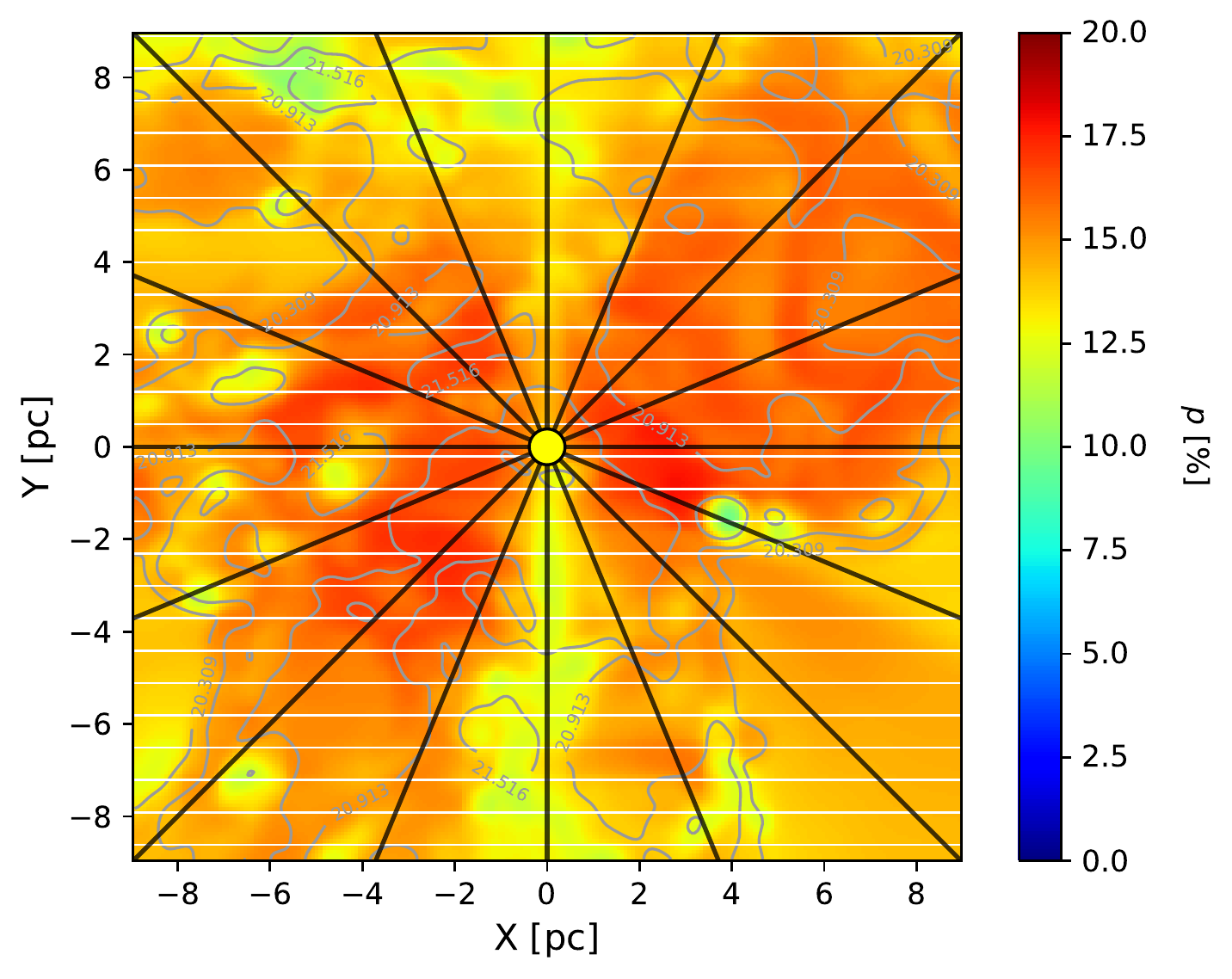}
\caption{The same case as setup STAR-RAT but with a uniform magnetic field along the $X$ direction. Top panel: projected alignment radius $\aalig$. Middle panel: projected average angle $\left\langle\cos{\vartheta}\right\rangle$. Bottom panel: linear polarization fraction overlaid with polarization vectors (white) rotated by $90^\circ$ tracing the magnetic field orientation $\vec{B}$.}
\label{fig:ConstField}
\end{figure}


\begin{figure}
\includegraphics[width=.49\textwidth]{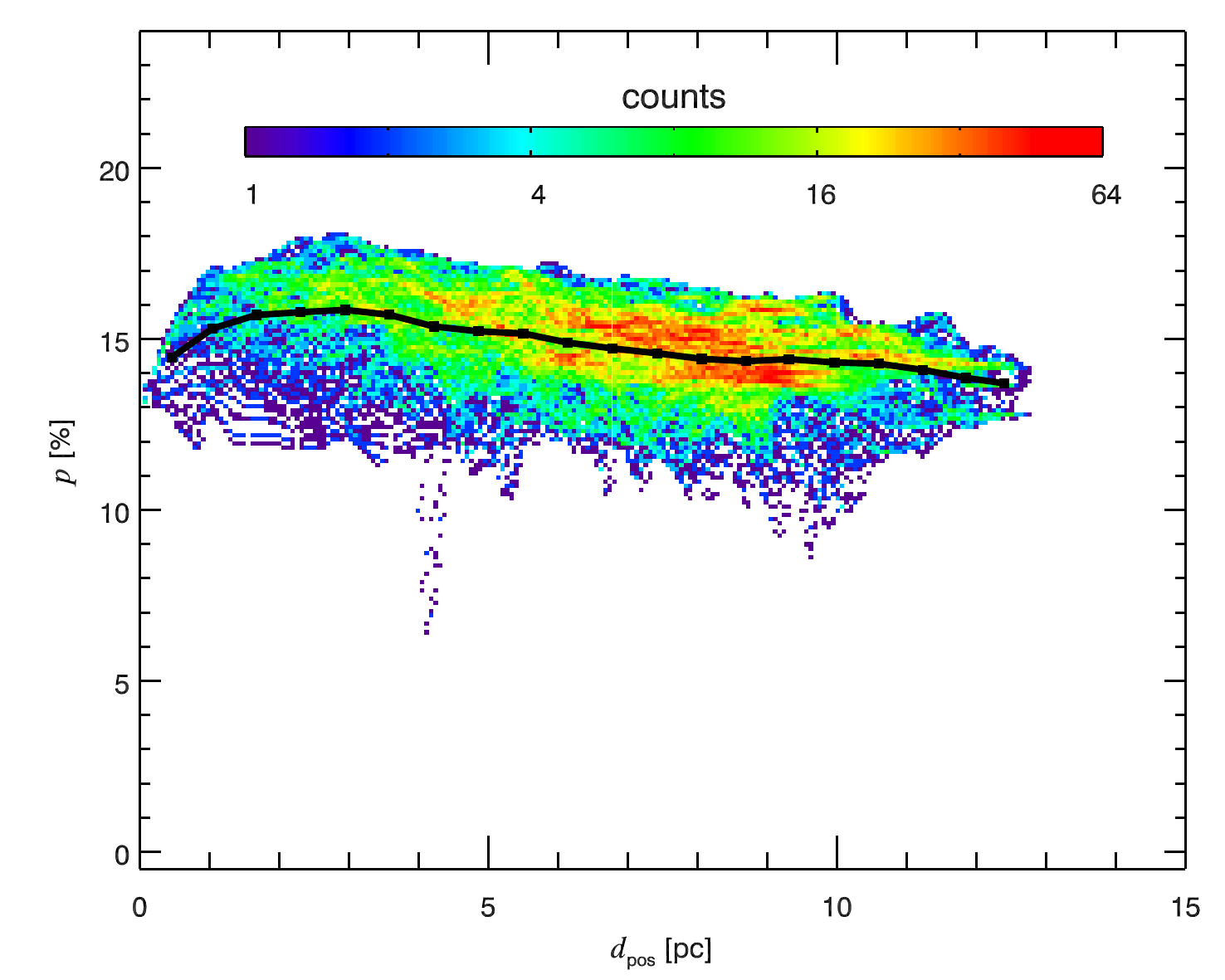}
\includegraphics[width=.49\textwidth]{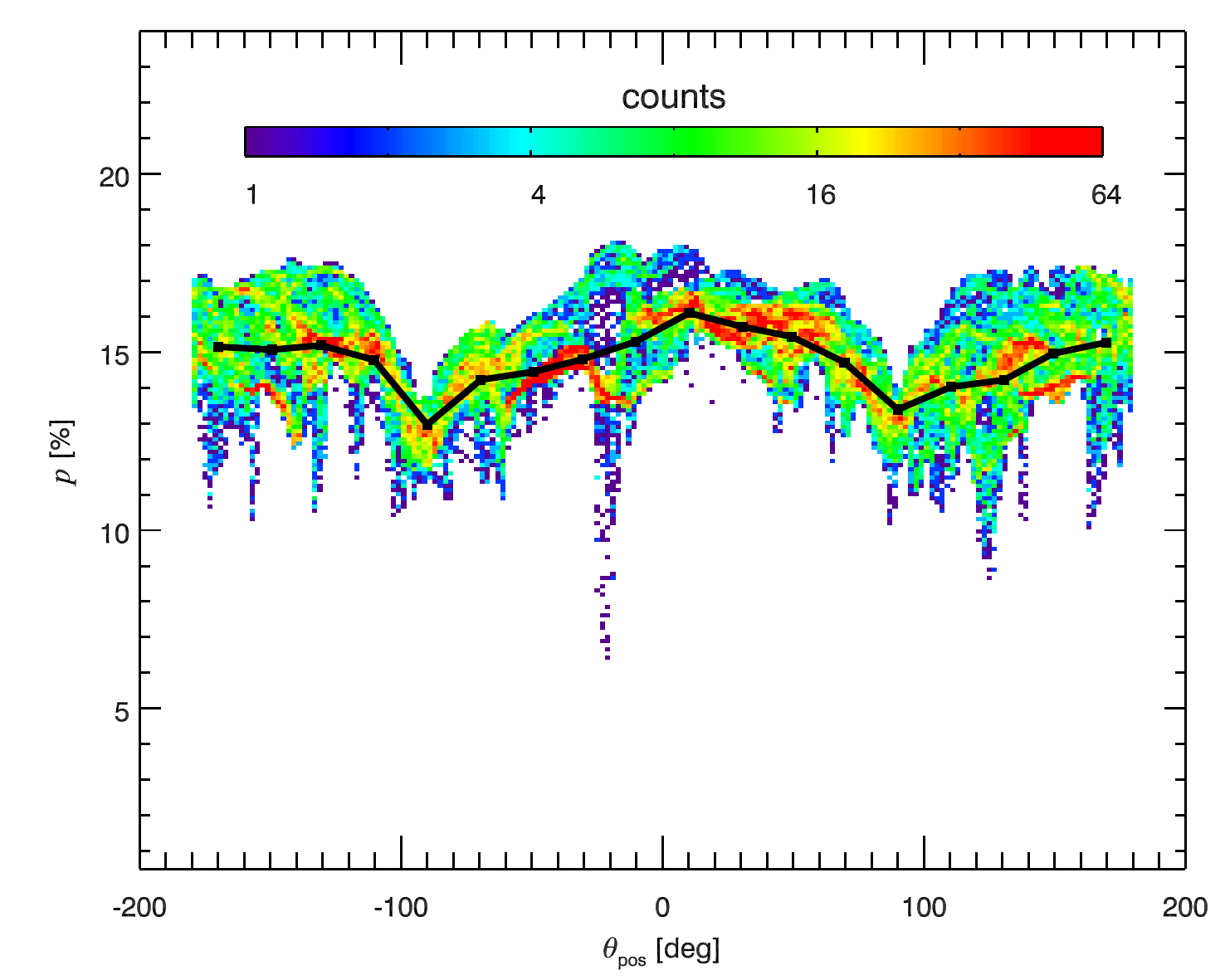}
\caption{Polarization fraction $p$ as a function of the distance in the plane of the sky (top) and as a function of the projected angle $\theta_{\mathrm{pos}}$ between the magnetic field and starlight (bottom), for the STAR-RAT case with a uniform magnetic field in the plane of the sky. The image corresponds to  Figure \ref{fig:ConstField} and is to be compared with Figure \ref{fig:PsI_STAR_VG}.} 
\label{fig:ConstField_VG}
\end{figure}

We pursue our analysis of the STAR case by studying a simple configuration where the distance and angle-dependence effect of RATs should be optimal. We run \POLARIS\ for our \RAMSES\  simulation, still with a star at the center, but replacing the magnetic field from the \RAMSES\ simulation by a magnetic field direction everywhere uniform in the $X$ direction in the plane of the sky. 

Figure~\ref{fig:ConstField} presents the resulting maps of $\aalig$, anisotropy $\left\langle \gamma \right\rangle$, and polarization $\PsI$. Comparing these maps to the ones of the ISRF-RAT setup and the STAR-RAT setup presented in Figure \ref{fig:POLARIS_ISRF_RAT} and Figure \ref{fig:POLARIS_STAR_RAT}, respectively, the dust grains along $X = 0\ \mathrm{pc}$ become severely depolarized with alignment radii $\aalig \gtrapprox 100\ \mathrm{nm}$ while we find $\aalig \lessapprox 90\ \mathrm{nm}$ for the rest of the map. Yet again we observe the characteristic shadowing effect at $X = 4\ \mathrm{pc}$ and $Y = -2\ \mathrm{pc}$ caused by the densest clump in the \RAMSES\  simulation (see Figure \ref{fig:MHD_input}). The average angle between the direction of radiation and magnetic field orientation $\left\langle \gamma \right\rangle$, is also characteristic of RAT alignment with lower values along the line $X=0\ \mathrm{pc}$. However, this influence is less obvious in the map of $\PsI$ which results from physical quantities integrated along the LOS and is also dependent on other quantities such as the magnetic field orientation (see Appendix \ref{app:RTequation}). Overall, the magnitude of $\PsI$ in Figure \ref{fig:ConstField} shows less variations compared to those of in Figure \ref{fig:POLARIS_ISRF_RAT} and Figure \ref{fig:POLARIS_STAR_RAT}. This demonstrates that a good part of depolarization is a result of the turbulent component of the magnetic field and not grain alignment physics itself. This finding is also consistent with the interpretation of synthetic dust polarization maps presented in \cite{Seifried2019}.

The dependence of $\PsI$ on the distance and on the angle $\theta_{\mathrm{pos}}$, presented in  Figure~\ref{fig:ConstField_VG}, do indeed present small trends expected from RATs. However, the decrease of $\PsI$ with the distance, as well as its sinusoidal modulation by $\theta_{\mathrm{pos}}$, are so small (by $1\%$ and $2\%$, respectively), that they would most probably not be observable once noise and background contamination are added, even in this optimal configuration of the magnetic field.

\section{Discussion}
\label{sec:discussion}
In this section, we discuss the implications and limits of our model as well as the observational possibilities of testing alignment theories.
\subsection{Impact of the fitted size distributions on our results}\label{sec:sizedist}
In Section \ref{sec:fitdustmodel}, we mentioned that 
our simple oblate\footnote{Using prolate grains instead of oblate grains imposes to compute the grain optical properties integrated over the grain spinning dynamics \citep[see][for a detailled description]{Guillet2018}.} grain shape and size distributions (power-laws) do not allow for a precise fit to the polarization and extinction curves (see Figures~\ref{fig:p_lambda} and \ref{fig:DiffuseDust}).
Let us first discuss the NIR extinction, which is not well reproduced by our dust model for $\lambda > 1.5\,\mu$m. According to Figure~\ref{fig:DiffuseDust}, we systematically underestimate the NIR extinction by a factor $\sim2$. With the same figure, we see that NIR extinction is significant ($\tau \ge 1$) only for column densities higher than $10^{22}$\,cm$^{-2}$ at $\lambda=2\,\mu$m, and higher than $5\,10^{22}$\,cm$^{-2}$  at $\lambda=4\,\mu$m. For these LOS, our calculations \emph{overestimate} the number of NIR photons that are present. Our model tends therefore to overestimate grain alignment at the highest column densities, in the densest clumps of our simulation, which are rare. 
Second, we inferred a maximal size $a_{\rm max}^{\rm S}=400\,$nm for the silicate distribution from a fit of the polarization curve, particularly its NIR part. A lower (resp. higher) value for $a_{\rm max}^{\rm S}$ would have increased (resp. decreased) the mass of dust grains above the mean alignment radius $\aalig$ in the diffuse ISM, which is of the order of 100\,nm. As a consequence, a loss of alignment would have had more (resp. less) impact on the local polarization fraction, i.e. the relation between $p$ and $\aalig$ would have been steeper (resp. less steep) than described in Figure~\ref{fig:pst_PsI_lambda}. All together, our model may slightly overestimate the alignment of grains by RATs, certainly not underestimate it.





\subsection{Can the angle-dependence of the RATs alignment effiency be tested observationally ?}
\label{sect:RAT_test}

Section \ref{sec:RT_STAR} has demonstrated that one of the characteristic effects expected from the RATs theory, namely the angle-dependence of the grain alignment efficiency, is too weak to be observed in realistic conditions. However, \cite{VA15} claimed to detect this effect, analyzing polarization data for the OMC-1 ridge with a star at its center: the IRc2 source. Their Figure~2, which shows how the polarization fraction varies with the angle $\theta_{\mathrm{pos}}$, indeed exhibits a sinusoidal variation that looks like what we expect from the RATs theory.

We propose an alternative explanation for these observations. It has been established long ago that the maximal polarization fraction, whether in extinction or in emission, tends to systematically decrease with the column density \citep[\eg][]{Jones1989}. The origin for this effect, whether it is due to the magnetic field tangling or to a drop in the alignment efficiency, is still debated and depends on the authors. More recently, \cite{Planck2016XXXV} demonstrated a systematic variation of the orientation of the magnetic field with respect to the gas structures, from parallel in the diffuse ISM to rather perpendicular to dense filaments. Such variations were observed in almost all regions of the Gould Belt. Both of these effects, which are observed all through the ISM, are present in the OMC-1 polarization maps of \cite{VA15}. If we combine these two effects and start our analysis at the position of the heating source IRc2, we can predict, without invoking any RAT physics, that the polarization fraction observed along the direction of the ridge will be weak and will correspond to $\theta_{\mathrm{pos}}$ close to $90^\circ$, while the polarization fraction observed perpendicular to the ridge, and therefore toward the less dense ISM, will be higher and correspond to  $\theta_{\mathrm{pos}}$ closer to $0^\circ$. 
 
We speculate that such a correlation should also be observed toward dense filaments even without embedded stars, as long as the external magnetic field is observed to be perpendicular to the filaments, as is the case for the Musca filament \citep{Pereyra2004} or for the B213 filament in Taurus \citep{Chapman2011}.
In summary, the characteristic effects of RAT alignment seem to be usually too weak to be observed, and can be mimicked by other physical effects, in particular those deriving from the orientation of the magnetic field with respect to the gas filaments.

We note, that there is an additional factor that needs to be taken into account for testing the RAT theory observationally. In \cite{HoangLazarian2016} it was demonstrated that for superparamegntic grains of size $a>100\ \mathrm{nm}$ the angular-dependency with $\vartheta$ may get lost completely. The criterion for the loss of angular dependency of RAT alignment goes with $1/(\ngas \Tgas^{1/2}) > C$ where $C$ is some constant \citep[see][for details]{HoangLazarian2016}. Hence, a dependency with $\vartheta$ can still be expected in dense molecular clouds while in the DISM it may become void. However, we already can barely report any angular-dependency in our setups ISRF-RAT as well as STAR-RAT (see Fig. \ref{fig:ConstField_VG}) so that this additional criterion is of minor relevance within the scope of this paper.

In essence, to test the angle-dependency of RATs, one should use optimal conditions such as a uniform $\vec{B}$ in the plane of the sky around a hot star and avoid dense regions where other effects may dominate. We also suggest to test this effect in the optical, where dust models predict steeper variations of the polarization fraction with the grain alignment efficiency (see Fig~\ref{fig:pst_PsI_lambda}).

\subsection{Testing grain alignment theories in dense cores}
\label{sect:TestingAlignment}

This article aims at demonstrating that it is necessary to provide quantitative tests of the RATs theory, and not only qualitative evidence as is usually done. The dependence of the polarization fraction on the dust properties or on the magnetic field structure is so degenerate that it is hard to disentangle between the different effects at work using only maps of the polarization fraction. In \cite{Planck2018XII}, we have advocated that using the statistics of the polarization angles, through the quantities $\S$ and $\S\times p$ could be useful to that purpose. 

In the present article, we have demonstrated that the efficiency of the radiative torques is constant in the diffuse and translucent ISM, and that all variations of the alignment efficiency are solely due to variations of the disalignement by gas collisions measured by the gas pressure, and not to the decrease of the radiation field intensity by dust extinction. The ISRF is dominated in energy by NIR ($\sim 1\,\mu$m) photons \citep[\textit{e.g.}][their Figure~1]{Mathis1983}. Comparing equation~\ref{eq:JradoverJth} with equation~\ref{eq:JRAT} especially its factor $\lambda\times u_{\lambda}$, shows that it is the total number of photons, not their total energy, that is involved in grain alignment by RATs. UV photons are unimportant for RAT alignment in the diffuse ISM, both in energy and - even more so - in numbers. \
As a consequence, the efficiency of the aligning torque will be rather constant under the ISRF radiation field as long as extinction in the NIR is not important. 
This could justify why no dependence of alignment on the grain temperature could be found in \Planck\ diffuse ISM data~\citep{PlanckXIX2015,Planck2018XII}. On the contrary, the disaligning torques exerted by gas collisions will vary a lot through the diffuse and translucent ISM, because of pressure variations. In particular, the pressure increases by orders of magnitude between the diffuse and dense ISM, as soon as the gas temperature gets stabilized around a few tens of Kelvin. This dimension is underestimated when one interprets the difference in the polarization patterns in distinct environments through the prism of the radiation field alone.

To test the decrease of the RATs efficiency, we therefore need to move to very dense environments where extinction in the NIR starts to be significant ($\NH \gg 10^{22}$ cm$^{-2}$). 
The key issue will remain to explain the level of polarization observed in dense cores, where we expect a huge increase in pressure combined with a severe drop in the RAT efficiency due to extinction of optical and NIR photons which are driving the grain alignment. 
Such data analysis is not possible with the 5 arcmin resolution of \Planck, but is accessible to the new generation of polarization instruments working at subarcmin resolutions such as JCMT/SCUBA-2/POL-2 \citep[][]{Holland2013}, SOFIA/HAWC+ \citep[][]{Dowell2010,Harper2018}, or NIKA2 \citep[][]{Monfardini2011,Monfardini2014,Calvo2016}.
Maintaining a high level of grain alignment by RATs in cores requires a significant grain growth \citep[\eg][]{Pelkonen2009}. This hypothesis, which is indeed reasonable, ignores however that grain growth will automatically change both the grains' shapes and optical properties, and therefore their polarization capabilities. Altogether, modeling such scenarios requires to complete our understanding of grain alignment physics and dust evolution (in particular grain-grain coagulation), and a comparison of observations with numerical results obtained with MHD simulations and tools like \POLARIS.

In this paper we focused on the spin-up of dust grains by RATS. Alternatively, irregularly-shaped dust grains may  spin up by means of mechanical torques \citep[MATs,][]{Hoang2018A}. Originally, such a theory was proposed for regular grain shapes by \cite{Gold1952A,Gold1952B}. It was later extended to a magneto-mechanical alignment theory by \cite{Lazarian1995MNRASL,Lazarian1997CL}. Here, a supersonic gas-dust drift velocity is required. In principle such a drift may be driven by cloud-cloud collisions, winds \citep[e.g.][]{Habing1994}, or MHD turbulence \citep[][]{Yan2003}. Although cloud-cloud collisions and winds cannot account for the large scale alignment of grains, MHD turbulence seems to be ubiquitous in the ISM \citep[][]{XuZhang2016}. However, it remains to be seen if MHD turbulence can provide a supersonic drift.
More recent studies indicate that mechanical grain alignment may be efficient for helical grains even in the case of a subsonic drift  \citep[][]{Lazarian2007C,Das2016,Hoang2018A}. In the MAT theory, the mechanical torque efficiency is proportional to the gas pressure \citep{Das2016,Hoang2018A}. This means that the grain alignment radius will be independent of the gas pressure, therefore of the gas density, unlike for RATs. We suggest that this
property, which implies high level of grain alignment in dense cores (though not a systematically high level of polarization because of magnetic field tangling and possible dust coagulation), could be used to disentangle between alignment by RATs and alignment by MATs.

\section{Summary}
\label{sect:summary}
In this paper, we presented a quantitative analysis of the impact of RAT alignment on dust polarimetry. This particular alignment theory predicts a sensitivity of the grain alignment efficiency with respect to the magnitude of the radiation field as well as an angular dependency on the direction of the radiation with respect to the magnetic field orientation. We aimed to model these dependencies for the diffuse and translucent ISM. For this we used a MHD cube representative of the diffuse ISM simulated with the $\RAMSES$ code. We post-processed the MHD data with the RT code $\POLARIS$ to produce synthetic dust polarization observations.  The latest version of the $\POLARIS$ code solves the full four Stokes parameters matrix equation of the RT problem, including RAT alignment, simultaneously. For the dust, we developed a best-fit model consisting of two populations of silicate and graphite grains following a power-law size distribution, that reproduce the mean Serkowski's law as well as the mean extinction curve in the diffuse ISM.

We first performed Monte-Carlo dust heating and grain alignment calculations assuming a diffuse ISRF. 
The resulting radiation field and grain alignment efficiency is consistent with the alignment theory of RATs.  

We analyze the polarization maps and reproduce the anti-correlation of polarization fraction with gas column density as well as with the angular dispersion known from Planck observations. However, we cannot trace any of the characteristic predictions of RAT alignment in the synthetic polarization data. Our scientific findings are summarized as follows:
\begin{enumerate}[(i)]
  \item Correlating the different parameters relevant for RATs reveals that the grain alignment efficiency in the diffuse and translucent ISM is primarily driven by the gas pressure (which tends to disalign grains, and varies by orders of magnitude through the ISM), and not by the radiation field intensity (which varies only moderately in the diffuse and translucent ISM).
  \item Anisotropy $\left\langle \gamma \right\rangle$ of the radiation field and its orientation $\left\langle \cos(\vartheta) \right\rangle$ with respect to the magnetic field have only a minor effect on grain alignment in the diffuse ISM.
  \item Despite the local drop of grain alignment in denser regions due to the increase in the gas pressure, the RATs alignment mechanism leaves no trace in the anti-correlation of gas column density $\NH$ with polarization fraction $p$; nor in the anti-correlation of the angular dispersion $\S$ with $p$, the possible signposts of RATs being washed out by line of sight integration and variations of the magnetic field structure on the line of sight and within the beam.
\end{enumerate}
We then considered a second setup to investigate the RAT alignment behavior for different variations of the radiation field, by placing a B-type star in the very center of the $\RAMSES\ $ MHD cube in addition to the ISRF, and repeating our RT simulations. We find that grain alignment efficiency is highest in close proximity of the star in concordance with RAT theory. Our findings in that case are the following:
\begin{enumerate}[(i)]
  \item Even under optimal conditions, fingerprints of RATs would be barely observable. In particular, the predicted dependency of grain alignement by RATs with the angle between the radiation field and the magnetic field direction would not be detectable by observations of dust emission.
  \item Even close to a star, the variations in the magnetic structure along the LOS and within the beam are much more important for dust polarization than the variation in the characteristics of the radiation field. 
\end{enumerate}
Altogether, our modelling of synthetic dust polarization observations indicates that the effects of RAT alignment are barely detectable in the diffuse and translucent ISM, but are predicted to be stronger in the optical (i.e. on starlight polarization) than in submillimetre polarized emission.

\appendix
\section{Radiative transfer of polarized radiation with $\POLARIS$}
\label{app:RTequation}

Performing RT simulations with the full Stokes vector transforms the RT problem into a matrix equation \citep[][]{Martin1974,Lee1985,Whitney2002}. Rotating the reference frame of the polarized light from the lab frame into the frame of the dust grain by a matrix $\hat{R}(\varphi)$ allows to eliminate some of the transfer coefficients \citep[][]{Mishchenko1991}. It follows from the Stokes vector formalism for the rotation matrix
\begin{equation}
\hat{R}(\varphi)=\begin{pmatrix} 1 &  0 & 0 & 0 \\  0 & \cos(2\phi) & -\sin(2\phi) & 0 \\ 0 & \sin(2\phi) & \cos(2\phi) & 0 \\ 0 & 0 & 0 & 1 \end{pmatrix}\, ,
\end{equation}
where the angle $\phi$ is defined to be between the coordinate system of the Stokes vector and the magnetic field direction \citep[see][for details]{Reissl2016}. The full RT equation along a path element $\mathrm{d}\ell$ of the LOS reads then
\begin{equation}
\frac{\ud}{\ud\ell}\begin{pmatrix} I\\ Q \\ U \\ V \end{pmatrix}=-\begin{pmatrix} \kI &  \kQ & 0 & 0 \\  \kQ & \kI & 0 & 0\\ 0 & 0 & \kI & - \kV \\ 0 & 0 & \kV &  \kI \end{pmatrix}\begin{pmatrix} I\\ Q \\ U \\ V \end{pmatrix} + \begin{pmatrix} j_{\mathrm{I}}\\ j_{\mathrm{Q}} \\ 0 \\ 0 \end{pmatrix}\, .
\label{eq:MatrixRT}
\end{equation}
Here, $\kI$, $\kQ$, and $\kV$ are the transfer coefficients associated to extinction, linear , and circular polarization while $j_{\mathrm{I}}$ and $j_{\mathrm{Q}}$ are the coefficients of total and polarized emission, respectively. Note that the emission coefficient for the Stokes $U$ parameter is zero here. The polarization by emission of the Stokes $U$ comes with the back rotation $\hat{R}(-\phi)$ into lab frame. The \POLARIS\ code deals with Equation~\eqref{eq:MatrixRT} by applying the Runge-Kutta-Fehlberg (RFK45) solver. This solver selects the step size $\mathrm{d}\ell$ variably by comparing the fourth order solution to the fifth order solution. This takes care to keep the maximal error in each integration step for each of the Stokes parameters below a certain error level. In \POLARIS\ this allowed error level is defined to be $ \leq10^{-6}$ by default.

The RT coefficients are dependent on the dust cross sections in extinction $C_{\mathrm{ext}}$, absorption $C_{\mathrm{abs}}$, and circular polarization $C_{\mathrm{circ}}$. The cross sections are in turn dependent on grain material, wavelength, grain size, and shape. Dealing with oblate dust grains is sufficient to calculate the cross sections $C_{\mathrm{||}}$ and $C_{\mathrm{\bot}}$ parallel and perpendicular, respectively, to the minor principle axis $a_{||}$. We define the radiation to propagate along the $Z$ axis whereas the state of polarization is determined with respect to the $X-Y$ plane, perpendicular to $Z$. In this reference frame the radiation experiences an extinction cross section 
\begin{equation}
        C_{\mathrm{ext,X}} (\aeff) =  \left\langle C_{\mathrm{ext}}(\aeff) \right\rangle+\frac{1}{3} R(\aeff)\times \left( C_{\mathrm{ext,||}}(\aeff)-C_{\mathrm{ext,\bot}}(\aeff) \right)
\end{equation}
along the $X$ axis and an extinction cross section
\begin{equation}
\begin{split}
        C_{\mathrm{ext,Y}}(\aeff) =  \left\langle C_{\mathrm{ext}}(\aeff) \right\rangle \qquad\qquad\qquad\qquad\qquad\qquad \\\quad  +\frac{1}{3}R(\aeff) \times \left( C_{\mathrm{ext,||}}(\aeff)-C_{\mathrm{ext,\bot}}(\aeff) \right)\left( 1- 3 \sin^2(\vartheta) \right)
\end{split}
\end{equation}
along the $Y$ axis, corrected for grain orientation and grain incomplete alignment with the magnetic field direction. Here, $\vartheta$ is defined to be between the LOS and the magnetic field direction and $R$ is the Rayleigh reduction factor (see Section \ref{sect:RATAlignment}). The quantity 
\begin{equation}
        \left\langle C_{\mathrm{ext}}(\aeff) \right\rangle = \frac{1}{3}\left(2C_{\mathrm{ext,||}}(\aeff)+C_{\mathrm{ext,\bot}}(\aeff)\right),
\end{equation}
denotes the cross section of a randomized oblate dust grain. In this paper we perform the RT simulations with size-averaged cross sections for different materials (marked by the index i) for extinction 
\begin{equation} 
        \overline{C}_{\mathrm{i,ext}} =\int_{a_{\mathrm{min}}}^{a_{\mathrm{max}}} N(\aeff) \left(C_{\mathrm{i,ext,X}}(\aeff)+C_{\mathrm{i,ext,Y}}(\aeff)\right)\ud\aeff
\end{equation}
and 
\begin{equation}
        \Delta\overline{C}_{\mathrm{i,ext}} =\int_{a_{\mathrm{min}}}^{a_{\mathrm{max}}} N(\aeff) \left(C_{\mathrm{i,ext,X}}(\aeff)-C_{\mathrm{i,ext,Y}}(\aeff)\right)\ud\aeff
\end{equation}
weighted by the size distribution function $N(\aeff)$ (see Sect. \ref{sect:DustModel}). The same geometrical considerations apply also for the cross sections of absorption $\overline{C}_{\mathrm{i,abs}}$ and $\Delta\overline{C}_{\mathrm{i,abs}}$, respectively, as well as for circular polarization $\Delta\overline{C}_{\mathrm{i,circ}}$. Consequently, the total RT coefficient of extinction reads
\begin{equation}
        \kI=\frac{1}{2} \sum_{i=1}^2 \nidust \overline{C}_{\mathrm{i,ext}}
\end{equation}
and the coefficient of linear polarization by extinction is defined by
\begin{equation}
        \kQ=\sum_{i=1}^2  \nidust \Delta\overline{C}_{\mathrm{i,ext}}
\end{equation}
where the sum accounts for the distinct cross sections and number densities for silicate and graphite grains, respectively. An already polarized radiation may also accumulate a small amount of circular polarization due to the differential phase lag along the distinct grain axes leading to a transfer coefficient of
\begin{equation}
        \kV=\sum_{i=1}^2 \nidust \Delta\overline{C}_{\mathrm{i,circ}}\, .
\end{equation}

For the RT coefficient of emission we account also for individual dust temperatures for each of the grain materials. Assuming the dust grain to be in equilibrium with its environment leads to the following emission coefficients:
\begin{equation}
\begin{split}
j_{\mathrm{I}} = \frac{1}{2}  \sum_{i=1}^2 \nidust B_{\lambda}\left(\Tidust\right)\overline{C}_{\mathrm{i,abs}}\, ,
\end{split}
\end{equation}
and
\begin{equation}
j_{\mathrm{Q}} =  \sum_{i=1}^2 \nidust B_{\lambda}\left(\Tidust\right) \Delta\overline{C}_{\mathrm{i,abs}}\, .
\end{equation}

\section{Monte-Carlo noise estimation}
\label{app:MCNoise}
\begin{figure}[]
\centering
\begin{minipage}[c]{1.0\linewidth}
   \begin{flushleft}
      \includegraphics[width=1.0\textwidth]{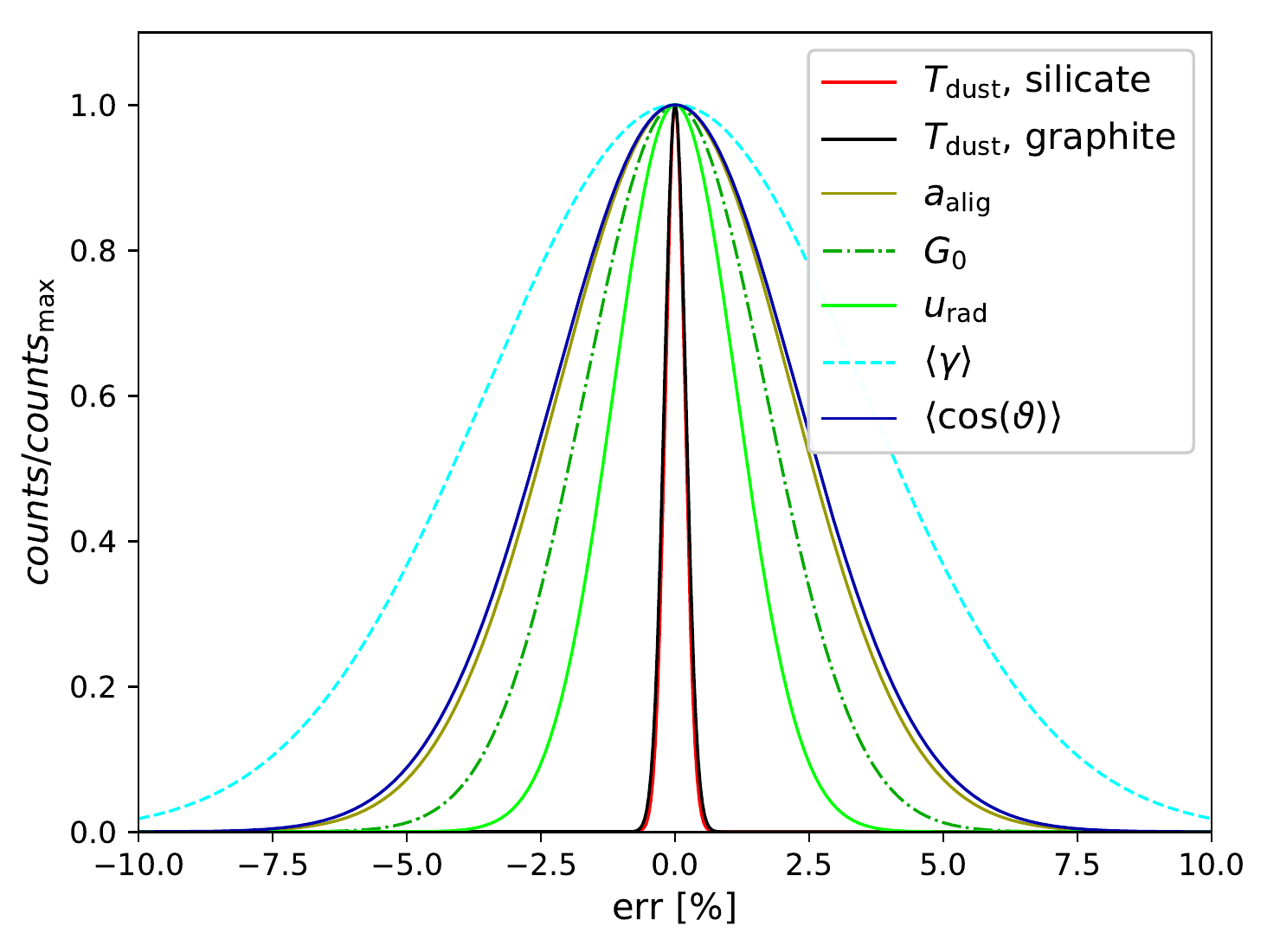}
 \end{flushleft}
\end{minipage}
\caption{Histogram of the MC error distribution.}
\label{fig:HistNoise}
\end{figure}
Due to its stochastic nature, a certain amount of noise is the inevitable drawback in MC RT simulations. In this section we quantify the noise in the MC runs for dust heating and grain alignment. The MC noise depends on the number of applied photons and the quality of the random number generator. In \POLARIS\ we implemented the random number generator scheme KISS \citep[][]{Marsaglia1993,Marsaglia2003} with a period of roughly $10^{75}$. We repeated the MC runs as outlined in Section \ref{sect:RTPostProcessing} with ten different random seeds. From these runs we calculated the average $\langle E \rangle$ for each of the grid cells where $E$  can stand for each of the quantities $\Tdust$, $\aalig$, $\Gzero$, $\urad$, $\left\langle \gamma \right\rangle$, and $\left\langle \cos(\vartheta) \right\rangle$ derived from the \POLARIS\ MC run. Consequently, $\langle E \rangle$   provides a noise reduced baseline for the error estimation. We use an error based on the quantity $E$ per run and grid cell with respect to the average over all ten runs defined to be 
\begin{equation}
\mathrm{err} = \left( E-\langle E \rangle \right) / \langle E \rangle \, .
\end{equation}
\begin{table}
\centering
   \begin{tabular}[]{ | l | c | c |}
   \hline
      parameter & mean [$\%$] & STD [$\%$] \\ 
   \hline  
   \hline 
      $\Tdust$, silicate                & $-6.33\times 10^{-6}$  &  $0.18$\\
      $\Tdust$, graphite                & $-8.44\times 10^{-7}$  &  $0.20$\\ 
      $\aalig$                          & $5.43\times  10^{-5}$ &  $2.18$\\ 
      $\Gzero$                             & $-9.07\times 10^{-5}$  &  $1.69$\\
      $\urad$                             & $1.07\times 10^{-6}$  &  $1.15$\\ 
      $\left\langle \gamma \right\rangle$          & $-7.22\times 10^{-5}$  &  $3.53$\\ 
      $\left\langle \cos(\vartheta) \right\rangle$ & $-5.65\times 10^{-5}$  &  $2.27$\\ 
   \hline
  \end{tabular}
  \caption{Mean values and standard deviations (STD) of the MC noise for the different parameters derived MC RT simulations.}
  \label{tab:Error}
\end{table}

In Figure \ref{fig:HistNoise} we show the distribution of deviations from the average as a measurement of the MC noise and the corresponding mean values and the standard deviations (STD) are listed in Tab. \ref{tab:Error}. The dust temperatures are the least affected by the MC noise with a mean close to zero and a STD less than a quarter of a percent. However, even the anisotropy factor $\left\langle \gamma \right\rangle$, although it is the quantity most affected by MC noise, barely exceeds a STD of $3.5\ \%$. The subsequent ray-tracing scheme of \POLARIS\ has an excellent signal to noise ratio. Thus, we estimate the maximal numerical error to be not larger than  $3.6\ \%$ for the entire \POLARIS\ RT pipeline and subsequent polarization maps presented in this paper.

\section{Histograms of Monte-Carlo quantities}
\label{app:3DDistribution}
In this section we briefly present the 3D distributions of the physical quantities derived with our \POLARIS\ MC simulations.
\begin{figure*}
\centering
\begin{minipage}[c]{1.0\linewidth}
  \includegraphics[width=.49\textwidth]{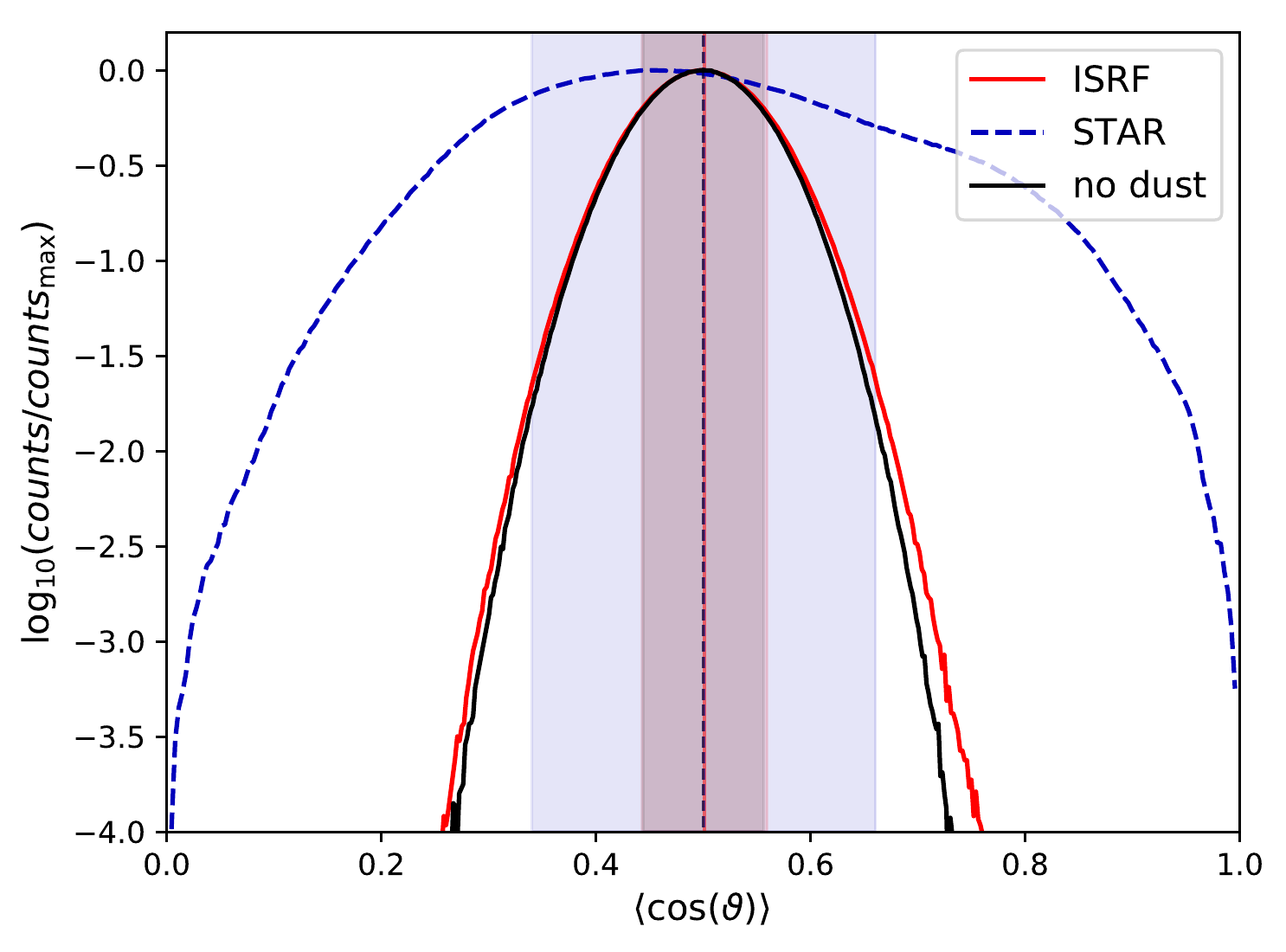}
  \includegraphics[width=.49\textwidth]{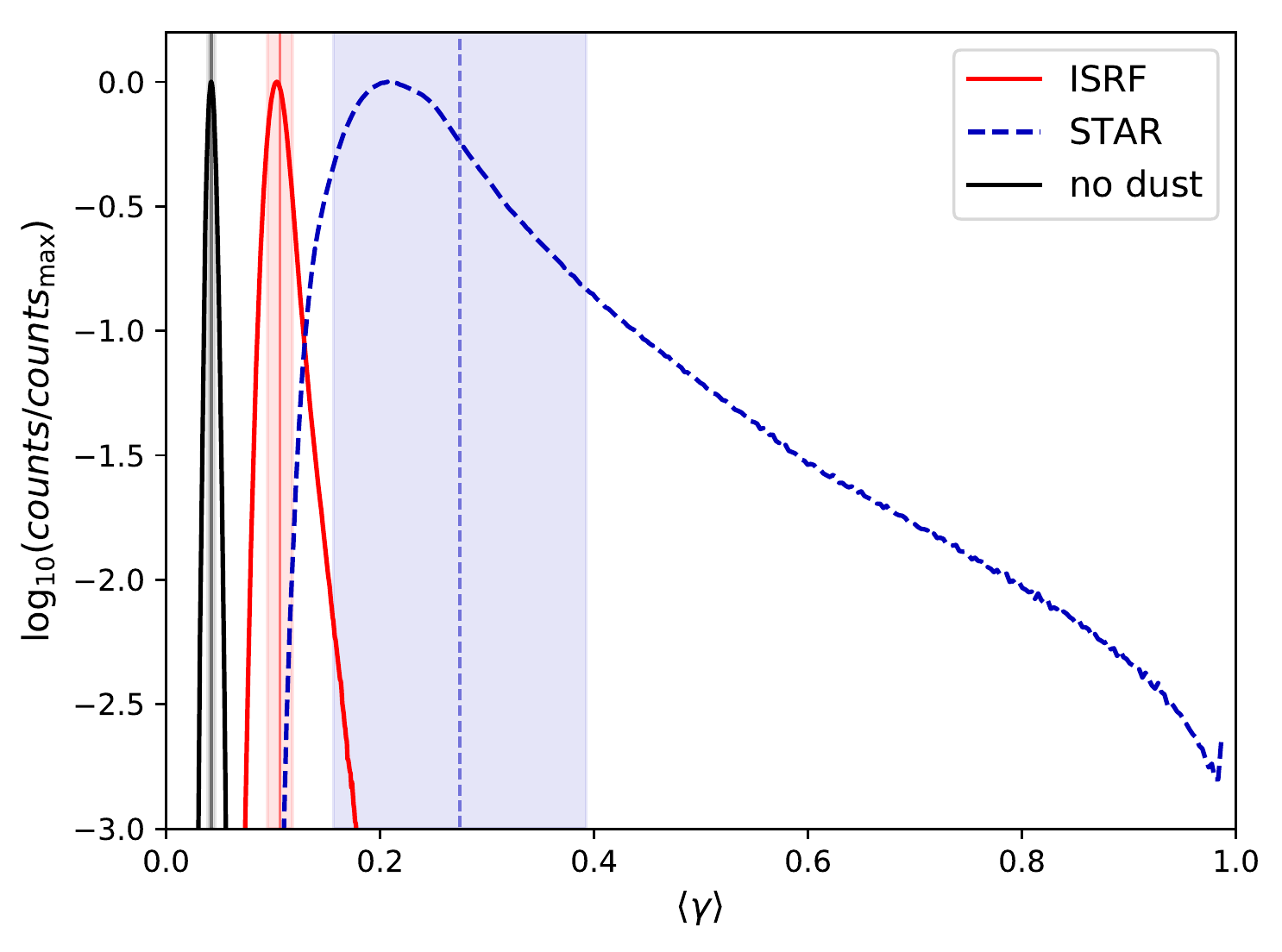}
\end{minipage}

\begin{minipage}[c]{1.0\linewidth}
  \includegraphics[width=.49\textwidth]{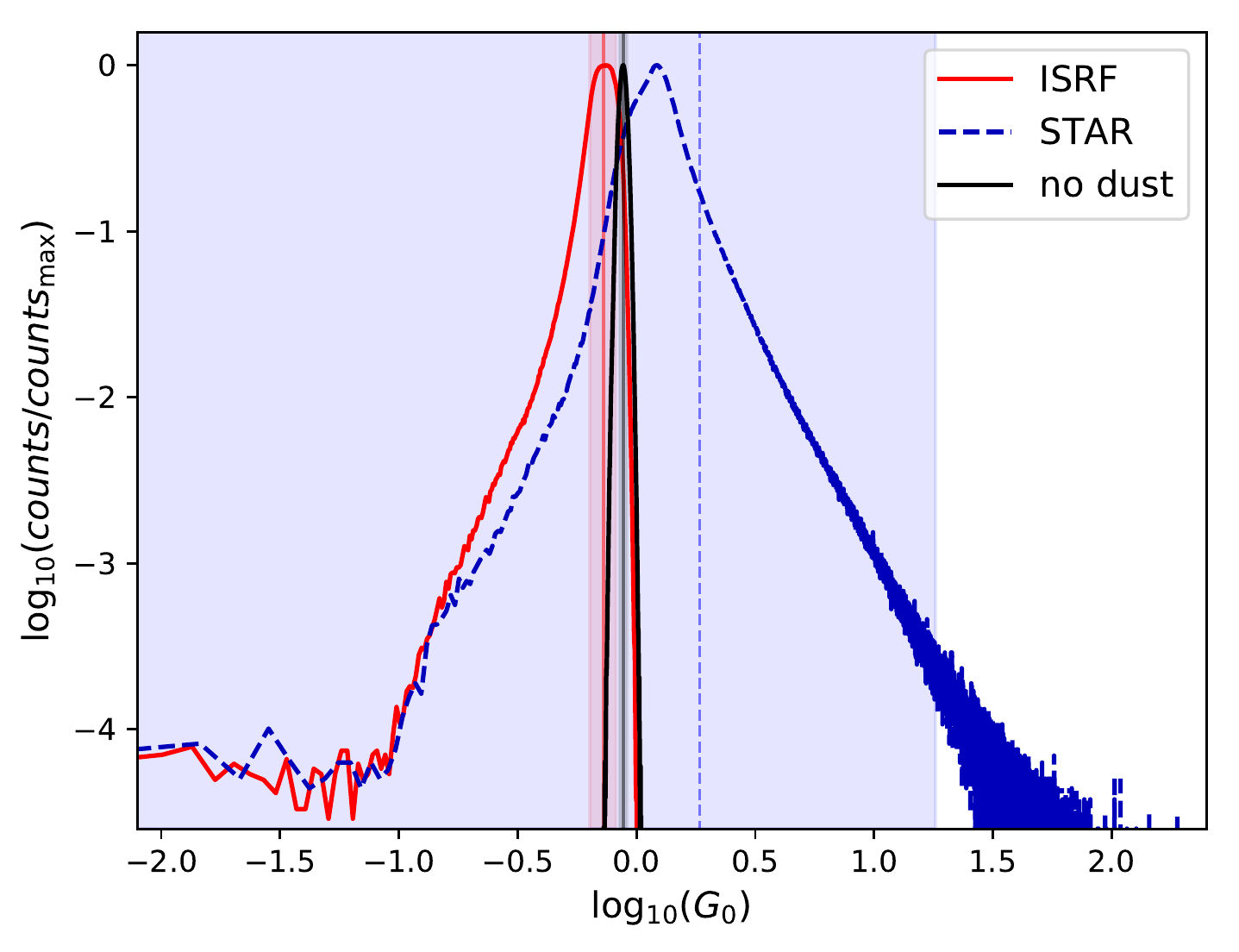}
  \includegraphics[width=.49\textwidth]{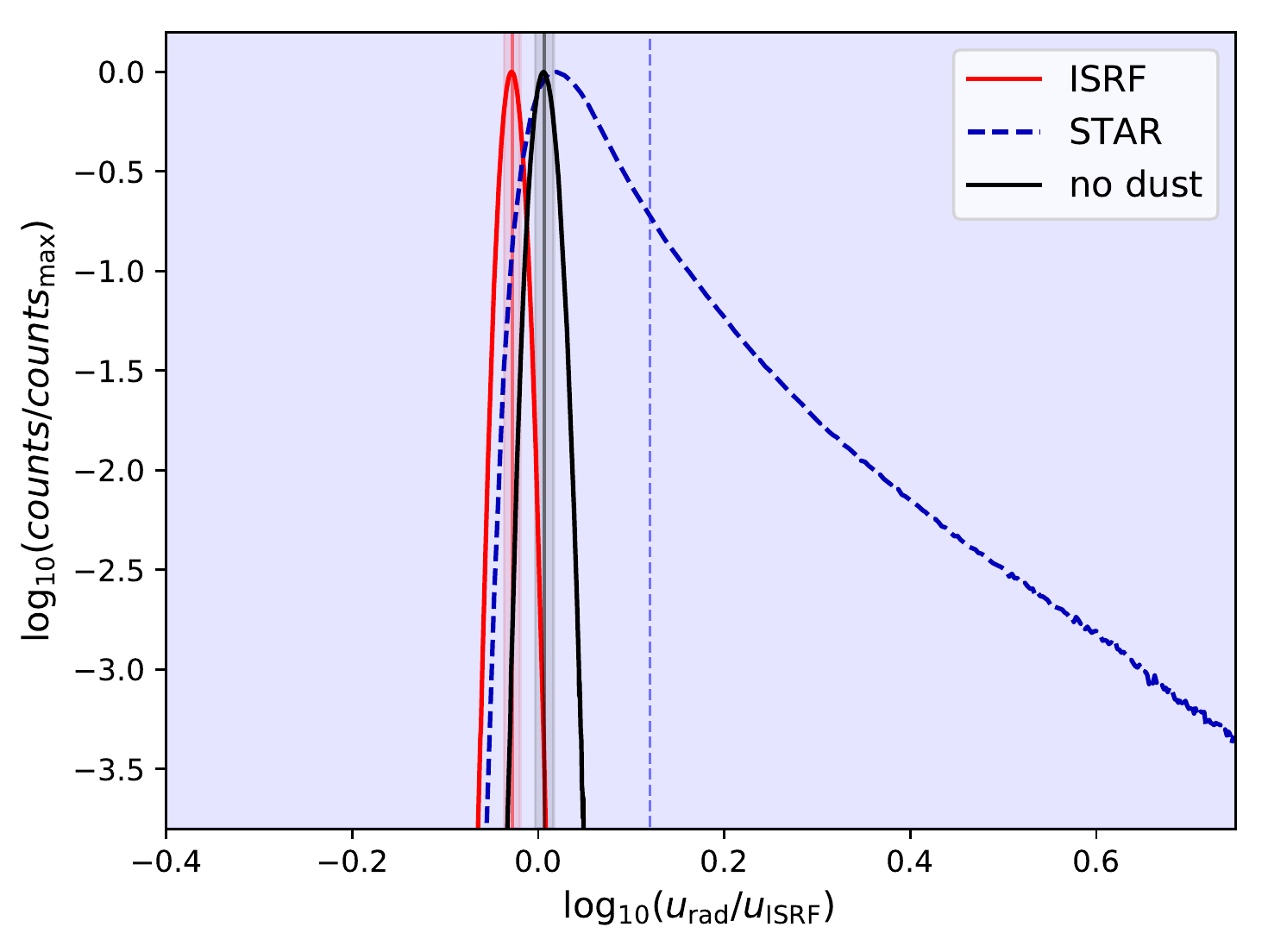}
\end{minipage}
\caption{Histograms considering all cells within the $\RAMSES$ cube for the setups ISRF and STAR, respectively, as well as a test run with the ISRF and no dust. The individual panels are the average angle $\left\langle \cos(\vartheta) \right\rangle$  between magnetic field direction and radiation field (upper left) and the anisotropy factor $\left\langle \gamma \right\rangle$ (upper right), $\Gzero$ (bottom left) and the average energy density $\urad$ of the radiation field (bottom right). All histograms are normalized to their respective peak values. Vertical lines and bars represent the corresponding mean values and the standard deviations, respectively.}
\label{fig:3DHistRad}
\end{figure*}

\subsection{The radiation field}

 In Figure \ref{fig:3DHistRad} we show the histograms of the 3D distributions of different quantities. The average angle $\left\langle \cos(\vartheta) \right\rangle$ for both the ISRF  and STAR cases shows a similar distribution as the projected images (see Sects. \ref{sec:RT_ISRF} and \ref{sec:RT_STAR}). The average values for ISRF, STAR, as well as the case with no dust at all in the cube are almost identical at $0.5$, indicating a large degree of random orientation between radiation field and magnetic field direction. The distributions in the ISRF and "no dust" cases are almost the same, while values down to zero and up to unity for the STAR case and the regions surrounding the central star are present.

The anisotropy factor $\left\langle \gamma \right\rangle$ clusters around mean values of $0.09$ and $0.27$ for the ISRF  and the STAR setups, respectively. However, the "no dust" case does not reach lower values than $0.04$ with an average of about $0.05$.  As outlined in Section \ref{sect:RTPostProcessing} we inject photons with random directions into the grid to mimick a completely isotropic ISRF. Hence, one could expect a value of $0.0$ for this case, indicating a minor numerical bias in the MC method.

The magnitude of the radiation field quantified by $\Gzero$ and $U_{\mathrm{rad}}$ peaks around unity for the "no dust" case, whereas the ISRF setup reaches unity only at the very borders of the grid. For the STAR case we find the radiation field on average to be increased by roughly a factor of two with peak values up to $100$ times larger than the $\Gzero=1$ ISRF, close to the star.

\subsection{Dust temperatures and characteristic grain alignment radii}
\label{app:Alignment}
\begin{figure*}
\centering
\begin{minipage}[c]{1.0\linewidth}
      \includegraphics[width=.327\textwidth]{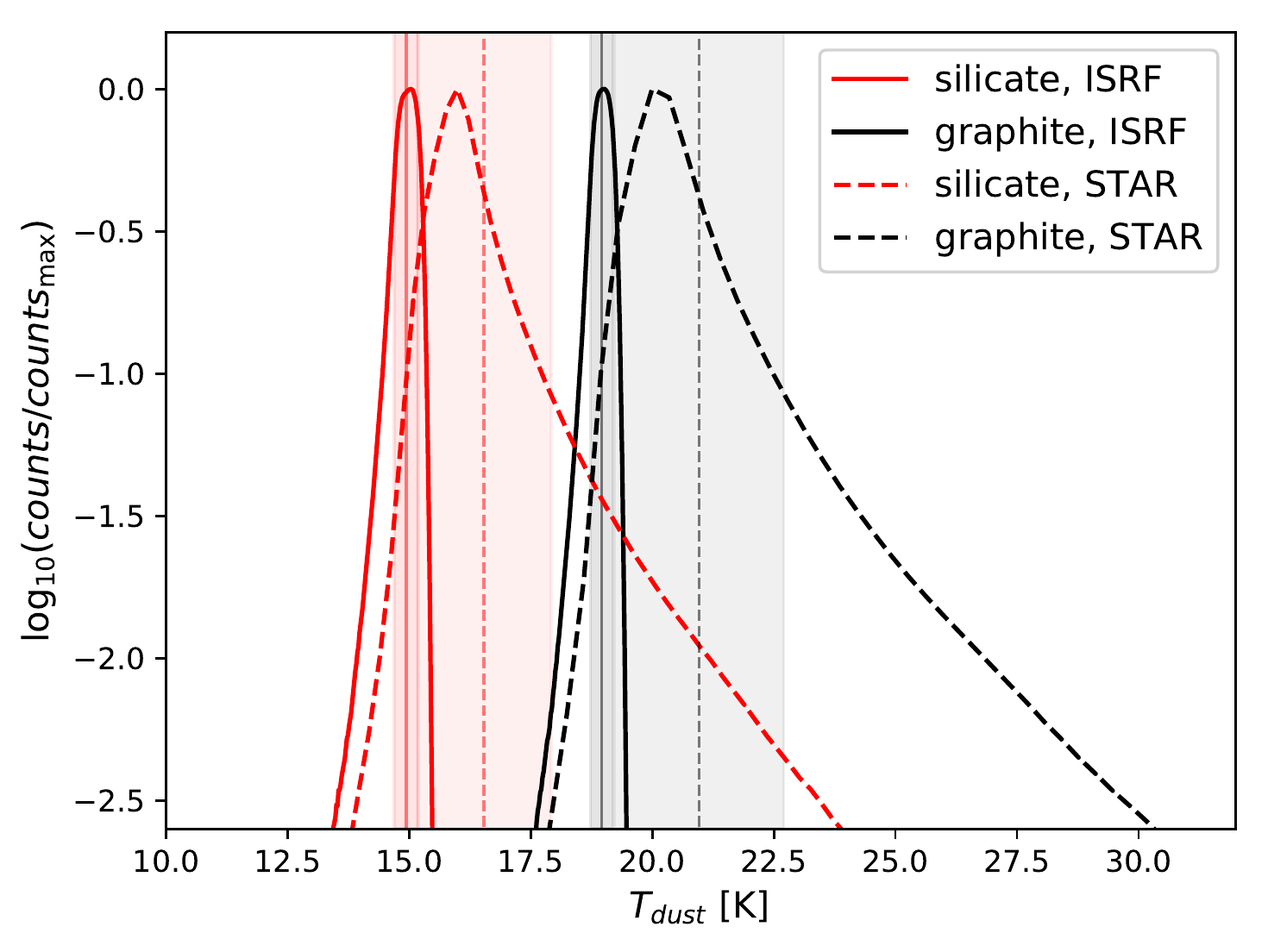}
      \includegraphics[width=.32\textwidth]{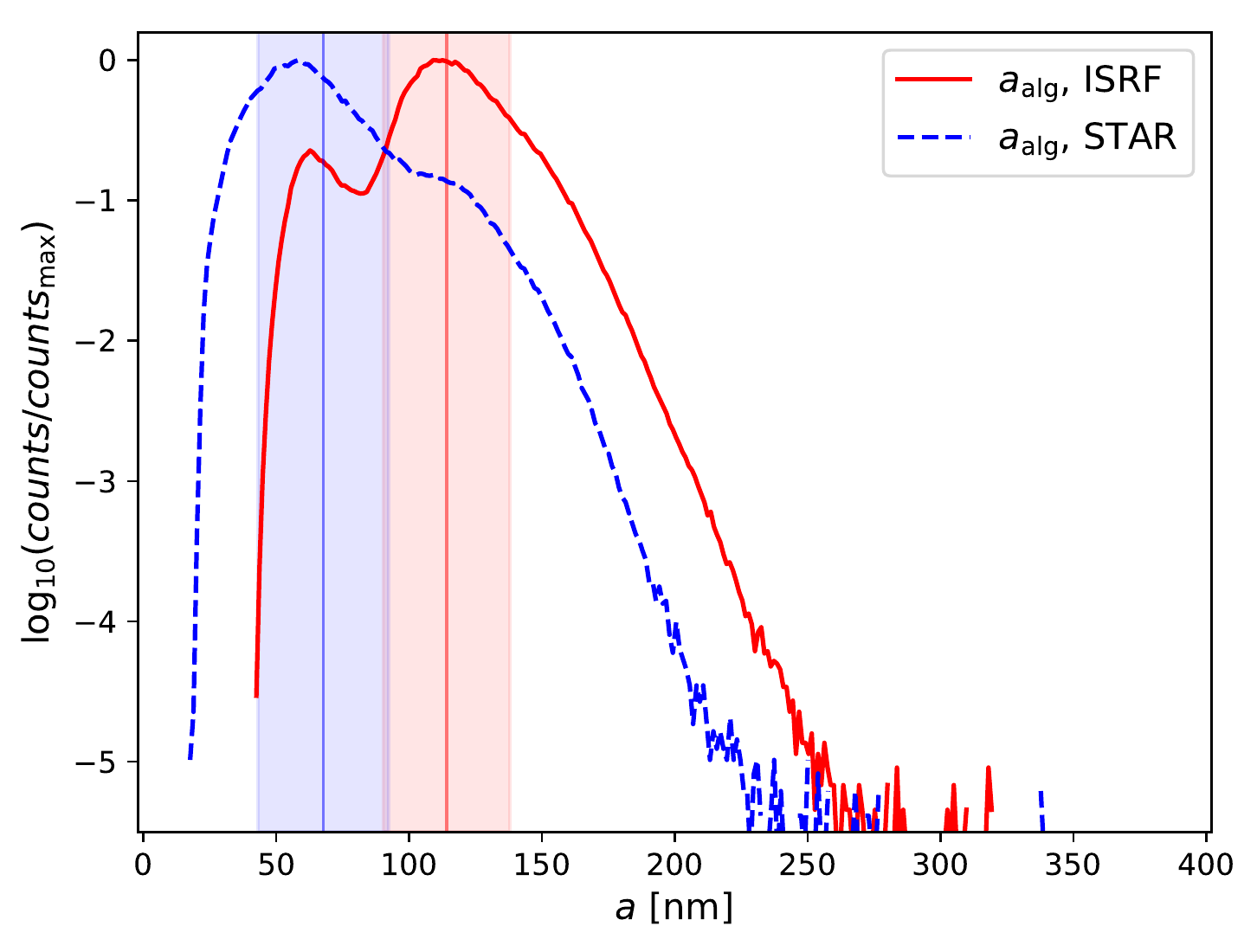}
      \includegraphics[width=.32\textwidth]{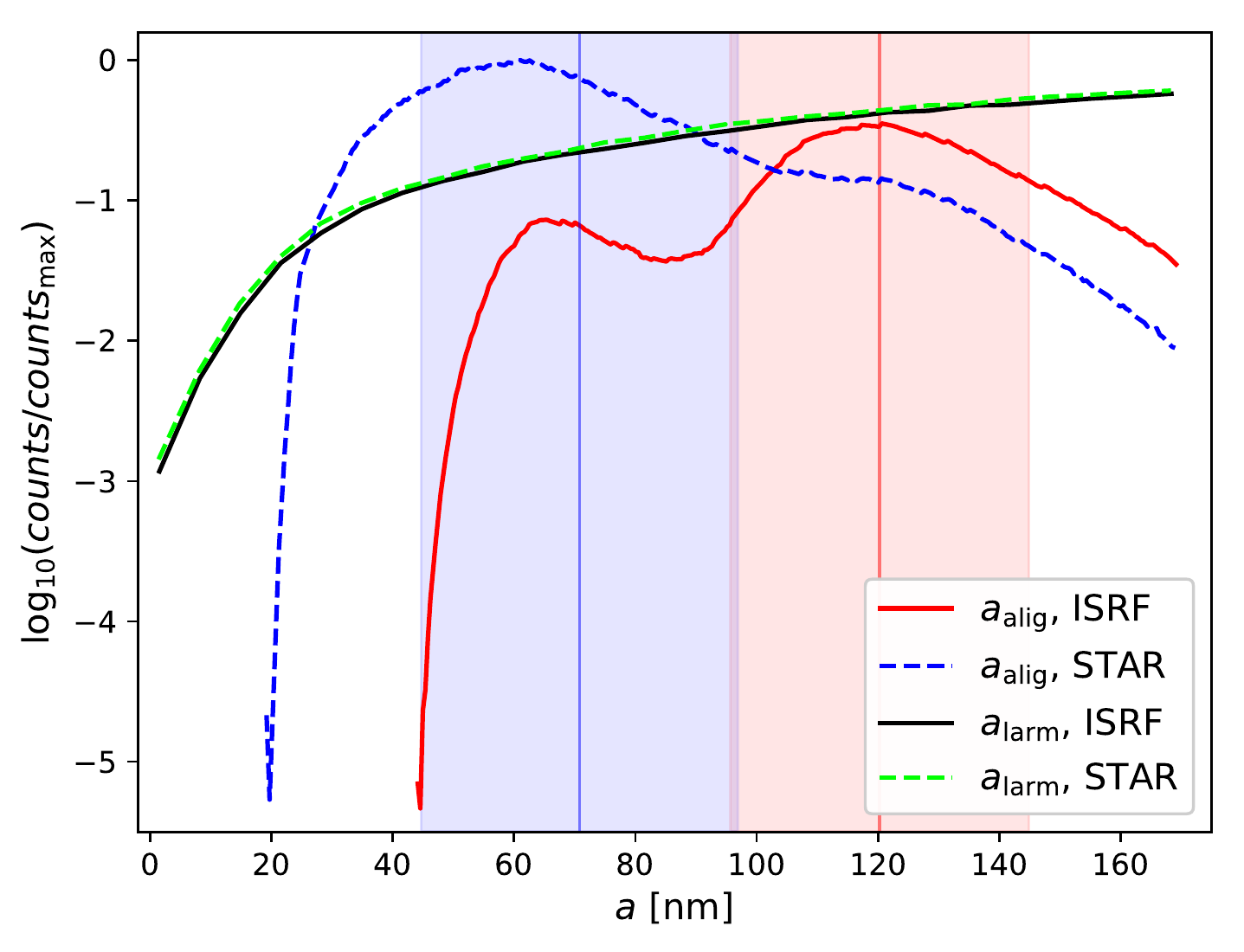}
\end{minipage}
\caption{The same as Figure \ref{fig:3DHistRad}. Left panel: Individual dust temperatures $\Tdust$ of silicate and graphite grains for the setups ISRF and STAR, respectively. Center panel: The alignment radius $\aalig$ for silicate grains. The Larmor limit is not plotted because $\alarm \gg 400\ \mathrm{nm}$. Right panel: Distribution of the radii $\aalig$ and $\alarm$ for graphite grains.}
\label{fig:3DHistTempRAT}
\end{figure*}
In Figure \ref{fig:3DHistTempRAT} we show the histograms of dust temperatures and grain alignment radii. For the ISRF case, the dust temperatures for silicate and graphite reach mean values of $15\ \mathrm{K}$ and $18\ \mathrm{K}$, respectively. For the STAR case we get mean temperatures of about $17\ \mathrm{K}$ for silicates and $18\ \mathrm{K}$ for graphite, while we report temperatures up to $118\ \mathrm{K}$ for a few grid cells in close proximity to the star.

For silicates, the mean values of the alignment radius are $\aalig=68\ \mathrm{nm}$ for the STAR setup and $\aalig=114\ \mathrm{nm}$ for the ISRF setup. Only a marginal amount of all grid cells  reaches the upper grain size of $a_{\rm amax}^{\rm S}=400\ \mathrm{nm}$. Furthermore, the Larmor limit (see Equation \ref{eq:TimeLarmor}) is $\alarm \gg a_{\rm amax}^{\rm S}$ and is therefore not shown in Figure \ref{fig:3DHistTempRAT}. Consequently,  the window $\aalig < a < \alarm$ of stable RAT alignment with the magnetic field direction (see also Equation \ref{eq:RayleighRAT}) falls within the size range of silicate grains. Hence, all cells of the MHD grid do contribute to dust polarization.

In this paper we assume that graphite grains are completely randomized, independently of local conditions. In Figure \ref{fig:3DHistTempRAT} we show also the radius $\aalig$ of graphite. For these grains, the mean values of the alignment radius are $\aalig=71\ \mathrm{nm}$ for the STAR setup and $\aalig=120\ \mathrm{nm}$ for the ISRF setup. However, we find the condition $\aalig<a_{\rm amax}^{\rm G}$ to be fulfilled for some rare cases in the MHD grid. Hence, a marginal amount of graphite grains can in principle spin-up to a stable alignment. Moreover, the Larmor limit $\alarm$ is of the same order as the graphite size range. Thus, we note that the condition $a < \alarm$ is fl{reached} for a small size range of the parameter set provided by the \RAMSES\ simulation. 


Indeed, we report that graphite grains may possibly align within a small range of grain sizes  for  about $0.7\ \%$ of all cells for the ISRF setup and $2.0\ \%$ for the STAR setup. For the ISRF an alignment with the magnetic field is in principle possible in some regions of the diffuse ISM. However, these regions are sparsely distributed over the entire grid and should not influence the polarization pattern in a detectable way. 
For the STAR setup, the regions of possible graphite alignment are clustered around the very position of the central star. Hence, graphite might also trace the magnetic field in close proximity to the star. A second possibility is the alignment of graphite with the radiation field. \cite{LazarianHoang2007} reported that graphite might align with the predominant direction of the radiation field for a distance  several $\mathrm{AU}$ away from the star. However, this distance is smaller than the resolution of the $\RAMSES$ simulation. Furthermore, charged dust grains may be randomized while drifting with respect to the magnetic field. This effect affects carbonaceous grainw more than silicate grains \citep[][]{Weingartner2006}. Overall, the assumption that graphite grains do not align at all remains justified within the scope of this paper.

\begin{acknowledgements}
We thank the anonymous referee for comments that helped to improve the paper. S.R. and R.S.K. acknowledge funding from the Deutsche Forschungsgemeinschaft (DFG, German Research Foundation) -- Project-ID 138713538 -- SFB 881 ``The Milky Way System'' (subprojects A06, B01, B02, and B08) and from the Priority Program SPP 1573 ``Physics of the Interstellar  Medium''  (grant  numbers  KL  1358/18.1,  KL  1358/19.2). S.R. and R.S.K. also acknowledge support from the DFG via the Heidelberg Cluster of Excellence {\em STRUCTURES} in the framework of Germany’s Excellence Strategy (grant EXC-2181/1 - 390900948).
F.B, F.L. and V.G. acknowledge support from the Agence Nationale de la Recherche (project BxB: ANR-17-CE31-0022). The authors thank T. Hoang for useful discussions on the physics of the grain spin-up process.
\end{acknowledgements}

\bibliographystyle{aa}
\bibliography{bibtex}

\begin{thebibliography}{105}
\expandafter\ifx\csname natexlab\endcsname\relax\def\natexlab#1{#1}\fi

\bibitem[{{Alves} {et~al.}(2014){Alves}, {Frau}, {Girart}, {Franco}, {Santos},
  \& {Wiesemeyer}}]{Alves2014}
{Alves}, F.~O., {Frau}, P., {Girart}, J.~M., {et~al.} 2014, \aap, 569, L1

\bibitem[{{Andersson} {et~al.}(2015){Andersson}, {Lazarian}, \&
  {Vaillancourt}}]{Andersson2015}
{Andersson}, B.~G., {Lazarian}, A., \& {Vaillancourt}, J.~E. 2015, \araa, 53,
  501

\bibitem[{{Andersson} {et~al.}(2011){Andersson}, {Pintado}, {Potter},
  {Strai{\v{z}}ys}, \& {Charcos-Llorens}}]{Andersson2011}
{Andersson}, B.~G., {Pintado}, O., {Potter}, S.~B., {Strai{\v{z}}ys}, V., \&
  {Charcos-Llorens}, M. 2011, \aap, 534, A19

\bibitem[{{Andersson} \& {Potter}(2007)}]{AnderssonPotter2007}
{Andersson}, B.~G. \& {Potter}, S.~B. 2007, \apj, 665, 369

\bibitem[{{Andersson} \& {Potter}(2010)}]{Andersson2010}
{Andersson}, B.~G. \& {Potter}, S.~B. 2010, \apj, 720, 1045

\bibitem[{{Barnett}(1917)}]{Barnett1915}
{Barnett}, S.~J. 1917, {Phys. Rev}

\bibitem[{{Bethell} {et~al.}(2007){Bethell}, {Chepurnov}, {Lazarian}, \&
  {Kim}}]{Bethell2007}
{Bethell}, T.~J., {Chepurnov}, A., {Lazarian}, A., \& {Kim}, J. 2007, \apj,
  663, 1055

\bibitem[{{Bjorkman} \& {Wood}(2001)}]{BjorkmanWood2001}
{Bjorkman}, J.~E. \& {Wood}, K. 2001, \apj, 554, 615

\bibitem[{{Brauer} {et~al.}(2017{\natexlab{a}}){Brauer}, {Wolf}, \&
  {Flock}}]{Brauer2017B}
{Brauer}, R., {Wolf}, S., \& {Flock}, M. 2017{\natexlab{a}}, \aap, 607, A104

\bibitem[{{Brauer} {et~al.}(2016){Brauer}, {Wolf}, \& {Reissl}}]{Brauer2016}
{Brauer}, R., {Wolf}, S., \& {Reissl}, S. 2016, \aap, 588, A129

\bibitem[{{Brauer} {et~al.}(2017{\natexlab{b}}){Brauer}, {Wolf}, {Reissl}, \&
  {Ober}}]{Brauer2017A}
{Brauer}, R., {Wolf}, S., {Reissl}, S., \& {Ober}, F. 2017{\natexlab{b}}, \aap,
  601, A90

\bibitem[{{Calvo} {et~al.}(2016){Calvo}, {Beno{\^\i}t}, {Catalano}, {Goupy},
  {Monfardini}, {Ponthieu}, {Barria}, {Bres}, {Grollier}, {Garde}, {Leggeri},
  {Pont}, {Triqueneaux}, {Adam}, {Bourrion}, {Mac{\'\i}as-P{\'e}rez}, {Rebolo},
  {Ritacco}, {Scordilis}, {Tourres}, {Adane}, {Coiffard}, {Leclercq},
  {D{\'e}sert}, {Doyle}, {Mauskopf}, {Tucker}, {Ade}, {Andr{\'e}}, {Beelen},
  {Belier}, {Bideaud}, {Billot}, {Comis}, {D'Addabbo}, {Kramer}, {Martino},
  {Mayet}, {Pajot}, {Pascale}, {Perotto}, {Rev{\'e}ret}, {Ritacco},
  {Rodriguez}, {Savini}, {Schuster}, {Sievers}, \& {Zylka}}]{Calvo2016}
{Calvo}, M., {Beno{\^\i}t}, A., {Catalano}, A., {et~al.} 2016, Journal of Low
  Temperature Physics, 184, 816

\bibitem[{{Chapman} {et~al.}(2011){Chapman}, {Goldsmith}, {Pineda}, {Clemens},
  {Li}, \& {Kr{\v{c}}o}}]{Chapman2011}
{Chapman}, N.~L., {Goldsmith}, P.~F., {Pineda}, J.~L., {et~al.} 2011, \apj,
  741, 21

\bibitem[{{Costantini} {et~al.}(2005){Costantini}, {Freyberg}, \&
  {Predehl}}]{Costantini2005}
{Costantini}, E., {Freyberg}, M.~J., \& {Predehl}, P. 2005, \aap, 444, 187

\bibitem[{{Das} \& {Weingartner}(2016)}]{Das2016}
{Das}, I. \& {Weingartner}, J.~C. 2016, \mnras, 457, 1958

\bibitem[{{Davis} \& {Greenstein}(1951)}]{Davis1951}
{Davis}, Jr., L. \& {Greenstein}, J.~L. 1951, ApJ, 114, 206

\bibitem[{{Davoisne} {et~al.}(2006){Davoisne}, {Djouadi}, {Leroux},
  {D'Hendecourt}, {Jones}, \& {Deboffle}}]{Davoisne2006}
{Davoisne}, C., {Djouadi}, Z., {Leroux}, H., {et~al.} 2006, \aap, 448, L1

\bibitem[{{Demyk} {et~al.}(2017){Demyk}, {Meny}, {Leroux}, {Depecker},
  {Brubach}, {Roy}, {Nayral}, {Ojo}, \& {Delpech}}]{Demyk2017}
{Demyk}, K., {Meny}, C., {Leroux}, H., {et~al.} 2017, \aap, 606, A50

\bibitem[{{Dobbs} {et~al.}(2006){Dobbs}, {Bonnell}, \& {Pringle}}]{Dobbs_2006}
{Dobbs}, C.~L., {Bonnell}, I.~A., \& {Pringle}, J.~E. 2006, \mnras, 371, 1663

\bibitem[{{Dolginov} \& {Mitrofanov}(1976)}]{Dolginov1976}
{Dolginov}, A.~Z. \& {Mitrofanov}, I.~G. 1976, \apss, 43, 291

\bibitem[{{Dowell} {et~al.}(2010){Dowell}, {Cook}, {Harper}, {Lin}, {Looney},
  {Novak}, {Stephens}, {Berthoud}, {Chuss}, {Crutcher}, {Dotson}, {Hildebrand
  }, {Houde}, {Jones}, {Krejny}, {Lazarian}, {Moseley}, {Tassis},
  {Vaillancourt}, \& {Werner}}]{Dowell2010}
{Dowell}, C.~D., {Cook}, B.~T., {Harper}, D.~A., {et~al.} 2010, in Society of
  Photo-Optical Instrumentation Engineers (SPIE) Conference Series, Vol. 7735,
  \procspie, 77356H

\bibitem[{{Draine} \& {Flatau}(2013)}]{DraineFlatau2013}
{Draine}, B.~T. \& {Flatau}, P.~J. 2013, ArXiv e-prints

\bibitem[{{Draine} \& {Fraisse}(2009)}]{DraineFraisse2009}
{Draine}, B.~T. \& {Fraisse}, A.~A. 2009, \apj, 696, 1

\bibitem[{{Draine} \& {Lazarian}(1998)}]{DraineLazarian1998}
{Draine}, B.~T. \& {Lazarian}, A. 1998, \apj, 508, 157

\bibitem[{{Draine} \& {Weingartner}(1996)}]{Draine_Weingartner1996}
{Draine}, B.~T. \& {Weingartner}, J.~C. 1996, \apj, 470, 551

\bibitem[{{Draine} \& {Weingartner}(1997)}]{DraineWeingartner1997}
{Draine}, B.~T. \& {Weingartner}, J.~C. 1997, \apj, 480, 633

\bibitem[{{Fromang} {et~al.}(2006){Fromang}, {Hennebelle}, \&
  {Teyssier}}]{Fromang2006}
{Fromang}, S., {Hennebelle}, P., \& {Teyssier}, R. 2006, \aap, 457, 371

\bibitem[{{Gold}(1952{\natexlab{a}})}]{Gold1952B}
{Gold}, T. 1952{\natexlab{a}}, \nat, 169, 322

\bibitem[{{Gold}(1952{\natexlab{b}})}]{Gold1952A}
{Gold}, T. 1952{\natexlab{b}}, \mnras, 112, 215

\bibitem[{{Greenberg}(1968)}]{Greenberg1968}
{Greenberg}, J.~M. 1968, {Interstellar Grains}, ed. B.~M. {Middlehurst} \&
  L.~H. {Aller} (the University of Chicago Press), 221

\bibitem[{{Guillet} {et~al.}(2018){Guillet}, {Fanciullo}, {Verstraete},
  {Boulanger}, {Jones}, {Miville-Desch{\^e}nes}, {Ysard}, {Levrier}, \&
  {Alves}}]{Guillet2018}
{Guillet}, V., {Fanciullo}, L., {Verstraete}, L., {et~al.} 2018, \aap, 610, A16

\bibitem[{{Habing}(1968)}]{Habing1968}
{Habing}, H.~J. 1968, \bain, 19, 421

\bibitem[{{Habing} {et~al.}(1994){Habing}, {Tignon}, \& {Tielens}}]{Habing1994}
{Habing}, H.~J., {Tignon}, J., \& {Tielens}, A.~G.~G.~M. 1994, \aap, 286, 523

\bibitem[{{Hall}(1949)}]{Hall1949}
{Hall}, J.~S. 1949, Science, 109, 166

\bibitem[{{Harper} {et~al.}(2018){Harper}, {Runyan}, {Dowell}, {Wirth},
  {Amato}, {Ames}, {Amiri}, {Banks}, {Bartels}, {Benford}, {Berthoud},
  {Buchanan}, {Casey}, {Chapman}, {Chuss}, {Cook}, {Derro}, {Dotson}, {Evans},
  {Fixsen}, {Gatley}, {Guerra}, {Halpern}, {Hamilton}, {Hamlin}, {Hansen},
  {Heimsath}, {Hermida}, {Hilton}, {Hirsch}, {Hollister}, {Hostetter}, {Irwin},
  {Jhabvala}, {Jhabvala}, {Kastner}, {Kov{\'a}cs}, {Lin}, {Loewenstein},
  {Looney}, {Lopez-Rodriguez}, {Maher}, {Michail}, {Miller}, {Moseley},
  {Novak}, {Pernic}, {Rennick}, {Rhody}, {Sandberg}, {Sand ford}, {Santos},
  {Shafer}, {Sharp}, {Shirron}, {Siah}, {Silverberg}, {Sparr}, {Spotz},
  {Staguhn}, {Toorian}, {Towey}, {Tuttle}, {Vaillancourt}, {Voellmer},
  {Volpert}, {Wang}, \& {Wollack}}]{Harper2018}
{Harper}, D.~A., {Runyan}, M.~C., {Dowell}, C.~D., {et~al.} 2018, Journal of
  Astronomical Instrumentation, 7, 1840008

\bibitem[{{Hennebelle} {et~al.}(2008){Hennebelle}, {Banerjee},
  {V{\'a}zquez-Semadeni}, {Klessen}, \& {Audit}}]{hennebelle_08}
{Hennebelle}, P., {Banerjee}, R., {V{\'a}zquez-Semadeni}, E., {Klessen}, R.~S.,
  \& {Audit}, E. 2008, \aap, 486, L43

\bibitem[{{Hennebelle} \& {Falgarone}(2012)}]{Hennebelle-Falgarone-2012}
{Hennebelle}, P. \& {Falgarone}, E. 2012, \aapr, 20, 55

\bibitem[{{Herranen} {et~al.}(2019){Herranen}, {Lazarian}, \&
  {Hoang}}]{Herranen2019}
{Herranen}, J., {Lazarian}, A., \& {Hoang}, T. 2019, \apj, 878, 96

\bibitem[{{Hildebrand} {et~al.}(2009){Hildebrand}, {Kirby}, {Dotson}, {Houde},
  \& {Vaillancourt}}]{Hildebrand2009}
{Hildebrand}, R.~H., {Kirby}, L., {Dotson}, J.~L., {Houde}, M., \&
  {Vaillancourt}, J.~E. 2009, \apj, 696, 567

\bibitem[{{Hiltner}(1949)}]{Hiltner1949}
{Hiltner}, W.~A. 1949, Science, 109, 165

\bibitem[{{Hoang} {et~al.}(2018){Hoang}, {Cho}, \& {Lazarian}}]{Hoang2018A}
{Hoang}, T., {Cho}, J., \& {Lazarian}, A. 2018, \apj, 852, 129

\bibitem[{{Hoang} \& {Lazarian}(2008)}]{Hoang2008}
{Hoang}, T. \& {Lazarian}, A. 2008, \mnras, 388, 117

\bibitem[{{Hoang} \& {Lazarian}(2014)}]{Hoang2014}
{Hoang}, T. \& {Lazarian}, A. 2014, \mnras, 438, 680

\bibitem[{{Hoang} \& {Lazarian}(2016)}]{HoangLazarian2016}
{Hoang}, T. \& {Lazarian}, A. 2016, \apj, 831, 159

\bibitem[{{Hoang} {et~al.}(2014){Hoang}, {Lazarian}, \& {Martin}}]{Hoang2014A}
{Hoang}, T., {Lazarian}, A., \& {Martin}, P.~G. 2014, \apj, 790, 6

\bibitem[{{Holland} {et~al.}(2013){Holland}, {Bintley}, {Chapin},
  {Chrysostomou}, {Davis}, {Dempsey}, {Duncan}, {Fich}, {Friberg}, {Halpern},
  {Irwin}, {Jenness}, {Kelly}, {MacIntosh}, {Robson}, {Scott}, {Ade},
  {Atad-Ettedgui}, {Berry}, {Craig}, {Gao}, {Gibb}, {Hilton}, {Hollister},
  {Kycia}, {Lunney}, {McGregor}, {Montgomery}, {Parkes}, {Tilanus}, {Ullom},
  {Walther}, {Walton}, {Woodcraft}, {Amiri}, {Atkinson}, {Burger}, {Chuter},
  {Coulson}, {Doriese}, {Dunare}, {Economou}, {Niemack}, {Parsons},
  {Reintsema}, {Sibthorpe}, {Smail}, {Sudiwala}, \& {Thomas}}]{Holland2013}
{Holland}, W.~S., {Bintley}, D., {Chapin}, E.~L., {et~al.} 2013, \mnras, 430,
  2513

\bibitem[{Hunt {et~al.}(1995)Hunt, Moskowitz, \& Banerjee}]{Hunt1995}
Hunt, C.~P., Moskowitz, B., \& Banerjee, S. 1995, {Magnetic properties of Rocks
  and Minerals, In: Rock Physics and Phase Relations -A handbook of Physical
  constants} (AGU Ref. Shelf 3. Washington, ed. Ahrens, T.J), 189--204

\bibitem[{{Jones} \& {Spitzer}(1967)}]{JonesSpitzer1967}
{Jones}, R.~V. \& {Spitzer}, Lyman, J. 1967, \apj, 147, 943

\bibitem[{{Jones}(1989)}]{Jones1989}
{Jones}, T.~J. 1989, \apj, 346, 728

\bibitem[{{Jones} {et~al.}(2015){Jones}, {Bagley}, {Krejny}, {Andersson}, \&
  {Bastien}}]{Jones2015}
{Jones}, T.~J., {Bagley}, M., {Krejny}, M., {Andersson}, B.~G., \& {Bastien},
  P. 2015, \aj, 149, 31

\bibitem[{{Kandori} {et~al.}(2018){Kandori}, {Tamura}, {Nagata}, {Tomisaka},
  {Kusakabe}, {Nakajima}, {Kwon}, {Nagayama}, \& {Tatematsu}}]{Kandori2018}
{Kandori}, R., {Tamura}, M., {Nagata}, T., {et~al.} 2018, \apj, 857, 100

\bibitem[{{Kandori} {et~al.}(2020){Kandori}, {Tamura}, {Saito}, {Tomisaka},
  {Matsumoto}, {Kusakabe}, {Kwon}, {Nagayama}, {Nagata}, {Tazaki}, \&
  {Tatematsu}}]{Kandori2020}
{Kandori}, R., {Tamura}, M., {Saito}, M., {et~al.} 2020, \pasj, 72, 8

\bibitem[{{Kim} \& {Martin}(1995)}]{KimMartin1995}
{Kim}, S.-H. \& {Martin}, P.~G. 1995, \apj, 444, 293

\bibitem[{{Lazarian}(1995)}]{Lazarian1995MNRASL}
{Lazarian}, A. 1995, \mnras, 277, 1235

\bibitem[{{Lazarian}(1997)}]{Lazarian1997CL}
{Lazarian}, A. 1997, \apj, 483, 296

\bibitem[{{Lazarian} {et~al.}(2015){Lazarian}, {Andersson}, \&
  {Hoang}}]{Lazarian2015}
{Lazarian}, A., {Andersson}, B.~G., \& {Hoang}, T. 2015, {Grain alignment: Role
  of radiative torques and paramagnetic relaxation}, 81

\bibitem[{{Lazarian} \& {Hoang}(2007{\natexlab{a}})}]{LazarianHoang2007}
{Lazarian}, A. \& {Hoang}, T. 2007{\natexlab{a}}, \mnras, 378, 910

\bibitem[{{Lazarian} \& {Hoang}(2007{\natexlab{b}})}]{Lazarian2007C}
{Lazarian}, A. \& {Hoang}, T. 2007{\natexlab{b}}, \apjl, 669, L77

\bibitem[{{Lazarian} \& {Hoang}(2008)}]{Lazarian2008}
{Lazarian}, A. \& {Hoang}, T. 2008, \apjl, 676, L25

\bibitem[{{Lazarian} \& {Hoang}(2018)}]{LH18}
{Lazarian}, A. \& {Hoang}, T. 2018, arXiv e-prints

\bibitem[{{Lee} \& {Draine}(1985)}]{Lee1985}
{Lee}, H.~M. \& {Draine}, B.~T. 1985, \apj, 290, 211

\bibitem[{{Lopez-Rodriguez} {et~al.}(2019){Lopez-Rodriguez}, {Dowell}, {Jones},
  {Harper}, {Berthoud}, {Chuss}, {Dale}, {Guerra}, {Hamilton}, {Looney},
  {Michail}, {Nikutta}, {Novak}, {Santos}, {Sheth}, {Siah}, {Staguhn},
  {Stephens}, {Tassis}, {Trinh}, {Ward-Thompson}, {Werner}, {Wollack}, \&
  {Zweibel}}]{Lopez-Rodriguez2019}
{Lopez-Rodriguez}, E., {Dowell}, C.~D., {Jones}, T.~J., {et~al.} 2019, arXiv
  e-prints, arXiv:1907.06648

\bibitem[{{Lucy}(1999)}]{Lucy1999}
{Lucy}, L.~B. 1999, \aap, 344, 282

\bibitem[{Marsaglia(2003)}]{Marsaglia2003}
Marsaglia, G. 2003, JMASM, 2, 2

\bibitem[{Marsaglia \& Zaman(1993)}]{Marsaglia1993}
Marsaglia, G. \& Zaman, A. 1993, The {KISS} generator, Technical report,
  Department of Statistics, Florida State University, Tallahassee, FL, USA

\bibitem[{{Martin}(1974)}]{Martin1974}
{Martin}, P.~G. 1974, \apj, 187, 461

\bibitem[{{Mathis}(1986)}]{Mathis1986}
{Mathis}, J.~S. 1986, \apj, 308, 281

\bibitem[{{Mathis}(1990)}]{Mathis1990}
{Mathis}, J.~S. 1990, \araa, 28, 37

\bibitem[{{Mathis} {et~al.}(1983){Mathis}, {Mezger}, \& {Panagia}}]{Mathis1983}
{Mathis}, J.~S., {Mezger}, P.~G., \& {Panagia}, N. 1983, \aap, 128, 212

\bibitem[{{Mathis, Rumpl, \& Nordsieck}(1977)}]{Mathis1977}
{Mathis, Rumpl, \& Nordsieck}. 1977, ApJ, 217, 425

\bibitem[{{Mezger} {et~al.}(1982){Mezger}, {Mathis}, \& {Panagia}}]{Mezger1982}
{Mezger}, P.~G., {Mathis}, J.~S., \& {Panagia}, N. 1982, \aap, 105, 372

\bibitem[{{Mishchenko}(1991)}]{Mishchenko1991}
{Mishchenko}, M.~I. 1991, \apj, 367, 561

\bibitem[{{Monfardini} {et~al.}(2014){Monfardini}, {Adam}, {Adane}, {Ade},
  {Andr{\'e}}, {Beelen}, {Belier}, {Benoit}, {Bideaud}, {Billot}, {Bourrion},
  {Calvo}, {Catalano}, {Coiffard}, {Comis}, {D'Addabbo}, {D{\'e}sert}, {Doyle},
  {Goupy}, {Kramer}, {Leclercq}, {Macias-Perez}, {Martino}, {Mauskopf},
  {Mayet}, {Pajot}, {Pascale}, {Ponthieu}, {Rev{\'e}ret}, {Rodriguez},
  {Savini}, {Schuster}, {Sievers}, {Tucker}, \& {Zylka}}]{Monfardini2014}
{Monfardini}, A., {Adam}, R., {Adane}, A., {et~al.} 2014, Journal of Low
  Temperature Physics, 176, 787

\bibitem[{{Monfardini} {et~al.}(2011){Monfardini}, {Benoit}, {Bideaud},
  {Swenson}, {Cruciani}, {Camus}, {Hoffmann}, {D{\'e}sert}, {Doyle}, {Ade},
  {Mauskopf}, {Tucker}, {Roesch}, {Leclercq}, {Schuster}, {Endo}, {Baryshev},
  {Baselmans}, {Ferrari}, {Yates}, {Bourrion}, {Macias-Perez}, {Vescovi},
  {Calvo}, \& {Giordano}}]{Monfardini2011}
{Monfardini}, A., {Benoit}, A., {Bideaud}, A., {et~al.} 2011, \apjs, 194, 24

\bibitem[{{Panopoulou} {et~al.}(2019){Panopoulou}, {Hensley}, {Skalidis},
  {Blinov}, \& {Tassis}}]{Panopoulou2019}
{Panopoulou}, G.~V., {Hensley}, B.~S., {Skalidis}, R., {Blinov}, D., \&
  {Tassis}, K. 2019, \aap, 624, L8

\bibitem[{{Pelkonen} {et~al.}(2009){Pelkonen}, {Juvela}, \&
  {Padoan}}]{Pelkonen2009}
{Pelkonen}, V.~M., {Juvela}, M., \& {Padoan}, P. 2009, \aap, 502, 833

\bibitem[{{Pellegrini} {et~al.}(2019){Pellegrini}, {Reissl}, {Rahner},
  {Klessen}, {Glover}, {Pakmor}, {Herrera-Camus}, \& {Grand}}]{Pellegrini2019}
{Pellegrini}, E.~W., {Reissl}, S., {Rahner}, D., {et~al.} 2019, arXiv e-prints,
  arXiv:1905.04158

\bibitem[{{Pereyra} \& {Magalh{\~a}es}(2004)}]{Pereyra2004}
{Pereyra}, A. \& {Magalh{\~a}es}, A.~M. 2004, \apj, 603, 584

\bibitem[{{Planck Collaboration}(2016)}]{Planck2016XXXV}
{Planck Collaboration}. 2016, \aap, 586, A138

\bibitem[{{Planck Collaboration XII}(2018)}]{Planck2018XII}
{Planck Collaboration XII}. 2018, arXiv e-prints

\bibitem[{{Planck Collaboration XIX}(2015)}]{PlanckXIX2015}
{Planck Collaboration XIX}. 2015, \aap, 576, A104

\bibitem[{{Planck Collaboration XX}(2015)}]{Planck2015XX}
{Planck Collaboration XX}. 2015, \aap, 576, A105

\bibitem[{{Purcell}(1975)}]{Purcell1975}
{Purcell}, E.~M. 1975, {Interstellar grains as pinwheels.}, ed. G.~B. {Field}
  \& A.~G.~W. {Cameron}, 155--167

\bibitem[{{Purcell}(1979)}]{Purcell1979}
{Purcell}, E.~M. 1979, \apj, 231, 404

\bibitem[{{Reissl} {et~al.}(2019){Reissl}, {Brauer}, {Klessen}, \&
  {Pellegrini}}]{Reissl2019}
{Reissl}, S., {Brauer}, R., {Klessen}, R.~S., \& {Pellegrini}, E.~W. 2019,
  \apj, 885, 15

\bibitem[{{Reissl} {et~al.}(2018){Reissl}, {Stutz}, {Brauer}, {Pellegrini},
  {Schleicher}, \& {Klessen}}]{Reissl2018}
{Reissl}, S., {Stutz}, A.~M., {Brauer}, R., {et~al.} 2018, \mnras, 481, 2507

\bibitem[{{Reissl} {et~al.}(2016){Reissl}, {Wolf}, \& {Brauer}}]{Reissl2016}
{Reissl}, S., {Wolf}, S., \& {Brauer}, R. 2016, \aap, 593, A87

\bibitem[{{Roberge} \& {Lazarian}(1999)}]{Roberge1999}
{Roberge}, W.~G. \& {Lazarian}, A. 1999, \mnras, 305, 615

\bibitem[{{Sadavoy} {et~al.}(2018){Sadavoy}, {Myers}, {Stephens}, {Tobin},
  {Commer{\c{c}}on}, {Henning}, {Looney}, {Kwon}, {Segura-Cox}, \&
  {Harris}}]{Sadavoy2018}
{Sadavoy}, S.~I., {Myers}, P.~C., {Stephens}, I.~W., {et~al.} 2018, \apj, 859,
  165

\bibitem[{{Seifried} {et~al.}(2019){Seifried}, {Walch}, {Reissl}, \&
  {Ib{\'a}{\~n}ez-Mej{\'\i}a}}]{Seifried2019}
{Seifried}, D., {Walch}, S., {Reissl}, S., \& {Ib{\'a}{\~n}ez-Mej{\'\i}a},
  J.~C. 2019, \mnras, 482, 2697

\bibitem[{{Serkowski}(1958)}]{Serkowski1958}
{Serkowski}, K. 1958, \actaa, 8, 135

\bibitem[{{Serkowski} {et~al.}(1975){Serkowski}, {Mathewson}, \&
  {Ford}}]{Serkowski1975}
{Serkowski}, K., {Mathewson}, D.~S., \& {Ford}, V.~L. 1975, \apj, 196, 261

\bibitem[{{Siebenmorgen} {et~al.}(2014){Siebenmorgen}, {Voshchinnikov}, \&
  {Bagnulo}}]{Siebenmorgen2014}
{Siebenmorgen}, R., {Voshchinnikov}, N.~V., \& {Bagnulo}, S. 2014, \aap, 561,
  A82

\bibitem[{{Tazaki} {et~al.}(2017){Tazaki}, {Lazarian}, \&
  {Nomura}}]{Tazaki2017}
{Tazaki}, R., {Lazarian}, A., \& {Nomura}, H. 2017, \apj, 839, 56

\bibitem[{{Teyssier}(2002)}]{Teyssier2002}
{Teyssier}, R. 2002, \aap, 385, 337

\bibitem[{{Vaillancourt} \& {Andersson}(2015)}]{VA15}
{Vaillancourt}, J.~E. \& {Andersson}, B.~G. 2015, \apj, 812, L7

\bibitem[{{Vaillancourt} \& {Matthews}(2012)}]{Vaillancourt2012}
{Vaillancourt}, J.~E. \& {Matthews}, B.~C. 2012, \apjs, 201, 13

\bibitem[{{Voshchinnikov} {et~al.}(2016){Voshchinnikov}, {Il'in}, \&
  {Das}}]{Vosh2016}
{Voshchinnikov}, N.~V., {Il'in}, V.~B., \& {Das}, H.~K. 2016, \mnras, 462, 2343

\bibitem[{{Weingartner}(2006)}]{Weingartner2006}
{Weingartner}, J.~C. 2006, \apj, 647, 390

\bibitem[{Weingartner \& Draine(2001)}]{Weingartner2001}
Weingartner, J.~C. \& Draine, B.~T. 2001, The Astrophysical Journal, 548, 296

\bibitem[{{Whitney} \& {Wolff}(2002)}]{Whitney2002}
{Whitney}, B.~A. \& {Wolff}, M.~J. 2002, \apj, 574, 205

\bibitem[{{Whittet} {et~al.}(2008){Whittet}, {Hough}, {Lazarian}, \&
  {Hoang}}]{Whittet2008}
{Whittet}, D.~C.~B., {Hough}, J.~H., {Lazarian}, A., \& {Hoang}, T. 2008, \apj,
  674, 304

\bibitem[{{Whittet} {et~al.}(1992){Whittet}, {Martin}, {Hough}, {Rouse},
  {Bailey}, \& {Axon}}]{Whittet1992}
{Whittet}, D.~C.~B., {Martin}, P.~G., {Hough}, J.~H., {et~al.} 1992, \apj, 386,
  562

\bibitem[{{Xu} \& {Zhang}(2016)}]{XuZhang2016}
{Xu}, S. \& {Zhang}, B. 2016, \apj, 824, 113

\bibitem[{{Yan} \& {Lazarian}(2003)}]{Yan2003}
{Yan}, H. \& {Lazarian}, A. 2003, \apjl, 592, L33

\end{thebibliography}
\end{document}